\documentclass[10pt,preprint]{aastex}
\usepackage{lscape}
\newcommand{\cnjwl}{$cn_{\rm JWL}$}

\newcommand{\caby}{$Ca\ by$}
\newcommand{\cacn}{$Ca$-CN}
\newcommand{\cas}{$Ca$-s}
\newcommand{\caw}{$Ca$-w}
\newcommand{\cnw}{CN-w}
\newcommand{\cns}{CN-s}
\newcommand{\vvhb}{$V - V_{\rm HB}$}
\newcommand{\hsthb}{F606W $-$ F606W$_{\rm HB}$}
\newcommand{\vhb}{$V_{\rm HB}$}
\newcommand{\vbump}{$V_{\rm bump}$}
\newcommand{\dvbump}{$\Delta V_{\rm bump}$}
\newcommand{\str}{Str\"omgren}
\newcommand{\kms}{km s$^{-1}$}
\newcommand{\dra}{$\Delta\alpha$}
\newcommand{\ddec}{$\Delta\delta$}
\newcommand{\nrgb}{$n$(\cnw):$n$(\cns)}
\newcommand{\nlardo}{$n$($UV$-blue):$n$($UV$-red)}
\newcommand{\njwl}{$n$($UV$-blue$_{\rm JWL}$):$n$($UV$-red$_{\rm JWL}$)}

\newcommand{\vby}{$(b-y)$ versus $V$}

\newcommand{\vcn}{\cnjwl\ versus $V$}

\newcommand{\hst}{{\it HST}}
\newcommand{\acs}{{\it ACS}}
\newcommand{\vlt}{{\it VLT}}
\newcommand{\sdss}{{\it SDSS}}
\newcommand{\mt}{{\it magic-trio}}
\newcommand{\cnwave}{$\lambda$ 3883 \AA}
\newcommand{\scn}{$S$(3839)}
\newcommand{\ds}{$\delta S$(3839)}
\newcommand{\dy}{$\Delta Y$}
\newcommand{\dcy}{$\delta cy$}
\newcommand{\dmo}{$\delta m1$}
\newcommand{\dug}{$\Delta^\prime_{u-g}$}

\newcommand{\dtrio}{$\Delta_{\rm C~F275W,F336W,F438W}$}
\newcommand{\vexcess}{$\Delta V_{\Sigma \rm nei}$}
\newcommand{\agbm}{AGB-manqu\'e}
\newcommand{\sigo}{$\sigma$[O]}
\newcommand{\signa}{$\sigma$[Na]}

\begin{document}

\title{MULTIPLE STELLAR POPULATIONS OF GLOBULAR CLUSTERS FROM 
HOMOGENEOUS \cacn\ PHOTOMETRY. 
II. M5 (NGC 5904) AND A NEW FILTER SYSTEM.\altaffilmark{1}}

\author{Jae-Woo Lee\altaffilmark{2, 3}}

\altaffiltext{1}{Based on observations made 
with the Cerro Tololo Inter-American Observatory (CTIO) 1m telescope, 
which is operated by the SMARTS consortium.}
\altaffiltext{2}{Department of Physics and Astronomy,
Sejong University, 209 Neungdong-ro, Gwangjin-Gu, Seoul, 05006, Korea;
jaewoolee@sejong.ac.kr, jaewoolee@sejong.edu}
\altaffiltext{3}{Visiting Astronomer, CTIO, 
National Optical Astronomy Observatories (NOAO), operated by the Association
of Universities for Research in Astronomy, Inc., under cooperative agreement
with the National Science Foundation.}

\begin{abstract}
Using our ingeniously designed new filter systems,
we investigate the multiple stellar populations (MSPs) 
of the red giant branch (RGB) and the asymptotic giant branch (AGB)
in the globular cluster (GC) M5.
Our results are the following.
(1) Our \cnjwl\ index accurately traces the nitrogen abundances in M5,
while other color indices fail to do so.
(2) We find bimodal CN distributions both in the RGB and the AGB sequences, 
with the number ratios between the CN-weak (\cnw) and the CN-strong (\cns) 
of \nrgb\ = 29:71 ($\pm$ 2) and 21:79 ($\pm$ 7), respectively.
(3) We also find the bimodal photometric [N/Fe] distribution for M5 RGB stars.
(4) Our \cnjwl-[O/Fe] and \cnjwl-[Na/Fe] relations show 
the clear discontinuities between the two RGB populations.
(5) Although small, the RGB bump of the \cns\ is slightly brighter,
\dvbump\ = 0.07 $\pm$ 0.04 mag. If real, the difference 
in the helium abundance becomes \dy\ = 0.028 $\pm$ 0.016,
in the sense that the \cns\ is more helium enhanced.
(6) Very similar radial but different spatial distributions with comparable
center positions are found for the two RGB populations. 
The \cns\ RGB and AGB stars are more elongated along the NW-SE direction.
(7) The \cns\ population shows a substantial net projected rotation,
while that of the \cnw\ population is nil.
(8) Our results confirm the deficiency of the \cnw\ AGB stars previously 
noted by others. We show that it is most likely due to the stochastic truncation
in the outer part of the cluster.
Finally, we discuss the formation scenario of M5.
\end{abstract}


\section{INTRODUCTION}
During the last decade, a drastic paradigm shift on the true nature
of the Galactic globular clusters (GCs) has emerged.
Almost all GCs exhibit variations in lighter elemental abundances,
for example bimodal CN distributions and Na-O anticorrelations 
\citep[e.g.,][]{smith87,carretta09}.
Understanding this ubiquitous nature of the multiple stellar populations (MSPs)
in GCs is one of the outstanding quest in the near field cosmology
\citep[e.g.,][]{jwlnat,lee15,piotto15,renzini15}.

The key feature of the MSPs in the peculiar GCs, 
like $\omega$ Cen and M22, is the discrete distributions 
in the heavy elemental abundances and these GCs are considered to be remnants 
of the dwarf galaxy related objects and accreted to the Milky Way later in time
\citep{ywlee99,jwlnat,lee15,lee16,johnson,marino11m22}.
On the other hand, the normal GCs exhibit the significant spread in 
the lighter elemental abundances, which is resulted from the proton capture
process at high temperature, most likely engraved during the multiple 
phases of star formation history\footnote{See \citet{lee10}
for the effect of the internal mixing in the RGB stars.} 
\citep[e.g.,][]{dercole08,carretta09}.

How normal GCs formed is still under debate.
By and large, four potential candidates of the source of the proton 
capture process at high temperature and helium  enhancement 
have been proposed, which are
(i) asymptotic giant (AGB) stars \citep{dercole10}; 
(ii) fast rotating massive stars \citep{decressin07};
(iii) interacting massive binaries \citep{demink09};  and
(iv) supermassive stars \citep{denissnkov14}.
Each candidate has pros and cons to explain the observational
lines of evidence in GCs.
In the context of the self-enrichment scenario for normal GCs,
\citet{renzini15} proposed that the AGB pollution scenario would be favored.
However, most of the proposed candidates cannot solve the so-called 
``mass-budget problem'' with satisfaction \cite[e.g., see][]{bl15}.
The GCs with MSPs formation scenario 
proposed by \citet{bastian13} can mitigate the mass-budget problem, 
but their model cannot explain the differences in the spatial
distributions and kinematics between the MSPs in M5, as we will discuss later.

The ultra-violet photometry from the {\it Hubble Space Telescope} (\hst) or 
the high-resolution spectroscopy using the {\it Very Large Telescopes}  (\vlt)
have long been the most reliable passages into the realm of the high precision 
MSP study of the Galactic or extra-Galactic 
GC systems \citep[e.g.][]{carretta09,piotto15}. 
The variations in the lighter elemental abundances, such as C, N and O,
greatly affect the UV regime through OH, NH, CN and CH molecular bands. 
These molecular bands are often very strong and the variations of lighter 
elemental abundances can be detected even with the wide-band photometry.
In this context, the photometric approach is still important for the sake
of completeness (in particular in the central part of GCs where
the spectroscopic method can not be applied) and easiness.\footnote{See 
\citet{lee16} for the non-trivial aspects of the LTE analysis of
high-resolution spectroscopy of GC RGB stars.}
Therefore, studies of MSPs in GCs through photometry and spectroscopy 
are complementary.

It is very unfortunate that only a few groups of astronomers around 
the globe are permitted to access these prestigious instruments, 
such as the \hst\ or the \vlt, for over decades, 
which made the competition unfair.
As a result of our decade long hard effort, 
we developed a new approach that small aperture telescopes 
empowered by ingeniously designed narrow-band filter systems can have 
a capability to measure not only the heavy but also the lighter elemental 
abundances of the RGB and AGB stars in GCs.
Our novel approach is very straightforward and easy to apply,
which can open a new era of prolific discoveries in the field of
the MSPs in our Galaxy with mere 1-m or sub 1-m class telescopes.
Also, our results presented here and in the future will complement 
the intrinsic weakness of the aforementioned instruments,
such as the \hst\ and the \vlt\ \citep{lee10,lee15,lee16}.

This is a part of the series of papers addressing the MSPs of 
Galactic GCs based on our homogeneous \cacn\ photometry.
In our previous study \citep{jwlnat,jwln1851,lee15}, 
we extensively demonstrated that the so-called extended \str\ photometry 
\citep{att91} can provide a very powerful means to study MSPs 
with heterogeneous metallicity.
In our current study, we explore the CN abundances of the RGB 
and the AGB stars in M5, one of the first archetype of the GC 
with the CN inhomogeneity \citep{osborn71},
using our new filter system, $JWL39$.

The outline of the paper is as follows.
We describe the Sejong \cacn\ survey program, 
our newly devised filter system, $JWL39$,
and new color index associated with it in Section 2.
We show observations and data reduction of M5 in Section 3.
Some very interesting results on the M5 RGB and AGB stars
will be presented in Section 4.
Finally, we address the summary and discussion in Section 5.
In appendix, we show the artificial star experiments to examine
the completeness of our photometry.
We also show a new strategy for the ground-based observations
by making use of the positional information from the \hst\ photometry,
which can significantly improve any of ground-based observations
of the very crowded region.

\section{A NEW CN FILTER SYSTEM: $JWL39$}
Sejong \cacn\ (formerly, Sejong \caby) survey program was launched in July, 2006.
As an official major partner of the SMARTS consortium, we acquired 
the guaranteed access to the small telescopes operated at the CTIO since 2006.
The summary of our survey program can be found in \citet{lee15}. 

Our survey program is consisted of three phases.
The Phase~I was from July, 2006 to November, 2010 \citep[see][]{jwlnat,jwln1851}.
During this period, we used \str\ $uvby$ and $Ca$ filters provided by the CTIO.
The CN band at \cnwave\ is often very strong in the RGB stars
and the CTIO $Ca$ filter was originally designed to avoid 
the CN band contamination.
However, it was suspected that the CTIO $Ca$ filter had undergone degradation
due to aging effect and the original transmission function had been altered
to the shorter wavelength.
As a result, the CTIO $Ca$ filter is suspected to suffer from 
the CN band contamination. Note that we do not use the $Ca$ photometry
from the Phase~I for our current work presented here.

\begin{figure}
\epsscale{1}
\figurenum{1}
\plotone{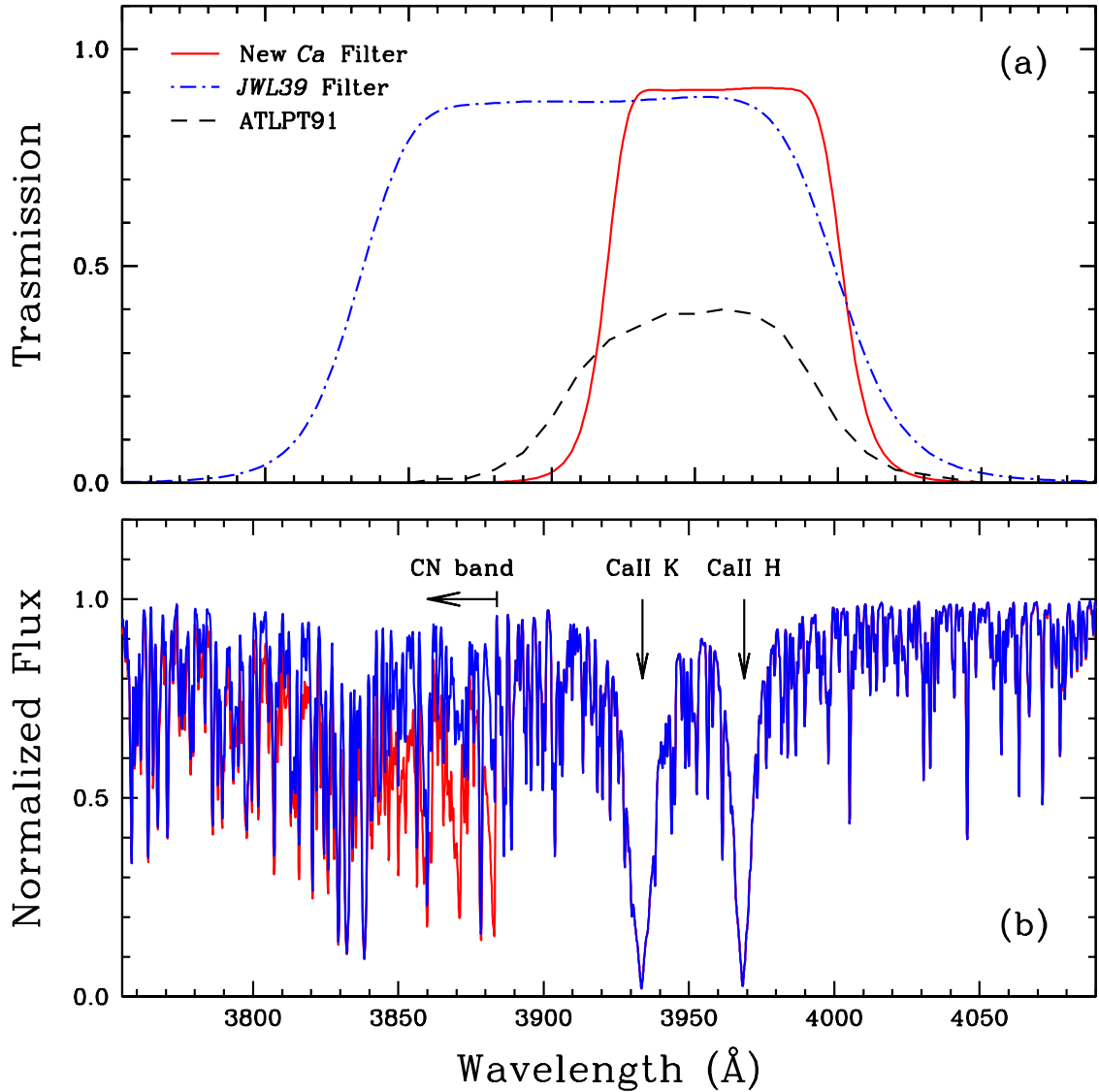}
\caption{(a) Filter transmission functions for our new $Ca$ and $JWL39$ filters,
measured with collimated beam.
The CN absorption band at \cnwave\ lies outside
of the lower boundary of our new $Ca$ filter.
Note that our new $Ca$ filter and that by \citet{att91} have similar FWHMs, 
approximately 90 \AA\, but our new $Ca$ filter has a more uniform and 
high transmission across the passband, dropping more rapidly at both edges.
Our new filter system, $JWL39$, is intended to measure CN band strength 
at \cnwave, in conjunction with our new $Ca$ filter.
(b) Normalized synthetic spectra for the \cnw\ (the blue line) and the \cns\
(the red line) RGB stars. 
}\label{fig:flt}
\end{figure}

Since March 2011 (Phase~II), we used our own \str\ $by$ and new $Ca$ filters, 
all of which were manufactured by Asahi Spectra, Japan.
Our new $Ca$ filter was carefully designed by the author of the paper
to have very similar filter bandwidth and pivot wavelength 
as those of F395N filter in the Wide-Field Camera 3 (WFC3) onboard the \hst.
In Figure~\ref{fig:flt}, we show a comparison of filter transmission functions 
between that in \citet{att91} and  that of our new $Ca$ filter 
measured with collimated beam by the manufacturer of the filter.
Both filter have similar full-width at the half maximums (FWHMs),
approximately 90 \AA\, but our new $Ca$ filter has a more uniform and 
high transmission across the passband, dropping more rapidly at both edges.
As shown in the figure, the CN band at \cnwave\ lies
on the lower tail of the $Ca$ filter by \citet{att91} while our new $Ca$ filter
is designed to be completely free from the CN band contamination.

In the Phase~III started in April 2013, we added a new filter system
also designed by the author of the paper, $JWL39$, 
which allows us to measure the CN band at \cnwave\
in combination with our new $Ca$ filter.
Note that our $JWL39$ filter is somewhat similar to the DDO38 filter \citep{ddo},
but our filter has the slightly different bandwidth and the pivot wavelength.
We show the transmission function of our $JWL39$ filter 
in Figure~\ref{fig:flt}.

\begin{figure}
\epsscale{1}
\figurenum{2}
\plotone{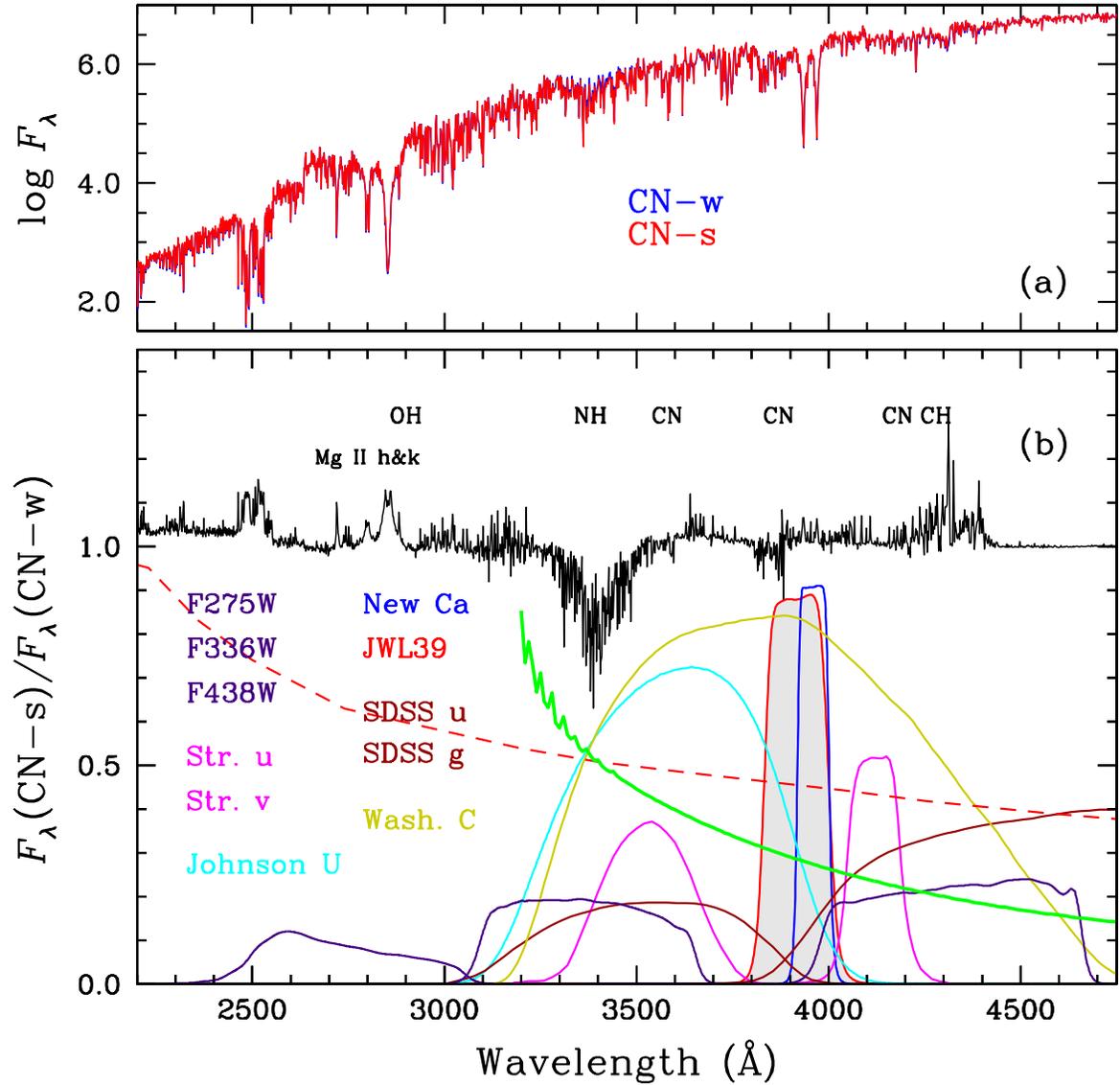}
\caption{
(a) A comparison of synthetic spectra for typical \cnw\ and \cns\
RGB stars in GCs \citep{sbordone11}.
(b) The flux ratio between the two spectra along with the
transmission functions for various filter systems used in the UV and 
in the blue part of the visible light.
The red dashed line denotes the interstellar extinction curve for
$E(B-V)$ = 0.1 mag \citep{mathis90} and the green solid line represents
the atmospheric extinction curve for Mauna Kea,
given in units of mag/airmass \citep{buton13}.
}\label{fig:fltcomp}
\end{figure}

In Table~\ref{tab:filter}, we show the pivot wavelengths, bandwidths and 
spectral resolutions of various filter system used in the ultra-violet (UV)
and in the blue part of the visible light.
Note that the \hst\ filter systems shown in the table are for the so-called
``\mt'', a combination of \hst\ WFC3/UVIS F275W, F336W and F438W filters
\citep{milone13}.
As \citet{piotto15} nicely demonstrated, the \hst\ magic-trio 
is known to  distinguish the multiple stellar populations in GCs.
However, the interpretation of the magic-trio can be somewhat complicated 
because it measures OH, NH, CN, CH and Mg (through \ion{Mg}{2} h \& k)
at the same time (see Figure~\ref{fig:fltcomp}).
Even worse is that the CN molecular bands are expected to show 
a positive luminosity effect while the OH, NH and CH bands are expected 
to show a negative luminosity effects due to the lower dissociation energies
of the hydrate molecules \citep[for example, see][]{gray09}.
The UV filters with very wide bandwidths ($\Delta\lambda >$ 50 nm),
such as the filters used in the \hst\ magic-trio, 
Johnson $U$, \sdss\ $u$ and Washington $C$, has some disadvantages, too.
These filters contain numerous very strong lines other than CNO and
small changes in the elemental abundances or in the stellar parameters
can affect the absorption strength and the continuum and, hence,
the photometric index using such wide band filters.
It is important to note that the NH band lies near the boundaries
of the steeply decreasing transmission functions of the Johnson $U$, 
Washington $C$\footnote{See \citet{cummings14} for the usage of 
the Washington $C$ filter for the MSPs in NGC~1851.}, \sdss\ $u$
and \str\ $u$ filters, where the degree of the atmospheric extinction
also increases very rapidly. Therefore these ill-matched combinations
of the transmission of these filters and the atmospheric extinction make 
their photometric systems less sensitive to the NH abundances,
as we will show later.
In sharp contrast, our \cnjwl\ index ( = $JWL39 - Ca_{\rm new}$)
is intended to measure the CN \cnwave\ band absorption strength only
and the atmospheric extinction does not vary significantly 
in the $JWL39$ passband.
We will show later that our \cnjwl\ index is as good as \scn\ or
\ds\ from low resolution spectroscopy and capable of distinguishing
MSPs in GCs with great satisfaction.

\section{OBSERVATIONS AND DATA REDUCTION}\label{s:reduction}
Observations for M5 were made in 21 nights, 
12 of which  were photometric, in 7 runs from 
May 2007 to May 2014 using the CTIO 1.0m telescope.
The CTIO 1.0m telescope was equipped with a STA 4k $\times$ 4k CCD camera,
providing a plate scale of 0\farcs289 pixel$^{-1}$ and 
a field of view (FOV) of about 20\arcmin\ $\times$ 20\arcmin.
During the whole seasons, 
the mean seeing from our M5 science frames is 1\farcs53 $\pm$ 0\farcs27.
In Table~\ref{tab:obs}, we show the total integration times for M5.

The detailed procedures for the raw data handling were described
in our previous works \citep{n6723,lee15,lp16}.
The photometry of the cluster and photometric standard frames were analyzed
using DAOPHOTII, DAOGROW, ALLSTAR and ALLFRAME, and  
COLLECT-CCDAVE-NEWTRIAL packages \citep{pbs87,pbs94,pbs95}.
In appendix~\ref{ap:s:acs}, 
we present a new strategy for the ground-based observations
to make use of the positional information from the \hst\ photometry
by \citet{anderson08}, which greatly improve the detection
rate in the central part of the cluster ($r \lesssim r_h).$\footnote{The 
half-light radius of M5 is $r_h$ $\approx$ 106\arcsec\ \citep{harris96}.}
The total number of stars measured in our M5 field
from our ALLFRAME run  was more than 60,000.

Astrometric solutions for individual stars in our field
were derived using the data extracted from the Naval 
Observatory Merged Astrometric Dataset \citep[NOMAD,][]{nomad}
and the IRAF IMCOORS package.
Then the astrometric solution was applied to calculate the equatorial 
coordinates for all stars measured in our science frames.

\section{RESULTS}

\subsection{Color-magnitude diagrams}\label{s:cmd}
\begin{figure}
\epsscale{1}
\figurenum{3}
\plotone{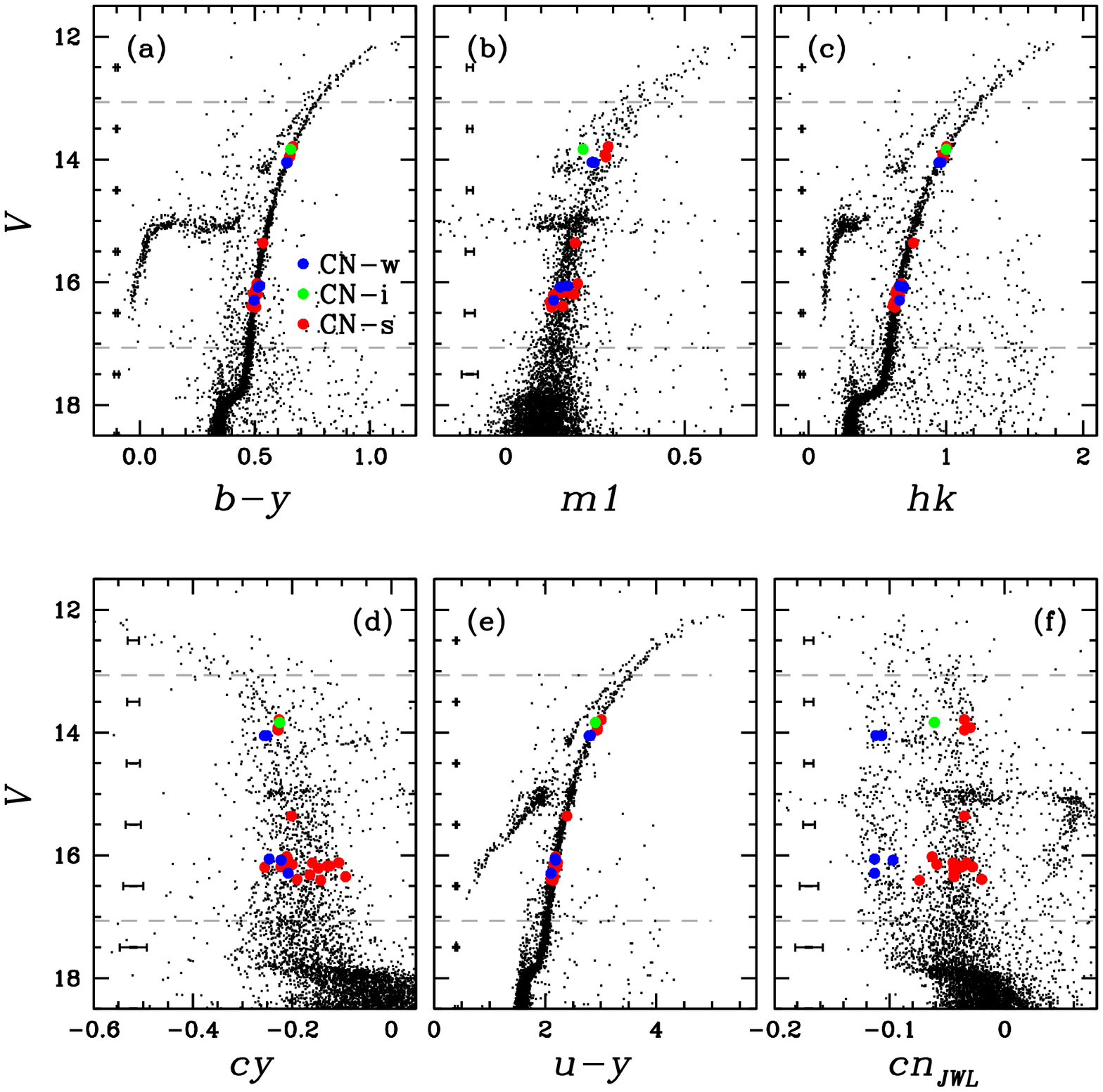}
\caption{CMDs for good quality stars in M5 based on the separation index
\citep{pbs03}. 
RGB stars studied by \citet{briley92} are also shown with filled circles. 
Blue filled circles denote the \cnw\ (the CN normal by \citealt{briley92}),
a green one the intermediate CN absorption, 
red ones the \cns\ RGB stars in M5.
We show the measurement error of individual stars at given magnitude bins.
The dashed horizontal lines are for \vvhb\ = $\pm$2.0 mag,
with \vhb(M5) = 15.07 mag \citep{harris96}.
Note that the narrow single RGB sequence in $b-y$, $u-y$ and $hk$ CMDs,
while the very broad RGB sequence in $m1$ and $cy$ CMDs, 
where the separation between different populations is ambiguous.
On the other hand, the RGB stars in our \cnjwl\ index show distinct 
double sequences, consistent with the bimodal nitrogen distribution in M5. 
}\label{fig:cmd}
\end{figure}

Figure~\ref{fig:cmd} shows the color-magnitude diagrams (CMDs) of 
bright stars in the M5 field. 
We used photometric results from our new filters for
$(b-y)$, $hk$ and \cnjwl\, which were taken after 2011,
while those from the CTIO filters for
$m1$, $cy$ and $(u-y)$, which were taken before 2011.
As can be seen in the figure, $(b-y)$ and $hk$ CMDs show 
rather narrow single RGB sequences without any significant spread, 
confirming previous results that M5 is a mono-metallic cluster
\citep[e.g.][]{ivans01,carretta09}.
On the other hand, $m1$, $cy$ and \cnjwl\ CMDs exhibit very broad or 
even discrete double RGB sequences, indicative of the spread 
in the lighter elemental abundances, more specifically, 
a significant spread or a bimodal distribution of the nitrogen abundance
because these color indices contain either NH or CN, or both, 
as already shown in Figure~\ref{fig:fltcomp}.
In particular, the clear split in the RGB sequence 
in the \cnjwl\ versus $V$ CMD is evident.
We will show later that M5 most likely has a bimodal nitrogen abundance 
distribution, which is consistent with our \cnjwl\ index showing 
the discrete double RGB sequences.
What makes our \cnjwl\ index great is that it is a photometric index so that 
one can measure very accurate CN \cnwave\ band absorption strengths of cool RGB
and AGB stars in the very crowded region.
On the other hand, the distinction between the different populations
is somewhat ambiguous in $cy$ and $m1$ indices in Figure~\ref{fig:cmd}. 
The results from these two color indices should be used to detect 
the spread in the lighter elemental abundances only and not be used
for the detailed chemical tagging.

\begin{figure}
\epsscale{1}
\figurenum{4}
\plotone{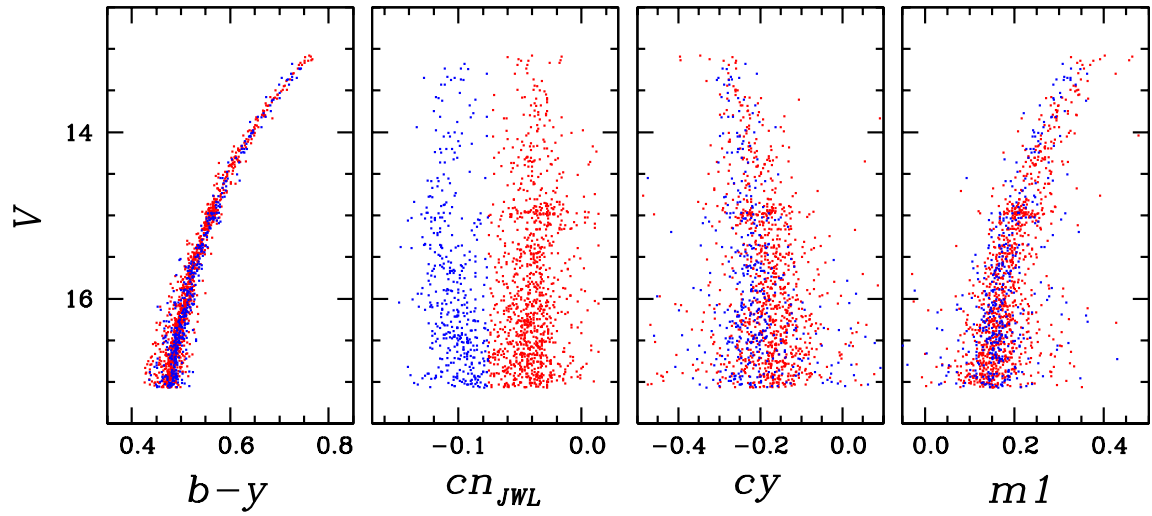}
\caption{CMDs for all RGB stars with \vvhb\ = $\pm$2.0 mag.
The blue dots denote the \cnw\ and the red dots the \cns\ RGB stars in M5,
based on the EM estimator for the two-component Gaussian mixture model
for the \cnjwl\ distribution of the RGB stars.
}\label{fig:rgbcmd}
\end{figure}

Based on our \vcn\ CMD, we derive the number ratio between the two groups 
of stars, the \cnw\ and the \cns\ (see below for the definitions).
We carefully chose RGB membership stars using our multi-color photometry
and we show our results in Figure~\ref{fig:rgbcmd}.
As elaborately shown in our previous work \citep{lee15}, our method 
is very effective to remove the off-cluster field stars
even for heavily field star contaminated GCs like M22.
Since M5 is located in a rather high Galactic latitude, $b = 46\fdg8$,
the field star contamination is expected to be not severe in our results.
Assuming a bimodal \cnjwl\ distribution, we applied the expectation
maximization (EM) algorithm for the two-component Gaussian mixture model
to distinguish the different RGB populations.
In an iterative manner, we calculated the probability of individual stars 
for being the \cnw\ and the \cns\ populations.
Stars with $P$(\cnw$|x_i) \geq$ 0.5 from the EM estimator correspond
to the \cnw\ population, where $x_i$ denotes the individual RGB stars,
while those with $P$(\cns$|x_i)$ $>$ 0.5  correspond to the \cns\ population. 
Through this process, we obtained the number ratio between the two populations,
\nrgb\ = 29:71 ($\pm$ 2).

\subsection{Red giant stars}
\subsubsection{A comparison with \citet{briley92}}
\citet{briley92} performed a spectroscopic study of M5 RGB star and 
they derived the nitrogen abundances of RGB stars.
As already shown in Figure~\ref{fig:cmd},
the $(b-y)$ and $hk$ CMDs show narrow single RGB sequences and
any significant differences in the positions of  the \cnw\ and 
the \cns\ RGB stars defined by \citet{briley92} can not be seen.
On the other hand, correlations between the nitrogen abundance
and color indices can be found in $m1$, $cy$ and \cnjwl\ CMDs,
although correlations involved with $m1$ and $cy$ indices
do not look as good as that with \cnjwl.

\begin{figure}
\epsscale{1}
\figurenum{5}
\plotone{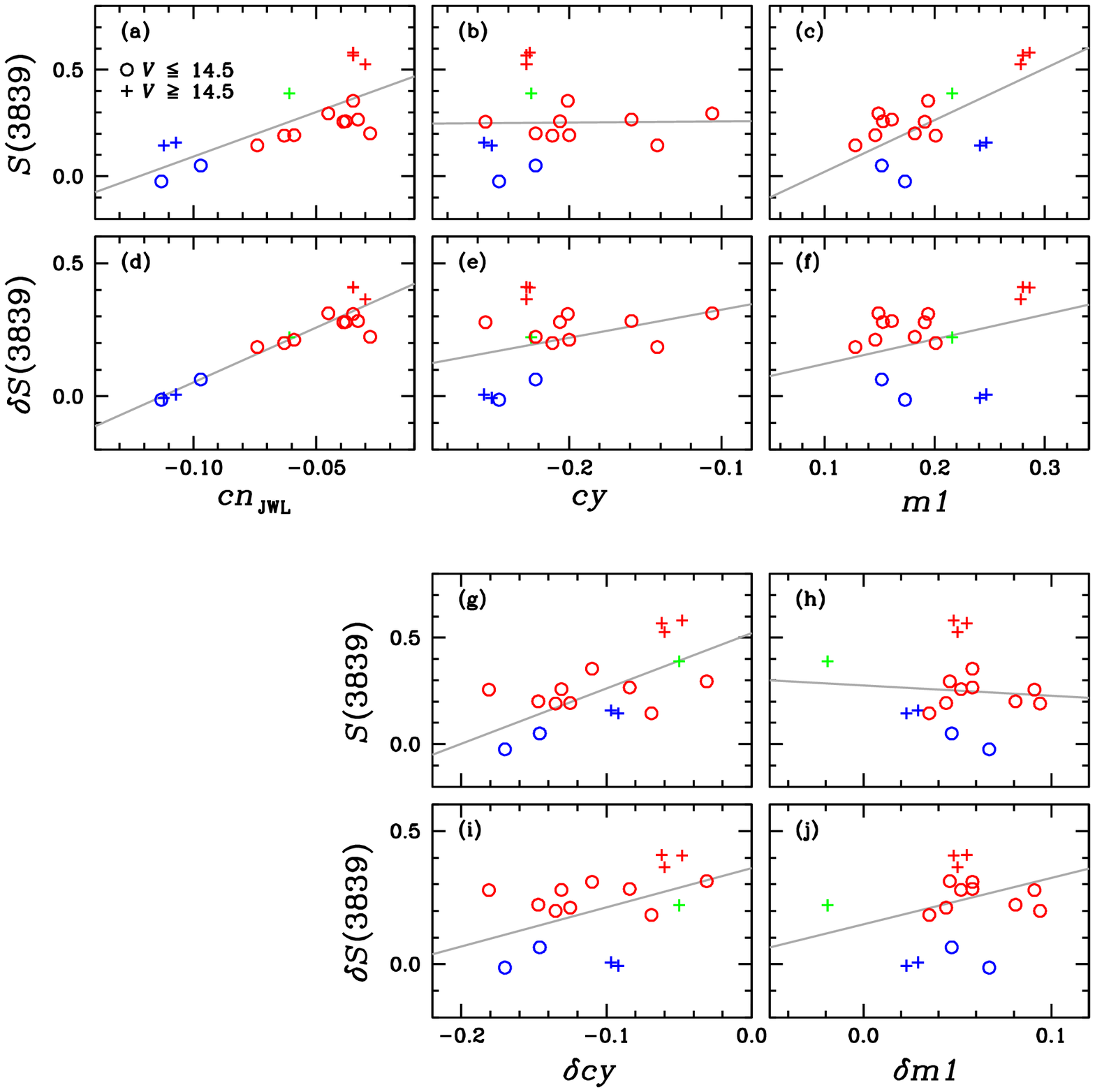}
\caption{(a) -- (c) Correlations between selected color indices and 
CN \cnwave\ strength, \scn.
The blue color denotes the \cnw, the green color the intermediate CN and 
the red color the \cns\ \citep{briley92}. 
The grey solid lines are linear fits to the data.
It can be clearly seen that our \cnjwl\ index is well correlated with \scn\ 
in the whole magnitude level, while $m1$ and $cy$ indices provide poor fit 
to the data (see Table~\ref{tab:deltas_briley}).
Also note that the scatters of the \scn\ distribution around the fitted line 
in the \scn\ versus \cnjwl\ relation are larger than 
those in the \ds\ versus \cnjwl\ relation, indicating that the gravity
and temperature effects should be removed.
(d) -- (f) Correlations between selected color indices and the CN excess, \ds.
The linear fit between the \cnjwl\ and \ds\ is greatly improved.
(g) -- (j) Correlations between excesses in the color indices and \scn\ and \ds.
Using color excesses $\delta cy$ and $\delta m1$ does not have a net gain.
}\label{fig:deltas_briley}
\end{figure}

First, we examined the correlations between the CN \cnwave\ absorption strengths 
by \citet{briley92}, \scn, and the color indices that we measured.
In Figure~\ref{fig:deltas_briley}(a)--(c), we show plots of \scn\ versus
each color index, and the goodness of the fits in Table~\ref{tab:deltas_briley}.
Note that the $p$-value of the fit for the $m1$ versus \scn\ is as good as
that for the \cnjwl versus \scn, 
but the figure clearly shows that the $m1$ index is not capable of discriminating
the two different populations.
On the $m1$ versus \scn\ or \ds\ planes, the \cnw\ and the \cns\ populations
by \citet{briley92} are superposed each other.
In panel (a), the distributions of the bright and the faint RGB stars
are clearly divided around the fitted line, indicating that the 
differences in the temperature and the surface gravity may
affect the \scn\ values at different $V$ magnitudes.
In order to correct these effects, we derive the CN excess, \ds,
which is the distance in \scn\ from the lower envelope
of the distribution of RGB stars in the $V$ magnitude versus \scn,
\begin{equation}
\delta S = S(3839) - (1.287 - 0.088\times V).
\end{equation}
Note that \citet{briley92} did not provide \ds.
Our results are shown in panels (d)--(f).
As shown, using \ds\ greatly improves the linear fit for our \cnjwl\ index,
with the correlation coefficient of 0.941 and the $p$-value of 0.000.
On the other hand, both the $cy$ and the $m1$ indices
do not accurately trace the CN excess.

In Figures~\ref{fig:cmd} and \ref{fig:rgbcmd}, 
the $cy$ and the $m1$ indices show considerable 
curvatures in the RGB magnitude level of our interest, 
indicative of the presence of the temperature and surface gravity effects 
upon these indices.
In order to correct these effects,
we derived the excess in the $cy$ and $m1$ indices, \dcy\ and \dmo,
which are the distances from the outer envelopes of RGB distributions
in each index at fixed $V$ magnitude, finding,
\begin{equation}
\delta cy = cy - (-4.030 - 0.480V - 0.015V^2)
\end{equation}
and
\begin{equation}
\delta m1 = m1 - (3.050 - 0.329V + 0.009V^2).
\end{equation}
Again, as shown in Figure~\ref{fig:deltas_briley}(g--j),
neither \dcy\ nor \dmo\ appear
to mitigate the discrepancy, suggesting that both indices
are not good CN tracers.

\begin{figure}
\epsscale{1}
\figurenum{6}
\plotone{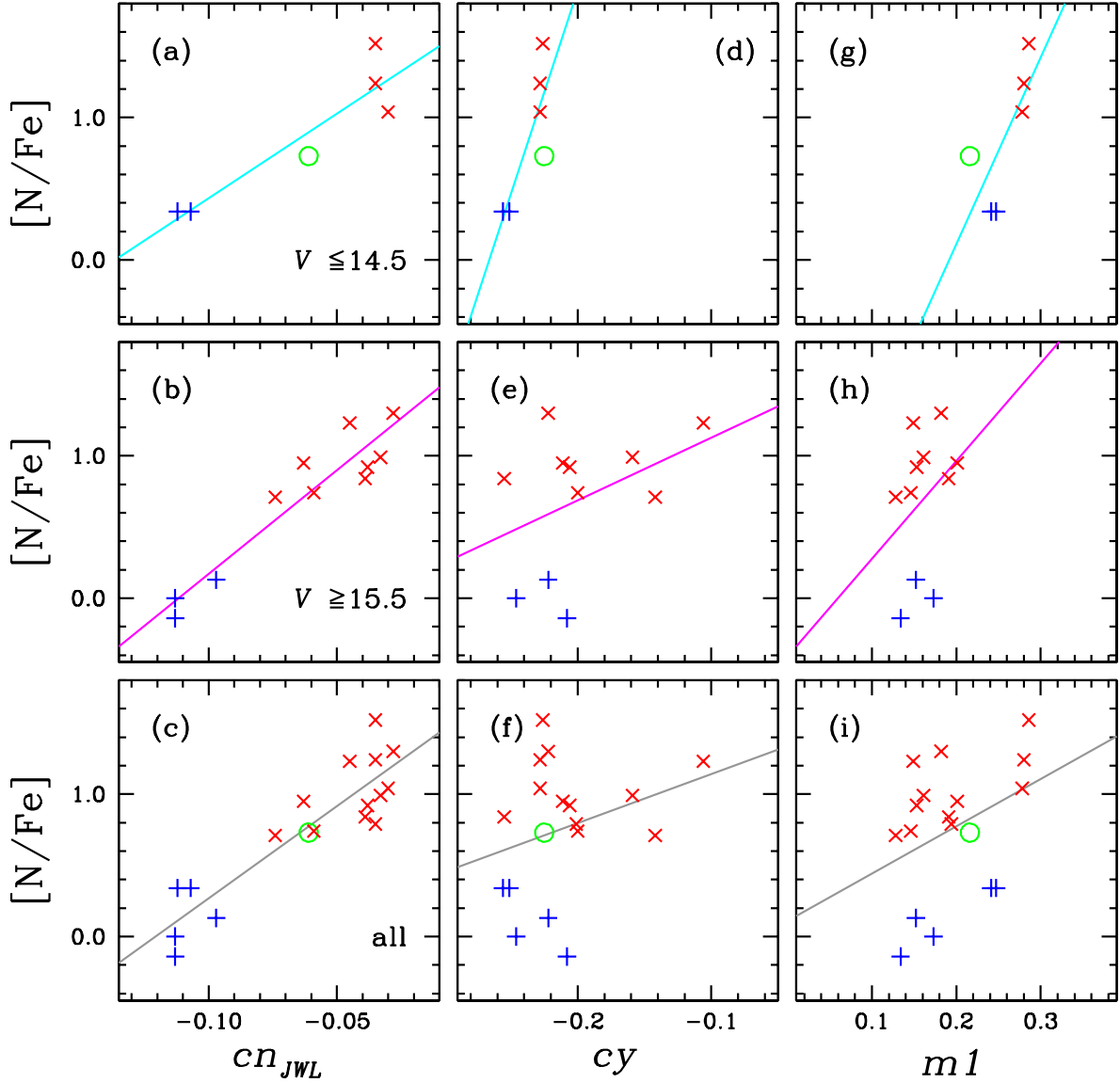}
\caption{Correlations between selected color indices and nitrogen 
abundances of M5 RGB stars studied by \citet{briley92}.
We derive the linear fits for bright ($V \leq$ 14.5 mag), 
faint ($V \geq$ 15.5 mag)
and all stars, and we show the fitted lines in each panel
(see also Table~\ref{tab:fit}).
Note that our \cnjwl\ index is well correlated with nitrogen abundances
in the whole magnitude level, while the $cy$ and the $m1$ provide poor fit 
to the data in the faint magnitude regime.
}\label{fig:cnvsphot}
\end{figure}

Next, we examine the correlations between the nitrogen abundances
and color indices.
In Figure~\ref{fig:cnvsphot}, we show least square fits between
the nitrogen abundance by \citet{briley92} and  individual color indices.
Since the visual magnitude of the RGB stars studied by \citet{briley92} 
has a gap, we divided the sample into two groups, 
the bright ($V \leq$ 14.5 mag) and the faint ($V \geq$ 15.5 mag) RGB groups.
We derived the linear correlations to the data and
we show our results in Table~\ref{tab:fit}.
The correlations between the nitrogen abundances and the \cnjwl\ index
are excellent for both magnitude bins, with consistent slopes 
and zero points.
The correlations between the nitrogen abundances and the $cy$ and $m1$ indices
are also excellent for the bright RGB stars.
However, those for the faint RGB stars are very poor.
Also importantly the least square fits for both magnitude bins are 
significantly different for $m1$ and $cy$ indices,
which make the both indices difficult to be good photometric indicators 
for nitrogen abundances.

\begin{figure}
\epsscale{1}
\figurenum{7}
\plotone{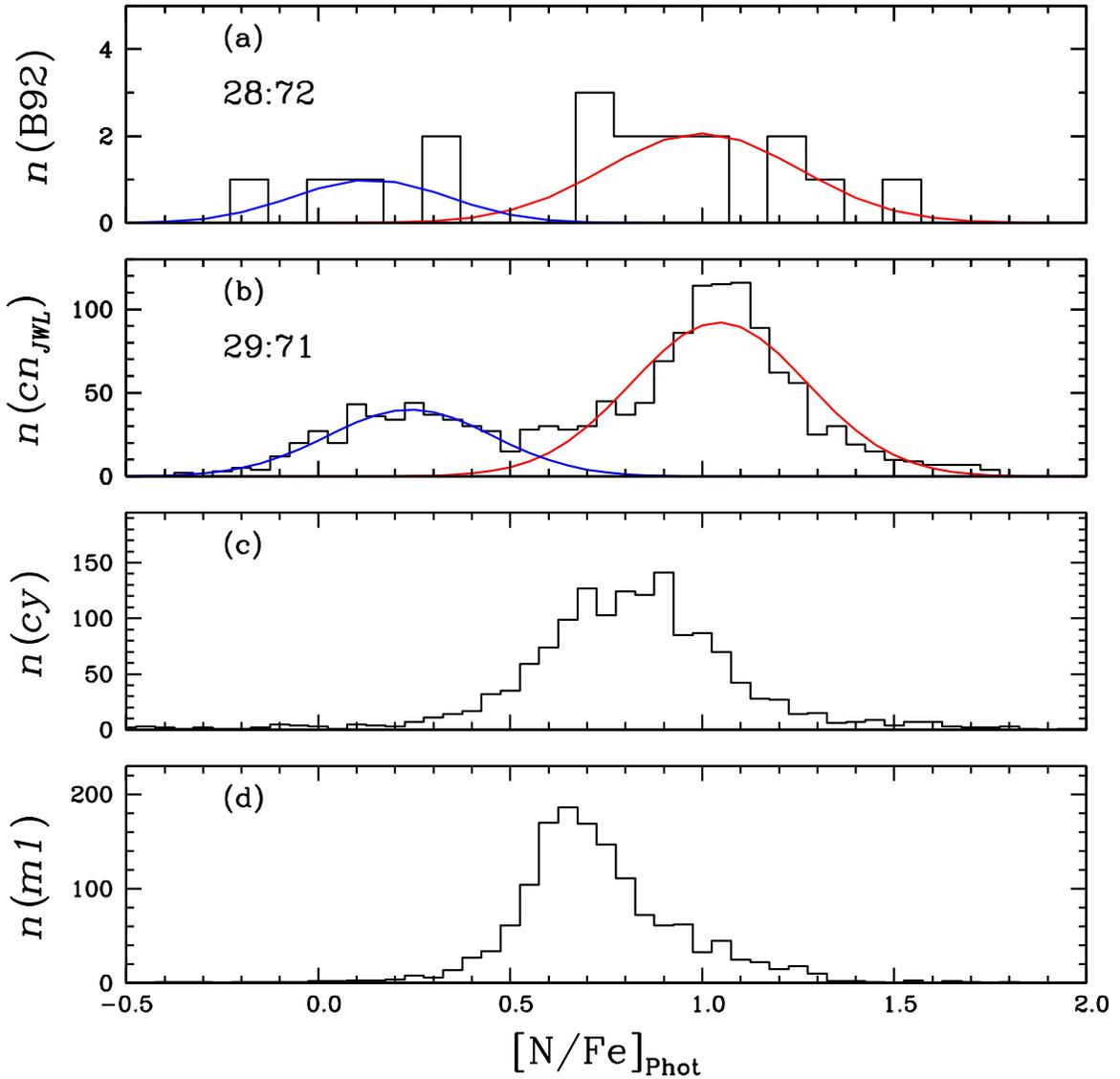}
\caption{(a) The histogram of the nitrogen abundance by \citet{briley92}.
Our Hartigan's dip test using the nitrogen abundance measurements
by \citet{briley92} shows that the [N/Fe] distribution of the M5
RGB stars in non-unimodal.
The blue and red lines are for the N-normal and N-enhanced populations
from the EM algorithm for the two-component
Gaussian mixture distribution model. 
(b) The photometric nitrogen abundance distribution from our \cnjwl\ index
using the relation given in Table~\ref{tab:fit}, which is in excellent
agreement with that of \citet{briley92}.
(c)--(d) The photometric nitrogen abundance distributions from $cy$ and
$m1$ indices, which fail to reproduce that of \citet{briley92}.
}\label{fig:cnhist}
\end{figure}

We derive the photometric nitrogen abundance for individual RGB stars
based on our linear correlations.
We performed a Hartigan's dip test to see if the nitrogen abundance
distribution of M5 RGB stars by \citet{briley92} is unimodal. 
We obtained $D$ = 0.061 and $p$-value = 0.870, suggesting that 
the nitrogen abundance distribution of the M5 RGB stars is non-unimodal.
Therefore, assuming a bimodal distribution of the nitrogen abundance for M5,
we applied the EM algorithm for the two-component Gaussian mixture distribution 
model to calculate the contributions from two populations.
In an iterative manner, we derive the probability of individual RGB stars
for being the \cnw\ and the \cns\ populations and we show our result
in Figure~\ref{fig:cnhist}(a), with the number ratio between
the \cnw\ and the \cns\ populations of \nrgb\ = 28:72 ($\pm$ 20).

Using the least square fits given in Table~\ref{tab:fit},
we calculate the photometric nitrogen abundances 
of the RGB stars with $-$2 $\leq$ \vvhb\ $\leq$ 2 mag.
The photometric nitrogen abundance from our \cnjwl\ index provides
a bimodal distribution as shown in Figure~\ref{fig:cnhist}(b).
The number ratio of the two populations based on
the photometric nitrogen abundance from the \cnjwl\ index is
\nrgb\ = 29:71 ($\pm$ 2), in excellent agreement with the spectroscopic
nitrogen abundance distribution by \citet{briley92}
and that from our \cnjwl\ index.
Our results strongly support that our \cnjwl\ index is truly 
a measure of the CN absorption strength at \cnwave, and furthermore
the nitrogen abundance.
On the other hand, the photometric nitrogen abundances from the $m1$ and $cy$
indices shown in Figure~\ref{fig:cnhist}(c--d) do not agree with the nitrogen
abundance measurements by \citet{briley92}, showing a conspicuous single peak
with a rather long tail.

We conclude that our \cnjwl\ index traces the nitrogen abundance of RGB stars
in M5, while the $m1$ and the $cy$ indices can provide some limited information
on the spread in the nitrogen abundances but fail to provide detailed 
substructures in the nitrogen abundance distribution.

\subsubsection{A comparison with \citet{smith13}}
\begin{figure}
\epsscale{1}
\figurenum{8}
\plotone{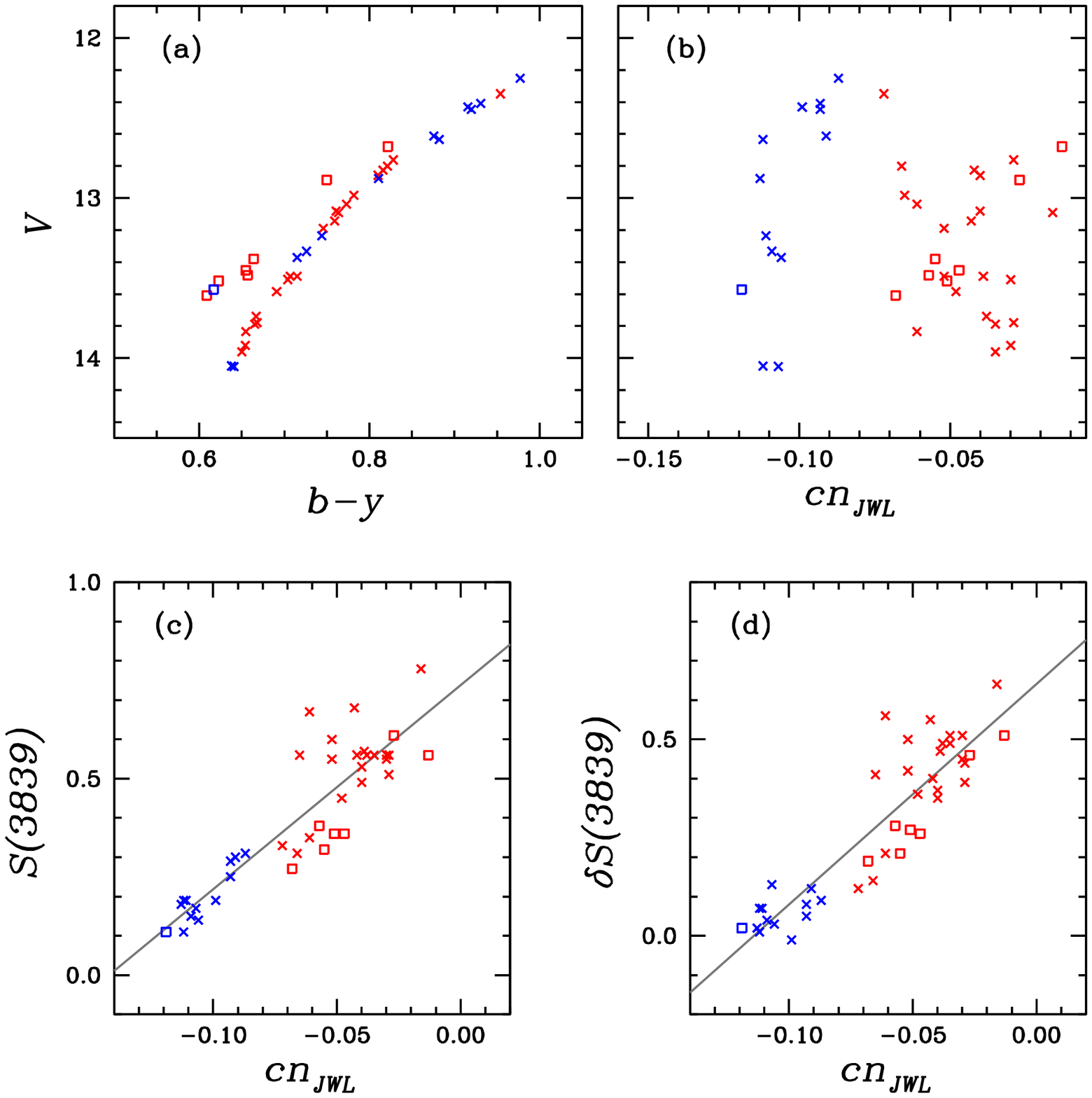}
\caption{(a) The \vby\ CMD for the stars studied by \citet{smith13}.
The crosses are for the RGB stars and the open squares for the AGB stars.
The blue color denotes the \cnw\ population, while the red color 
the \cns\ population.
(b) The \vcn\ CMD, where the distinct double AGB sequences are evident.
(c) A plot of \scn\ versus \cnjwl.
(d) A plot of \ds\ versus \cnjwl.
Note that our \cnjwl\ index correlates nicely with both \scn\ and \ds.
}\label{fig:deltas}
\end{figure}

In Figure~\ref{fig:deltas}(a)-(b), we show the $(b-y)$ and the \cnjwl\ CMDs for
M5 RGB and AGB stars used by \citet{smith13}, who compiled 
the \cnwave\ CN band strengths for the M5 RGB and  AGB stars from the literature.
As discussed earlier, there exists no difference between
the \cnw\ and \cns\ populations in the \vby\ CMD,
while a clear split between the two populations, including
the AGB stars, in the \vcn\ CMD can be seen.
We will discuss more about the AGB stars in \S\ref{s:agb}.

It should be emphasized that our \cnjwl\ index do not show any gradient 
against the $V$ magnitude in the region of our interest 
($-2 \leq$ \vvhb\ $\leq$ 2 mag) for M5, which suggests that 
no temperature and surface gravity corrections are required for our \cnjwl\ index.
We compare our \cnjwl\ index with the \scn\ or the \ds\ measurements 
by \citet{smith13} and we show our results in
Figure~\ref{fig:deltas}(c)-(d) and Table~\ref{tab:deltas}.
It can be clearly seen that our \cnjwl\ index correlates nicely with 
both \scn\ and \ds, indicating again that the \cnjwl\ index is 
a measure of the CN band strength at \cnwave.

It is interesting to note that using the CN excess, \ds, does
not improve the correlations as we showed in Figure~\ref{fig:deltas_briley}.
It is suspected that it may be due to the heterogeneous nature of the sample
stars used by \citet{smith13}.

\subsubsection{A comparison with \citet{carretta09}}
\begin{figure}
\epsscale{1}
\figurenum{9}
\plotone{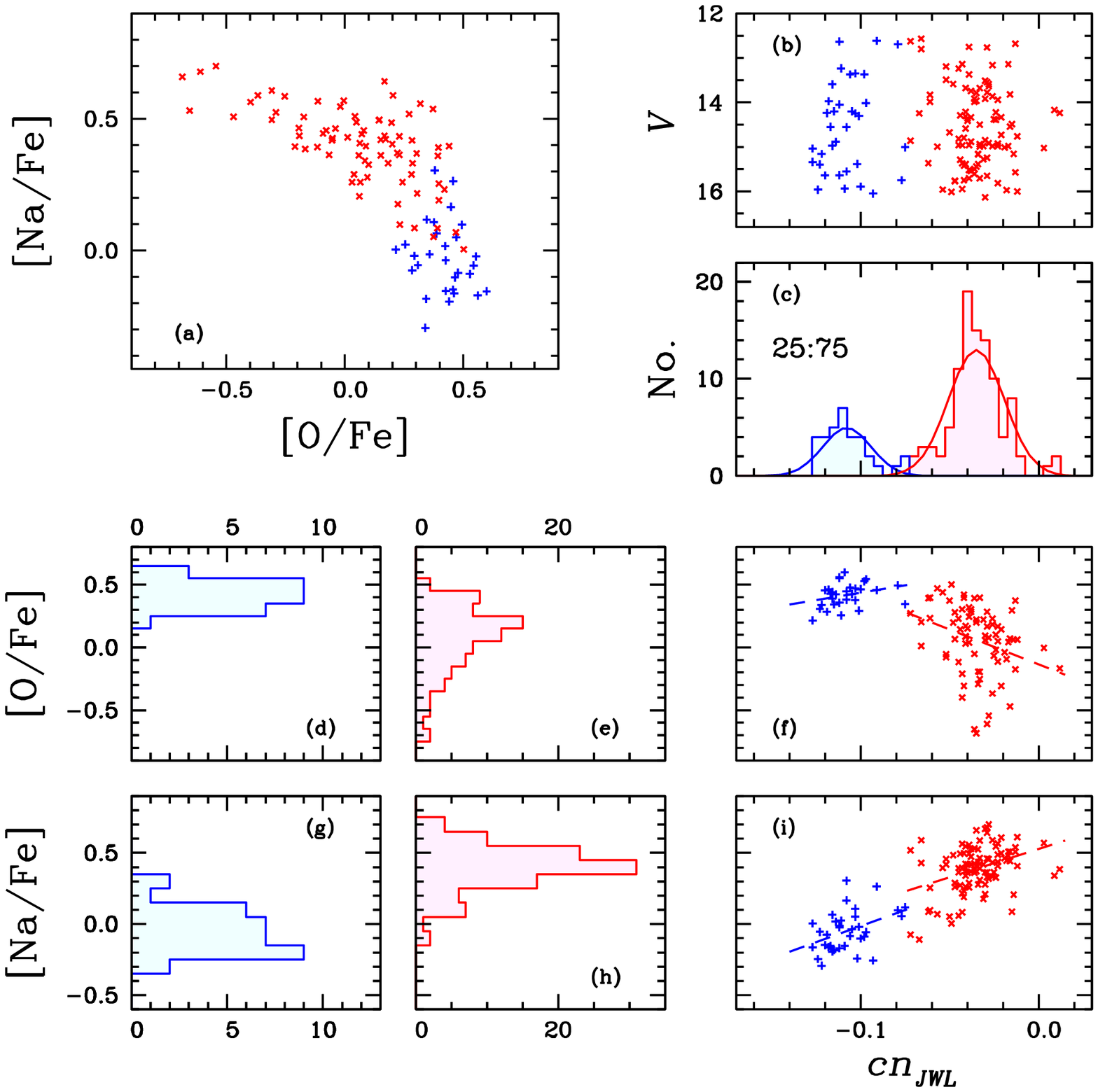}
\caption{(a) The Na-O anticorrelation of M5 RGB stars by \citet{carretta09}.
The \cnw\ and \cns\ stars based on the EM estimator
are shown with blue plus signs and red crosses.
(b) The \cnjwl\ versus $V$ CMD of RGB stars studied by \citet{carretta09}.
(c) The \cnjwl\ distribution.
(d)--(e) The [O/Fe] distributions of the \cnw\ and the \cns\ RGB stars.
(f) A plot of \cnjwl\ versus [O/Fe]. Note that there appear to exist
two separate \cnjwl-[O/Fe] (anti)correlations. 
The \cnjwl-[O/Fe] of the \cnw\ population is positively correlated,
while that of the \cns\ population is anticorrelated.
(g)--(h) The [Na/Fe] distributions of the \cnw\ and the \cns\ RGB stars.
(i) A plot of \cnjwl\ versus [Na/Fe].
The \cnjwl-[Na/Fe] of both populations are positively correlated, albeit 
the correlations in the two populations may not be continuous.
}\label{fig:nao}
\end{figure}

In Figure~\ref{fig:nao}(a), we show the Na-O anticorrelation of M5 RGB stars
studied by \citet{carretta09}, where one can find M5 has an extended and 
well defined Na-O anticorrelation \citep[e.g., see also,][]{ivans01}.
In Figure~\ref{fig:nao}(b), we show the \vcn\ CMD of the RGB stars 
studied by \citet{carretta09}, where the distinct double RGB sequences
can be clearly seen.
We show the histograms for the \cnw\ and  the \cns\ populations 
in Figure~\ref{fig:nao}(c) and we obtain the number ratio of 
\nrgb\ = 25:75 ($\pm$ 5) from the EM estimator, 
which is marginally in agreement with our results shown previously.
Note that this number ratio does not represent the complete RGB number ratio of M5
because the sample RGB stars chosen by \citet{carretta09} are
restricted to those adequate for the spectroscopic observations
in the outer part of the cluster.
In Figure~\ref{fig:nao}(d)-(i), we show the [O/Fe] and the [Na/Fe] distributions
for each population, where one can find that our \cnjwl\ index can nicely
distinguish the primordial\footnote{The definition of the primordial population
by \citet{carretta09} is the group of stars with [Na/Fe] $\leq$ 0.1 dex.}
and other (i.e.\ the intermediate and the extreme) populations.
As shown, the \cnw\ population from our \cnjwl\ index has
higher oxygen and lower sodium abundances while the \cns\ population
has lower oxygen and higher sodium abundances 
\citep[see also Figure~9 of][]{ivans01}.
Of particular interest is the two separate \cnjwl-[O/Fe] and 
the \cnjwl-[Na/Fe] relations for both populations, 
i.e., the two separate [N/Fe]-[O/Fe] and the [N/Fe]-[Na/Fe] relations.
In Figure~\ref{fig:nao}(f), the [O/Fe] abundance of the \cnw\ RGB stars
appear to increase (or, perhaps remains flat) against the \cnjwl\ index
and then the [O/Fe] abundance of the \cns\ RGB stars decreases
with the \cnjwl\ index.
In the plot, we also show the linear fits for each population.
\citet{smith13} also suspected non-continuous relations between
their \ds\ measurements and either [O/Fe] or [Na/Fe]. 
However, due to their small sample size, their conclusion was
somewhat provisional. 
We note that the same trend can also be seen in NGC~6752 
(see Figure~9 of \citealt{yong08}), which will be discussed
in our future work (Lee 2017a, in preparation).

It is also very interesting to note that the spread in the [O/Fe] abundance
of the \cns\ population, $\sigma$[O/Fe] $\approx$ 0.27 dex,
is very large compared to that of the \cnw\ population, 
$\sigma$[O/Fe] $\approx$ 0.09 dex,
while the spreads in the [Na/Fe] abundance for both populations are compatible,
$\sigma$[Na/Fe] $\approx$ 0.15 dex.
In the context of the chemical evolution of GC stars, however, 
what really matter would be the total number of atomic species.
We define the standard deviation of the numbers of the atom in GC RGB stars,
\begin{equation}
\sigma {\rm [X]} = \log [\sigma\langle10^{[\log({\rm X/H}) + 12]}\rangle].
\end{equation}
Despite the factor of 2 difference in the mean oxygen abundances between
the two populations, 
$\langle\log$(O/H) + 12 $\rangle$ = 7.78 for the \cnw\ stars
and 7.47 for the \cns\ stars, we obtained the very similar values 
for the standard deviations of the numbers  of the oxygen atom, 
\sigo\ = 7.11 for the \cnw\ stars and 7.18 for the \cns\ stars.
On the other hand, the standard deviations of the numbers of 
the sodium atom are quite different between the two populations, 
\signa\ = 4.43 for the \cnw\ stars while 4.86 for the \cns\ stars.

\begin{figure}
\epsscale{1}
\figurenum{10}
\plotone{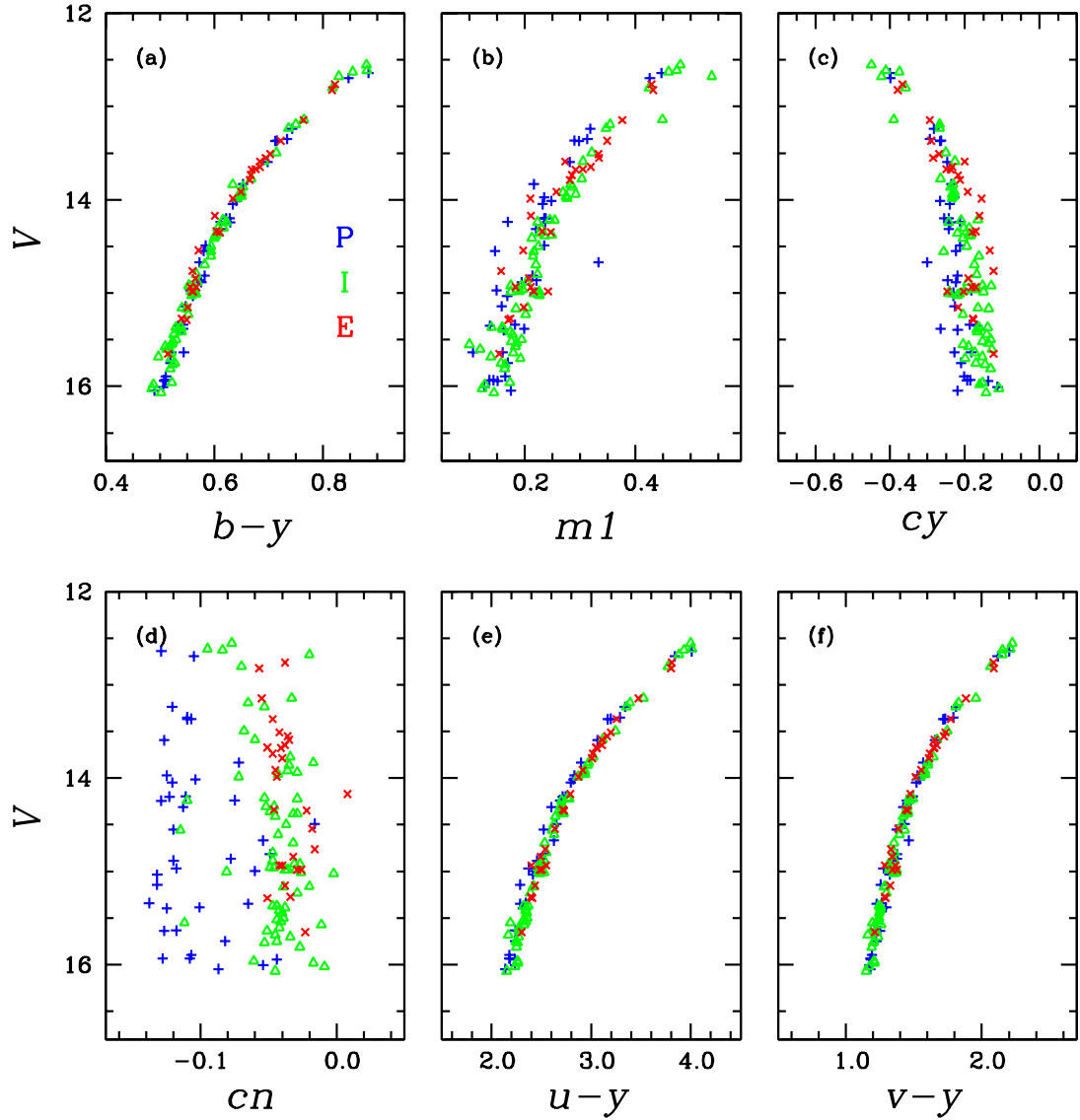}
\caption{CMDs of spectroscopic target RGB stars of \citet{carretta09}.
The blue plus sings denote the primordial, the green open triangles 
the intermediate, and the red crosses the extreme populations 
defined by \citet{carretta09}.
From the photometric point of view, there is no difference between
the intermediate and the extreme populations defined by \citet{carretta09}.
}\label{fig:pie}
\end{figure}

Finally, we show CMDs of spectroscopic target RGB stars of \citet{carretta09}
in Figure~\ref{fig:pie}.
As we mentioned before, the $m1$ and $cy$ indices are capable of detecting
non-uniform lighter elemental abundances.
However, the level of confusion is severe so that both indices 
cannot clearly separate different populations as we pointed out previously.
Only the \cnjwl\ index can fully separate 
the primordial population from the others,
where no clear distinction between the intermediate and the 
extreme populations can be seen from the photometric point of view
as shown in the figure.
This is also the case for the \ds, as noted by \citet{smith13}.
In the figure, the superpositions of the primordial and the intermediate
populations can be seen in the \cnjwl\ index.
It is most likely due to the arbitrary definition of the boundary
between the two populations set by \citet{carretta09} in the continuous
distribution of the RGB stars on the [O/Fe] versus [Na/Fe] plane.

\subsubsection{A comparison with \citet{lardo11}}
Using the \sdss\ archive data, \citet{lardo11} studied MSPs in GCs, 
including M5.
Based on their newly devised  normalized color spread, $\Delta^\prime_{\rm color}$,
they claimed that the radial distributions of the the UV-blue 
(i.e., the FG of the stars) and the the UV-red stars (the SG of stars) 
in M5 are distinctively different,
in the sense that the UV-red stars are more centrally concentrated.

It is thought that using the \sdss\ photometry in the study of the MSPs 
in GCs has some potential problems.
As already shown in Figure~\ref{fig:fltcomp}, the \sdss\ $u$ filter contains 
the NH and the CN bands, where the atmospheric extinction becomes stronger
at shorter wavelength. As a consequence, it is suspected that 
the \sdss\ $u$ magnitude or color indices involved with it
become less sensitive to changes in the nitrogen abundance.
Also importantly, the \sdss\ $g$ filter is very broad
($\Delta\lambda \approx$ 140 nm) and it contains many
strong molecular bands, such as CN, CH and MgH.
Therefore, the color index composed of the \sdss\ $u$ and $g$ filters
may behave in a complicated way against the changes in
the light elemental abundances.
Here we compare our photometry to that from the SDSS to see if
both photometric systems provide consistent results.

\begin{figure}
\epsscale{1}
\figurenum{11}
\plotone{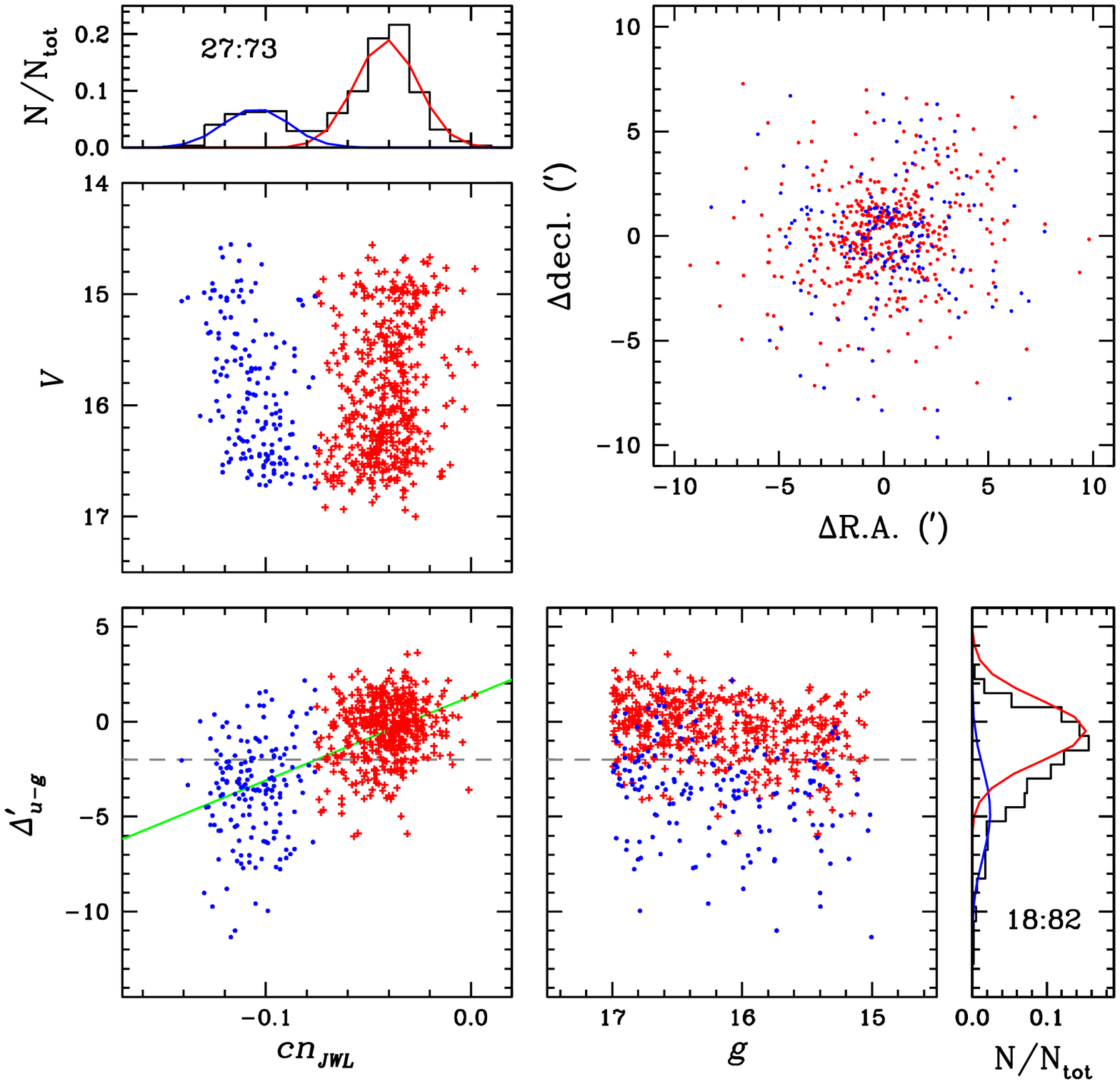}
\caption{ 
A comparison of our photometry with that of \citet{lardo11}.
The RGB stars with \dug\ $< -2$ are classified
as the $UV$-blue and those with $\geq$ $-$2 as the $UV$-red stars
by \citet{lardo11}. The green solid line in the lower left panel 
represents the linear fit to the data.
Our CMD shows bimodal RGB sequences, while that of \citet{lardo11}
does not show clear separation between MSPs, suggesting that 
the boundary between the two populations set by \citet{lardo11} 
is rater arbitrary. Also note the continuous transition, with severe
superpositions of the two populations, from the \cnw\ to the \cns\
along the \dug.
Our result strongly suggests that the SDSS photometric system
is not adequate to study the MSPs in GCs.
}\label{fig:sdss}
\end{figure}

In Figure~\ref{fig:sdss}, we show a comparison of our \cnjwl\ and 
\dug\ by \citet{lardo11}.
Our \cnjwl\ versus $V$ CMD nicely shows double RGB sequences with the
number ratio of \nrgb\ = 27:73 ($\pm$ 3).
Again, this number ratio is not a complete one, but the one restricted
by the sample RGB stars used by \citet{lardo11}.
On the other hand, the distributions of stars from both populations
are continuously superposed on the \dug\ versus $g$ CMD, showing
no clear separation between the two RGB sequences.
The color indices using the \sdss\ $u$ filter appears to suffer from
confusion in discriminating the MSPs in GCs, as we already
demonstrated for the $cy$ and $m1$ indices.
In particular, note that the \cnw\ RGB stars take larger range 
in the \dug\ index than the \cns\ stars.
As shown, the histogram for the \dug\ index exhibits a rather long asymmetric tail.
We performed a Hartigan's dip test to see if the bimodality
of the RGB stars along the \dug\ is real, and
this test suggests a bimodal \dug\ distribution.
Therefore, we applied an EM estimator, finding \njwl\footnote{ 
Note that the $UV$-blue$_{\rm JWL}$ and the $UV$-red$_{\rm JWL}$
denote the two groups of RGB stars based on our EM estimator and 
they are not the $UV$-blue and the $UV$-red populations 
originally defined by \citet{lardo11}.}
= 18:82 ($\pm$ 3), which is significantly different from that of our photometry.
Also note the very large standard deviation for the $UV$-blue$_{\rm JWL}$
population compared to that of the $UV$-red$_{\rm JWL}$.
Originally, \citet{lardo11} classified stars with \dug\ $<$ $-$2
as the $UV$-blue and stars with \dug\ $\geq$ $-$2 as the $UV$-red.
Using their classification scheme, we obtained the number ratio
of \nlardo\ = 32:68 ($\pm$ 3), apparently consistent with that of
our photometry.
However, both the $UV$-blue and the $UV$-red groups defined by \citet{lardo11}
contain mixed populations, i.e., the \cnw\ and the \cns.
It is thought that their boundary between the two populations is 
somewhat arbitrary in the continuous distribution of the two mixed populations 
in the \dug\ distribution and there is no astrophysical basis for it.
Since \citet{lardo11} used the RGB stars  with $r$ $\gtrsim$ 1\arcmin\
as shown in Figure~\ref{fig:sdss}, the confusion of the populations in
the \sdss\ photometric system can not be attributed to the crowding effect.
It is a rather intrinsic drawback of the \sdss\ $u$ filter,
in addition to its very broad passband in the $UV$;
Simply an ill-matched combination of the filter transmission and the atmospheric
extinction is not adequate to study the NH band absorption features
to characterize the MSPs in GCs.
We suggest that, for example, the \dug\ should be used only for
detection of the light elemental abundance variations,
but not be used for the detailed chemical tagging of individual stars,
same as the $cy$ and the $m1$ indices as we discussed above.

\subsubsection{A comparison with \citet{piotto15}}
We make a comparison of our photometry with 
the \hst\ $UV$ photometry by \citet{piotto15}.
Note that the \hst\ photometry is available only for the central part
of the cluster ($r \lesssim$ 1\arcmin).
In spite of our new data reduction strategy (see Appendix B),
our ground-based photometry may suffer from incompleteness due to
the very dense environment of the cluster's center.
This is an unavoidable weakness of any ground-based observations,
however, we emphasize that this should not affect our results shown here.

\begin{figure}
\epsscale{1}
\figurenum{12}
\plotone{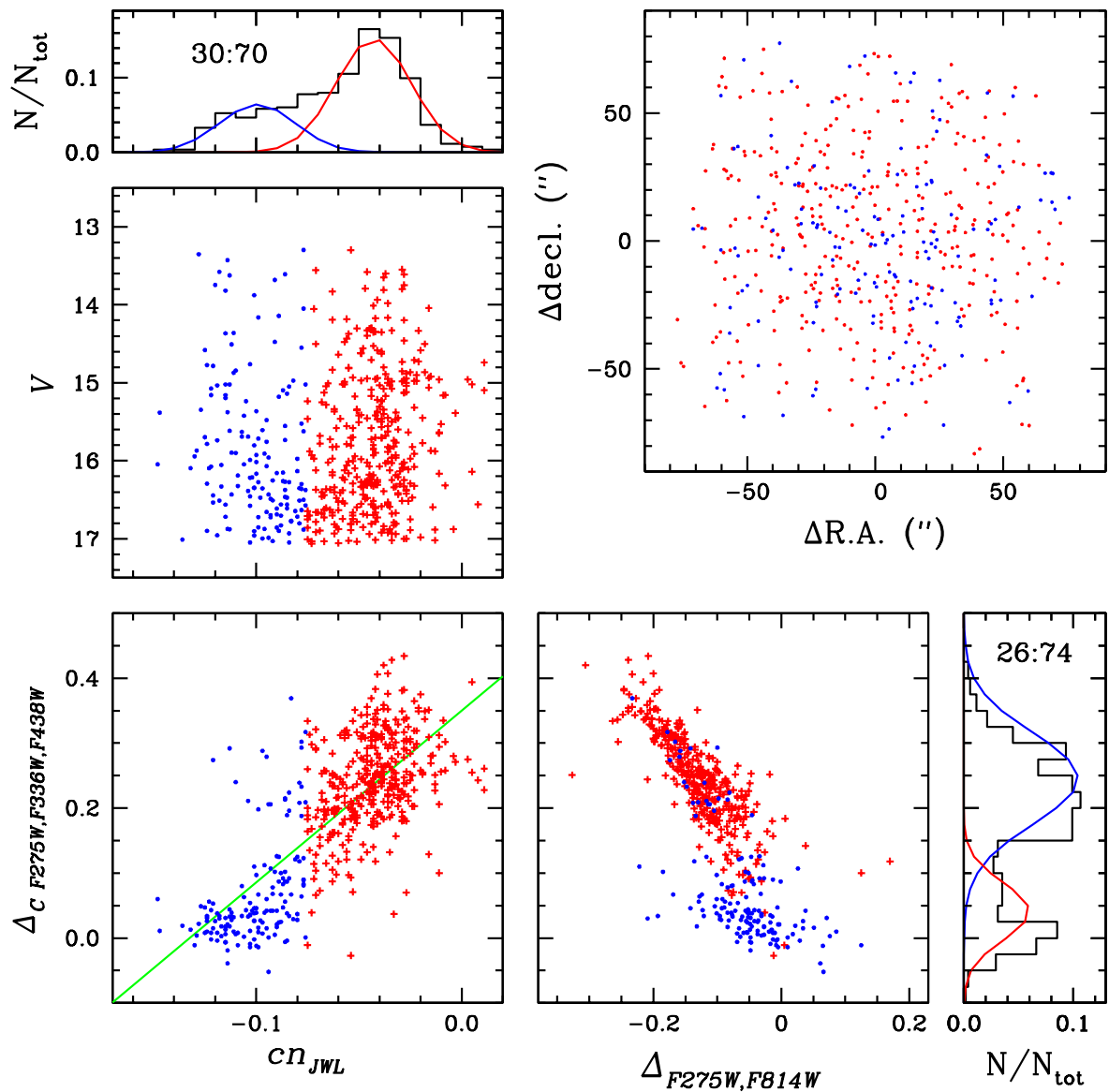}
\caption{
Same as Figure~\ref{fig:sdss}, but for \citet{piotto15}.
In general, our results are in good agreement with those of \citet{piotto15},
although some confusion in the \hst\ photometry can be seen.
}\label{fig:hst}
\end{figure}

From our \cnjwl\ versus $V$ CMD for common stars,
we obtained \nrgb\ = 30:70 ($\pm$ 3), consistent with
our previous results, and we show our results in Figure~\ref{fig:hst}.
On the other hand, the RGB number ratio 
based on the \dtrio\ is 26:74 ($\pm$ 4), still marginally consistent with 
that from our photometry.
It should be emphasized that there exists a good correlation between
our \cnjwl\ and the \dtrio.
As shown in Figure~\ref{fig:fltcomp}, the \hst\ F336W filter covers 
the whole NH band features and, furthermore unlike other ground-based
observations, it is free from the atmospheric extinction,
leading the \dtrio\ to be sensitive to the NH abundances.
Therefore, it is not a surprise that there is a tight correlation
between our \cnjwl\ and the \dtrio, which is
a photometric analogue of the NH-CN positive correlation
seen in GC RGB stars, although some confusion in the \hst\ photometry
can be seen in the figure.
We note that \citet{milone17} presented the fraction of the FG stars
in M5, finding 0.235 $\pm$ 0.013 (see their Table~2), 
and their value is significantly different from our value.
It is thought that the difference in the FG frequency arose from 
the different definitions of MSPs in both studies.
It is also important to note that their FOV for M5 covers only the radial 
distance of 0.9$r_h$ from the cluster's center, 
although M5 does not show any radial gradient in the population ratio.

We conclude that our \cnjwl\ index can outperform the \hst\ \dtrio.
Incompleteness and, perhaps, inaccuracy in any ground-based photometry
for the central part of the very dense GCs is unavoidable.
However, a definite advantage of our observations is a very large FOV
compared to the \hst\ observations. 
As mentioned before, the FOV of our instrument setup
can cover more than 55 times larger area than the \hst\ can.
The large FOV is one of the critical requirements to delineate
the complete picture of the formation of the MSPs in GCs.
We emphasize that our photometry presented here is adequate to such requirements.

\subsubsection{RGB bump magnitude: \vbump}
The helium abundance of GCs is very difficult to measure due to the absence
of helium absorption lines in the visible spectra of the cool RGB stars.
Instead, one can rely on the indirect methods to estimate the helium 
abundance using the helium sensitive features during the evolution of 
the low mass stars, which include the RGB bump luminosity 
\citep[e.g.][]{cassisi13} and the 
number ratio of the horizontal branch (HB) to RGB stars \citep{buzzoni83}.

\begin{figure}
\epsscale{1}
\figurenum{13}
\plotone{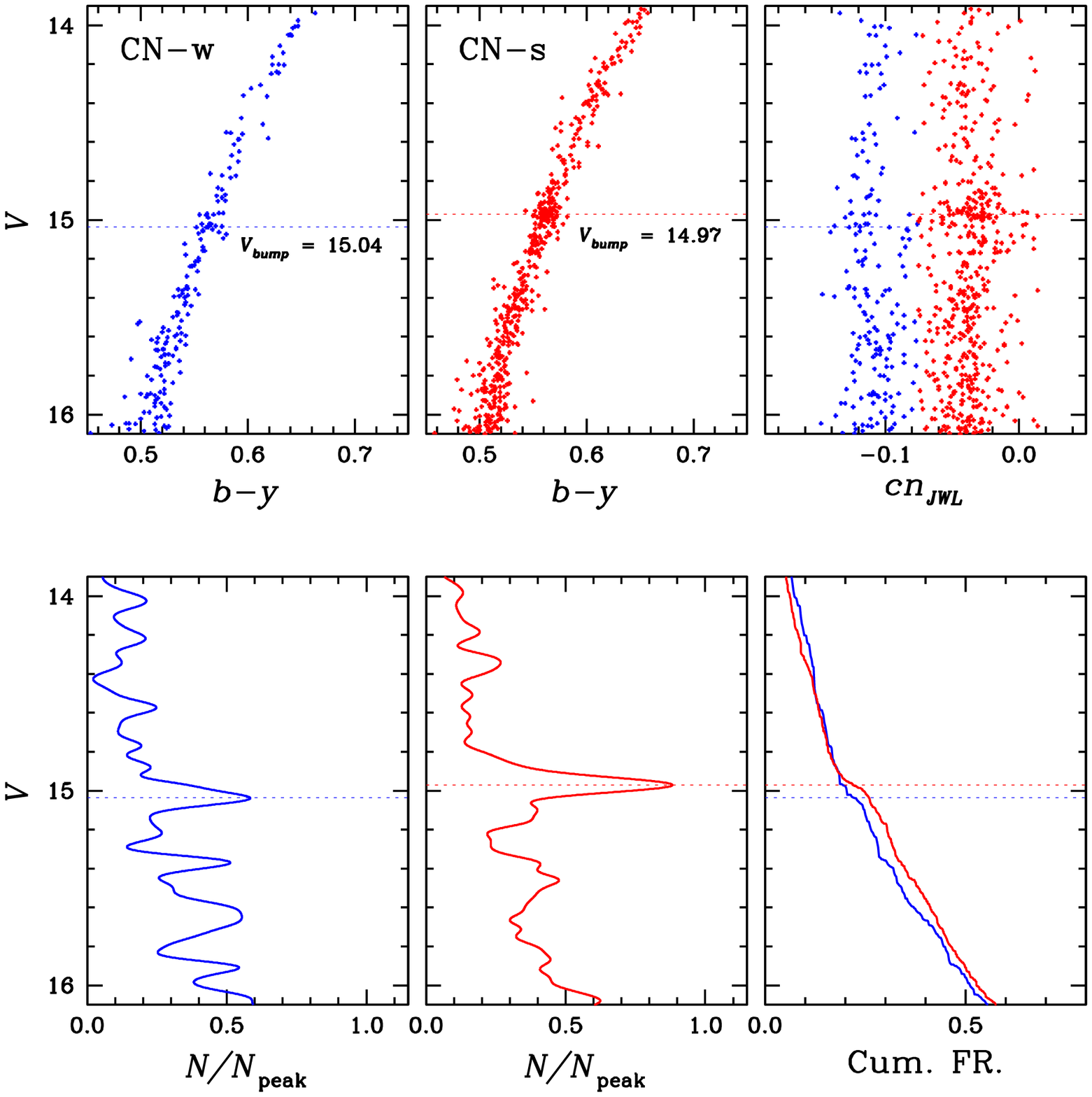}
\caption{(Upper panels)
Plots of \vby\ and \vcn\ CMDs of the \cnw\ and the \cns\ RGB stars in M5.
The dashed lines are the $V$ magnitude of the RGB bump, \vbump.
We obtained \vbump\ = 15.04 mag for the \cnw\ population 
and 14.97 mag for the \cns\ population.
Note that the \vbump\ magnitudes of each population do not show 
any radial gradient (see Table~\ref{tab:bump}).
(Lower panels) Generalized histograms for RGB stars against $V$ magnitude 
and the cumulative LFs for the \cnw\ and the \cns\ populations.
Although small, the \vbump\ of the \cns\ population is slightly brighter,
which may reflect the fact that the \cns\ population has an enhanced
helium abundance, \dy\ = 0.028.
}\label{fig:bump}
\end{figure}

We compare the RGB bump magnitudes of each populations in order to
explore the difference in the helium abundance in a relative sense.
In Table~\ref{tab:bump}, we present our measurements for RGB bump $V$ magnitudes
for each population.
We obtained \vbump\ = 15.038 $\pm$ 0.030 and 14.970 $\pm$ 0.030 for 
the \caw\ and the \cas\ RGB stars, respectively.
In particular, the \vbump\ of the \cas\ population, which is the major
component of the cluster, is in good agreement with that of
\citet{sandquist96}, who obtained 14.964 $\pm$ 0.007.
In the central part of M5 ($r \leq 1\arcmin$),
the \vbump\ magnitudes for both populations are slightly brighter than
those in the outer part of the cluster ($r > 1\arcmin$).
However, the differences are no larger than 0.01 $\pm$ 0.04 mag,
exhibiting no effective radial fluctuations in the \vbump\ magnitude
(see also Appendix~\ref{ap:s:acs}).
We show CMDs around the RGB bump for both populations
in the top panels of Figure~\ref{fig:bump}
and the differential and cumulative luminosity functions 
in the lower panels.
Although small, our measurements suggest that the visual magnitude
of RGB bump of the \cns\ population is slightly brighter by 
$\Delta$\vbump\ = 0.07 $\pm$ 0.04 mag than that of the \cnw\ population.

It is well known that at a given age the RGB bump becomes fainter 
with increasing metallicity and with decreasing helium abundance,
due to changes in the envelope radiative opacity
\citep[e.g., see][]{cassisi13}. 
As we have shown in our previous study \citep{lee15},
one can quantitatively estimate the effect of metallicity and
the helium abundance on the RGB bump luminosity.
In their Table~2, \citet{bjork06} presented how the absolute $V$ magnitude
of the RGB bump depends on both age and metallicity and we derived 
$\Delta M_{V,{\rm bump}}/\Delta$[Fe/H] $\approx$ 0.93 mag/dex for 13 Gyr
from their data.
It is very unlikely,\footnote{See Figures~\ref{fig:withoutACS} and 
\ref{fig:withACS} for the precision \hst\ \acs\ photometry of M5 \citep{anderson08},
where no discernible sign of age spread can be seen in its
main-sequence (MS) turn-off point.}
however, the age difference of about 2 Gyr between the two
populations can result in the \vbump\ magnitude of the younger population
to be 0.08 mag brighter.

The effect of helium abundance on the RGB bump can be found 
in \citet{valcarce12}. From their Figure~9, we obtained 
$\Delta m_{\rm bol} \approx  2.5\times\Delta Y$ for the isochrones
with $Z = 1.6\times10^{-3}$ and 12.5 Gyr.
The previous spectroscopic studies by others \citep{ivans01,carretta09}
and our current photometric study of the cluster suggest that
there is no metallicity spread in M5.
Assuming no age difference,
there should be no metallicity effect on the \vbump\ magnitude
and the bolometric corrections should be the same for both populations.
Hence, the difference in the  RGB bump magnitude of 0.07 $\pm$ 0.04 mag
can be translated into the difference in the helium abundance of
\dy\ = 0.028 $\pm$ 0.016, in the sense that the \cns\ RGB stars
are slightly more helium enhanced.

We conclude that the two RGB populations defined by our \cnjwl\ index
exhibit the different chemical compositions: the \cnw\ population has
high [O/Fe] and low [N/Fe], [Na/Fe] abundances, while the \cns\ population
has low [O/Fe] and high [N/Fe], [Na/Fe] abundances 
with the sign of a helium enhancement.
In the context of the self enrichment scenario, therefore, 
our \cnw\ population is equivalent to the first generation (FG) of the stars, 
while the \cns\ population is the second generation (SG) of the stars.

\subsubsection{Centers}
As a first step to compare the structural differences between the two populations,
we measured the centers of each population using three different methods; 
the arithmetic mean, the half-sphere and the pie-wedge methods
\citep[see also,][]{lee15}.

Using the coordinate of the center of the cluster measured by \citet{goldsbury}
as an initial value we chose RGB stars in each population within 
3$r_h$ ($\approx$ 320\arcsec) from the center 
of the cluster and we calculated the mean values for each group.

We obtained the offset values with respect to the coordinate of 
the cluster's center by \citeauthor{goldsbury},
(\dra, \ddec) = (8\farcs2, 5\farcs2) 
for the \cnw\ population and ($-$2\farcs5, 3\farcs2) for the \cns\ population, 
resulted in an angular separation of about 11\arcsec\ 
between the centers of both populations.
Note that the core radius of M5 is about 26\arcsec\ \citep{harris96},
and the angular separation between the centers of the two populations
is considered to be relatively small.

Similar to the simple mean calculation, we use the coordinate of the cluster by
\citeauthor{goldsbury} as an initial value, we chose RGB stars in each population 
within 3$r_h$ from the center of the cluster.
Then we divided the sphere into two halves by assuming 
the radial symmetry in the distribution of RGB stars in M5.
We compared the number of RGB stars between the two halves 
by rotating the position angle by 10 degree at a fixed coordinate of the center
and we obtained the differences in the number of RGB stars between both halves.
We repeated this calculation with varying coordinates of the center
and we derived the coordinates of the centers of each population 
with the minimum difference in the number of RGB stars between the two halves.
We obtained (9\farcs5, 0\farcs6) for the \cnw\ population and
(0\farcs6, 1\farcs5) for the \cns\ population, slightly different from
those from the simple mean method.
The angular separation between the two populations is about 9\arcsec\ and, 
again, it is relatively small compared with the core radius of the cluster.

Finally, we applied the pie-wedge method. 
We divided the sphere of a radius of 3$r_h$ into 12 different slices. 
Then we compared the number of stars in the opposing distribution.
We repeated this calculation with varying coordinates of the center
and calculated the center of each population with the minimum differences.
We obtained (6\farcs5, 0\farcs5) for the \cnw\ population and
($-$4\farcs2, 1\farcs3) for the \cns\ population.
Again, the angular separation between the two populations is about 11\arcsec\
and it is still relatively small compared with the core 
or the half-light radii of the cluster.

We summarize our results in Table~\ref{tab:cnt}.
We conclude that the coordinates of the centers of each population
from bright RGB stars can be slightly different. 
However the differences in the coordinates of the center
from various methods do not appear to be substantially large 
to claim that the centers of the two RGB populations 
are distinctively different.

\subsubsection{Radial distributions}
The radial distribution of the MSPs in GCs can provide crucial information
on the GC formation and early chemical enrichment history, although
the time scale required to homogenize the radial distributions of the MSPs
is not clear \citep[e.g., see][]{lee15}.

\begin{figure}
\epsscale{1}
\figurenum{14}
\plotone{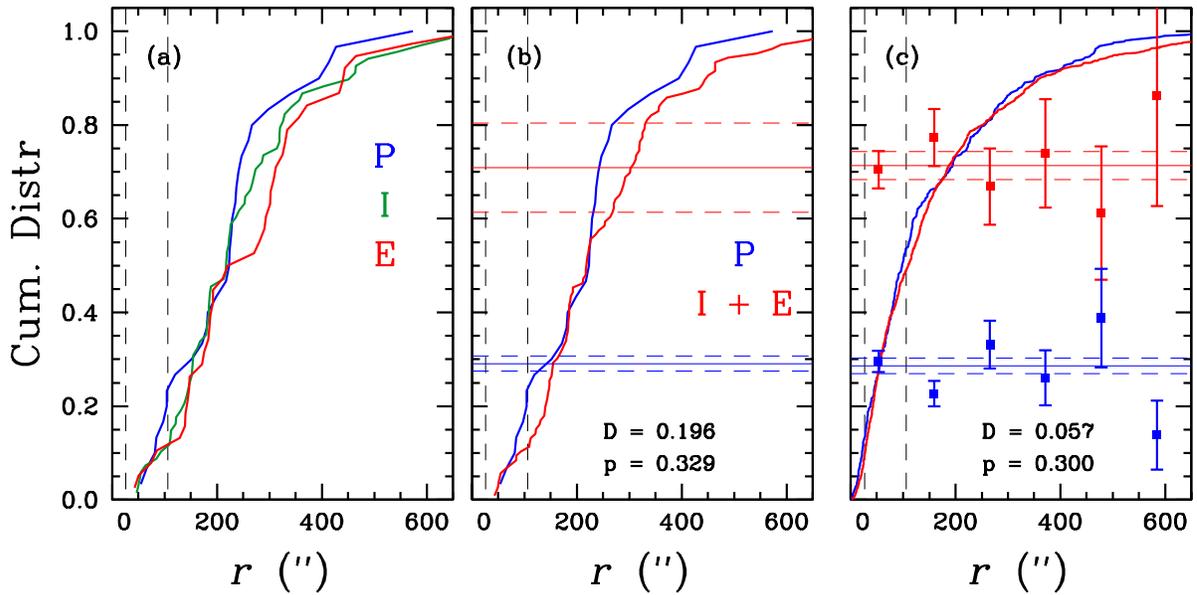}
\caption{(a)--(b) The radial distributions of the multiple stellar populations
based on the definition of individual populations by \citet{carretta09},
showing that the primordial population appears to be more centrally concentrated
than other populations.
In (b) the blue and red horizontal lines are for the fraction
of the primordial population, n(P)/n(P+I+E) = 0.291 $\pm$ 0.016, 
and the fraction of the intermediate and the extreme populations,
n(I+E)/n(P+I+E) = 0.709 $\pm$ 0.095.
The vertical black dashed lines are for the core and the half-light 
radii of the cluster.
(c) The radial distributions of the \cnw\ (blue) and the \cns\ (red)
populations in M5. The blue and the red horizontal lines
are for the fraction of the \cnw\ and the \cns\ populations,
n(\cnw)/n(\cnw\ + \cns) = 0.286 $\pm$ 0.017
and n(\cns)/n(\cnw\ + \cns) = 0.714 $\pm$ 0.030.
Our K-S test suggest that the radial distributions of the two populations
are likely drawn from the identical parent distribution.
Also note that the fractions of the \cnw\ and the \cns\ populations 
do not appear to vary against the radial distance from the center.
}\label{fig:rad}
\end{figure}

First, we examined the radial distributions of RGB stars by \citet{carretta09}.
In Figure~\ref{fig:rad}(a), we show the radial distributions of
the primordial, the intermediate and the extreme populations.
It can be clearly seen that the primordial population is the most centrally
concentrated, while the extreme population is the least centrally concentrated.
As discussed earlier, because there is no difference between the intermediate
and the extreme populations from the photometric point of view, we combine
the intermediate and the extreme populations together.
We show a comparison of their distribution with that of the primordial population
in Figure~\ref{fig:rad}(b). 
Not surprisingly, the primordial population does look to be more centrally 
concentrated than the others, however, it may be due to the small number
of sample in the spectroscopic study by \citet{carretta09}.
We performed a Kolmogorov-Smirnov (K-S) test and we obtained 
the significance level for the null hypothesis that the both distributions
are drawn from the same distribution, $p$ = 0.329, with a K-S discrepancy
of 0.196, indicating that they are likely drawn from identical parent distributions.
We also calculate the fraction of the primordial population,
n(P)/n(P+I+E) = 0.291 $\pm$ 0.016, and the fraction of the intermediate and 
the extreme populations, n(I+E)/n(P+I+E) = 0.709 $\pm$ 0.095,
consistent with our results based on the \cnjwl\ index.

Next, we examine the radial distribution of the \cnw\ and the \cns\
populations from our \cnjwl\ index and we show our result 
in Figure~\ref{fig:rad}(c),
where the radial distributions of both populations look very similar.
Our K-S test shows the significance level for the null hypothesis 
that the both distributions are drawn from the same distribution, 
$p$ = 0.300, with a K-S discrepancy of 0.057, suggesting that 
the radial distributions of the \cnw\ and the \cns\ populations
are likely drawn from the identical parent distribution. 

\begin{figure}
\epsscale{1}
\figurenum{15}
\plotone{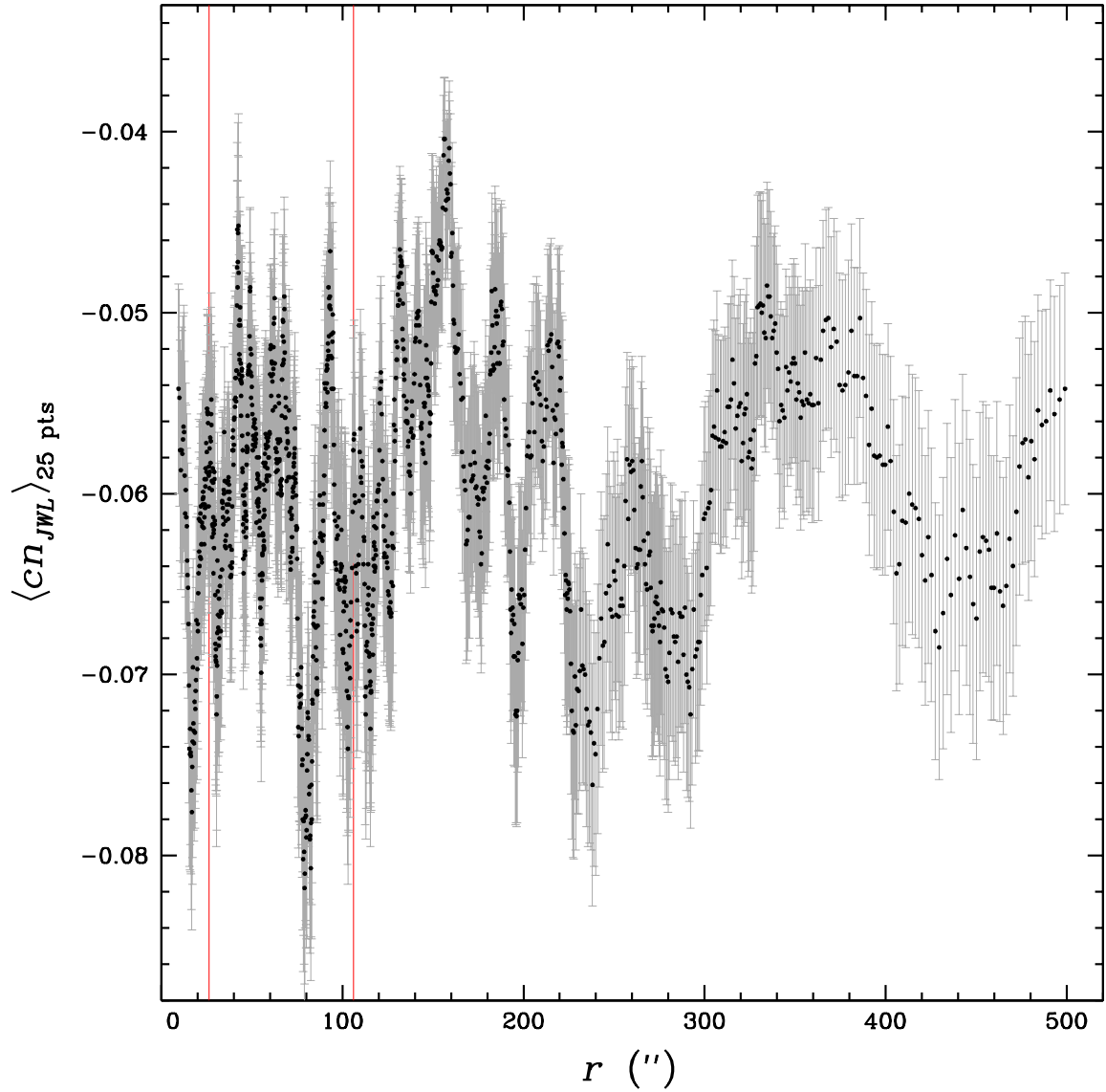}
\caption{
The moving average from the adjacent 25 points
for our \cnjwl\ index of the RGB stars with 
$-$2 $\leq$ \vvhb\ $\leq$ 2 mag against
the radial distance from the center.
The thin grey error bars represent the standard error of the mean
and vertical thin red solid lines are for the core and the half-light 
radii of the cluster.
The mean \cnjwl\ does not significantly vary against the radial distance,
consistent with Figure~\ref{fig:rad}(c).
}\label{fig:runavg}
\end{figure}

Finally, we note that the relative fractions of the \cnw\ and the \cns\
RGB populations do not appear to vary against the radial distance 
from the center, albeit the radial distribution of the \cnw\ RGB population 
show some mild fluctuations due to the small number statistics
in Figure~\ref{fig:rad}. 
In Figure~\ref{fig:runavg}, we show the moving average 
from the adjacent 25 points for the \cnjwl\ index of the all RGB stars 
with $-$2 $\leq$ \vvhb\ $\leq$ 2 mag 
against the radial distance from the center.
The moving average shows some small scale local fluctuations but 
does not show any large scale gradient in the figure,
i.e., no radial CN gradient in M5
\citep[e.g., see][for the radial CN variation in NGC~104]{chun79}.
It is thought that the flat number ratio against the radial distance 
between the two populations up to more than 5$r_h$
may suggest that M5 is already in the stage of 
complete mixing \citep[e.g., see][]{vesperini13}.

\subsubsection{Surface brightness profiles}
\begin{figure}
\epsscale{1}
\figurenum{16}
\plotone{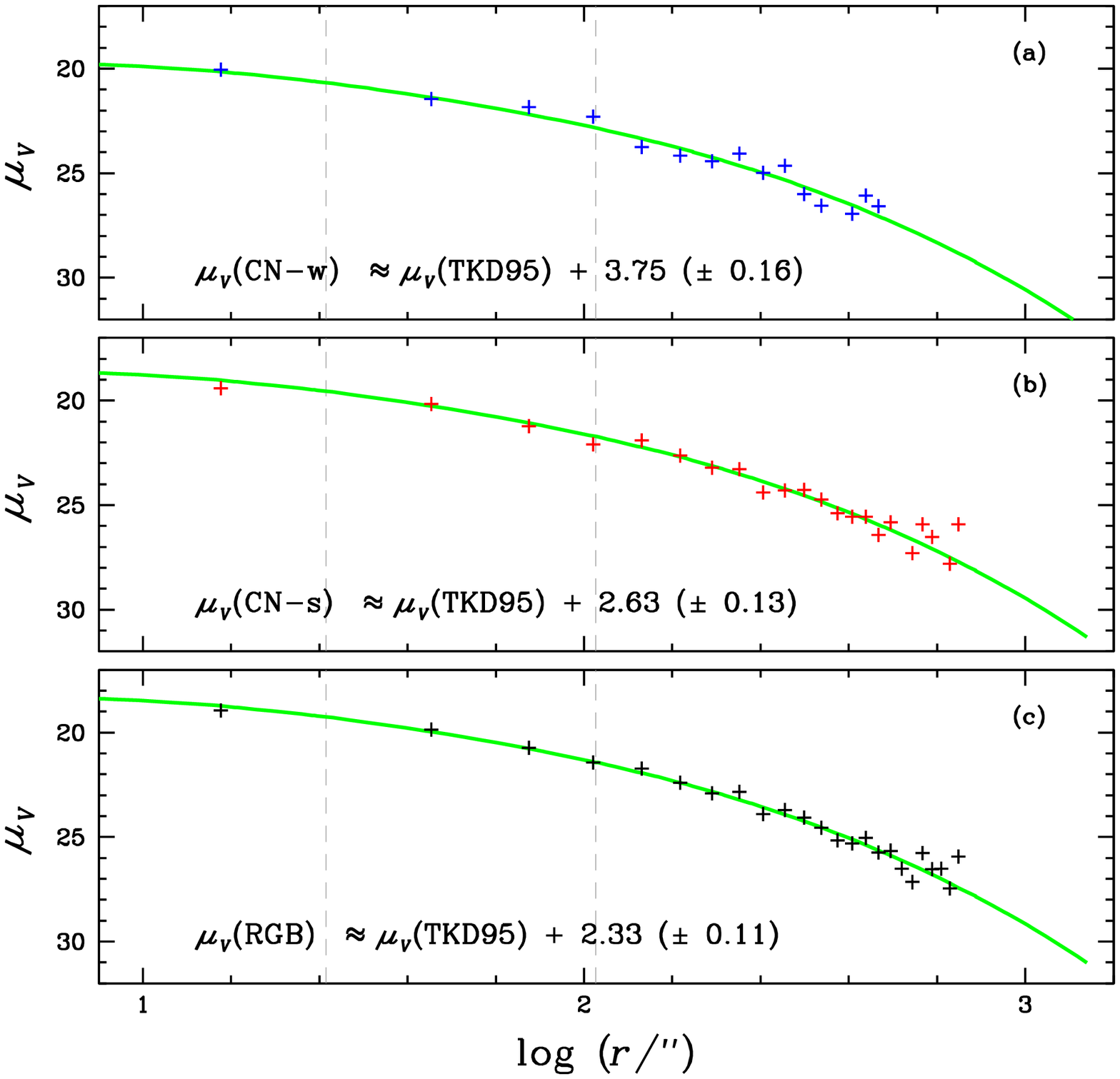}
\caption{
Surface-brightness profiles for the  RGB stars with 
$-$2 $\leq$ \vvhb\ $\leq$ 2 mag: (a) \cnw\ RGB stars only; 
(b) \cns\ stars only; and (c) all RGB stars in M5.
Green lines denote the Chebyshev polynomial fit of surface-brightness
profile of M5 by \citet{trager95}.
Our surface-brightness profiles for M5 are in excellent agreement with 
that of \citeauthor{trager95} up to 10\arcmin\ from the center of the cluster.
The vertical grey dashed lines are for the core and the half-light 
radii of the cluster.
}\label{fig:sbp}
\end{figure}

The surface brightness profile (SBP) also provides a useful means 
to compare the structural property and 
we explore the SBPs of M5 RGB stars in each population.
In Figure~\ref{fig:sbp}, we show the SBPs of the bright RGB stars
($-2 \leq$ \vvhb\ $\leq 2$ mag) in the \cnw\ and the \cns\ populations.
Also shown is the Chebyshev polynomial fit of the M5's SBP by \citet{trager95}.
Our SBP measurements for both populations are in excellent agreement
with that of \citet{trager95} up to 10\arcmin\ (more than 5$r_h$)
from the center.
Our SBP measurements also confirm our previous results
that both populations should have
very similar radial distribution and, furthermore, very similar structure.
The similarity in the SBP for both populations also suggests that
they are already in a well-mixed state.

\subsubsection{Spatial distributions}
In our previous study of M22 \citep{lee15}, we showed that the spatial
distributions can provide a very powerful diagnostics
to compare the structural properties between MSPs,
which are closely related to and affected
by their internal kinematics.
Here, we explore the spatial distributions of RGB stars in M5.

\begin{figure}
\epsscale{1}
\figurenum{17}
\plotone{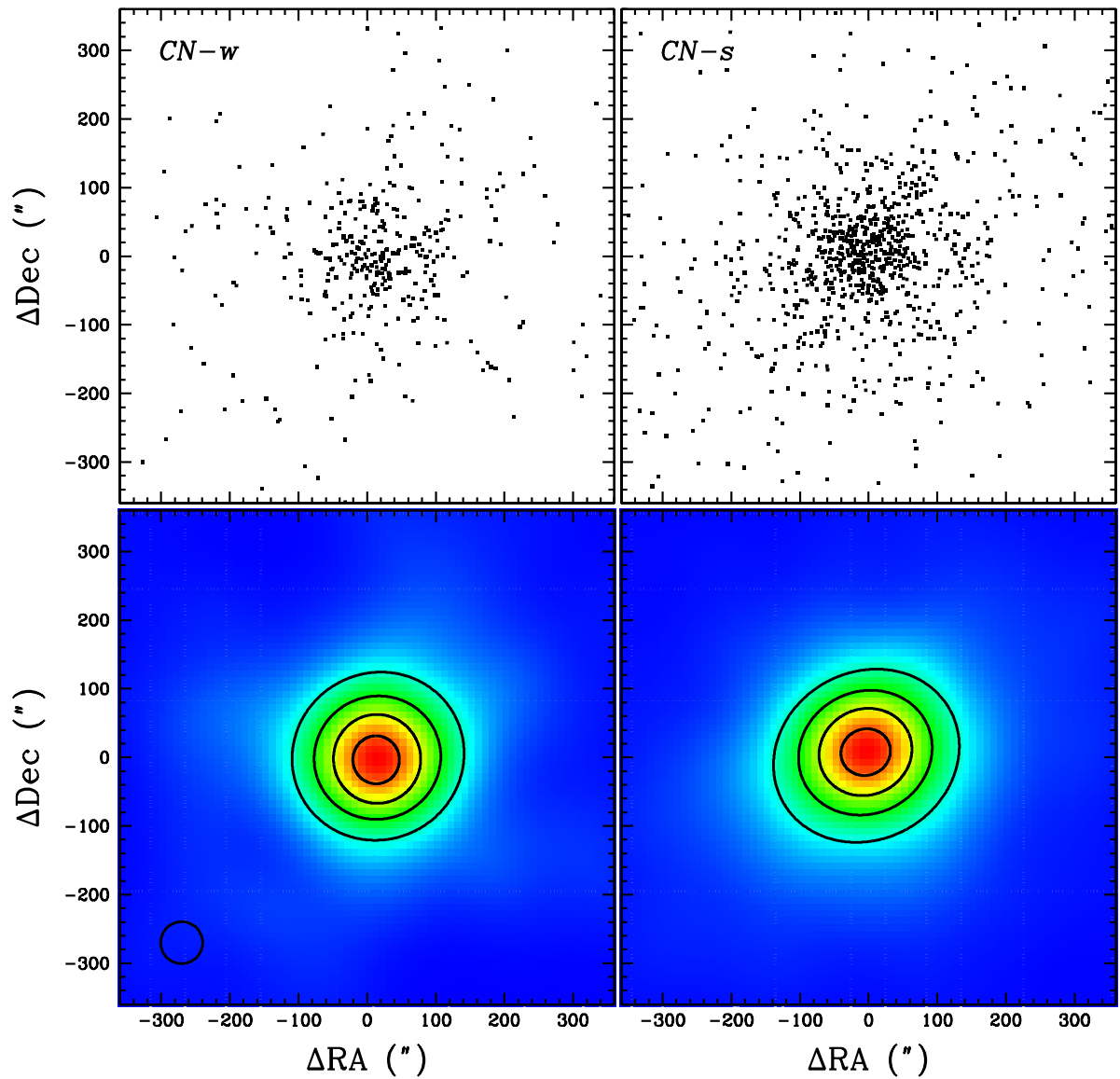}
\caption{
(Upper panels) Spatial distributions of the \cnw\ and the \cns\ RGB stars in M5.
(Lower panels) The smoothed contour maps using the fixed Gaussian kernel, where
we show the iso-density contour lines for 90, 70, 50, and 30\% 
of the peak values for both populations. We also show the FWHM of our adopted 
Gaussian kernel in the lower left panel of the figure. 
Note that the spatial distribution of the \cns\ RGB population 
is more elongated along the NE-SW direction.
}\label{fig:density}
\end{figure}

In the top panels of Figure~\ref{fig:density}, 
we show the projected spatial distributions 
of the \cnw\ and the \cns\ populations in M5.
In the figure, the offset values of the projected right ascension and 
the declination in the units of arcsec were calculated 
using the transformation relations given by \citet{vandeven},
\begin{eqnarray}
\Delta {\rm R.A.} &=& \frac{648000}{\pi}\cos\delta\sin\Delta\alpha, \\
\Delta {\rm decl.} &=& \frac{648000}{\pi}(\cos\delta\cos\delta_0 - 
\cos\delta\sin\delta_0\cos\Delta\alpha), \nonumber
\end{eqnarray}
where $\Delta\alpha = \alpha - \alpha_0$ and $\Delta\delta = \delta - \delta_0$,
and $\alpha_0$ and $\delta_0$ are the coordinate of the center.
For our calculations, we adopted the coordinate for the cluster center
measured by \citet{goldsbury}.

In the lower panels of the figure, we show smoothed density distributions of
RGB stars along with iso-density contours for each population.
For the smoothed density distribution of each population, 
we applied a fixed Gaussian kernel estimator algorithm 
with a full-width at the half-maximum (FWHM) of 52\arcsec\ \citep{silverman}.
We show the FWHM of our Gaussian kernel in the lower left panel of the figure.
To derive the iso-density contour for each population,
we applied the second moment analysis \citep{dodd,stone}
and we show the iso-density contour lines for 90, 70, 50, and 30\% 
of the peak values for both populations.

\begin{figure}
\epsscale{1}
\figurenum{18}
\plotone{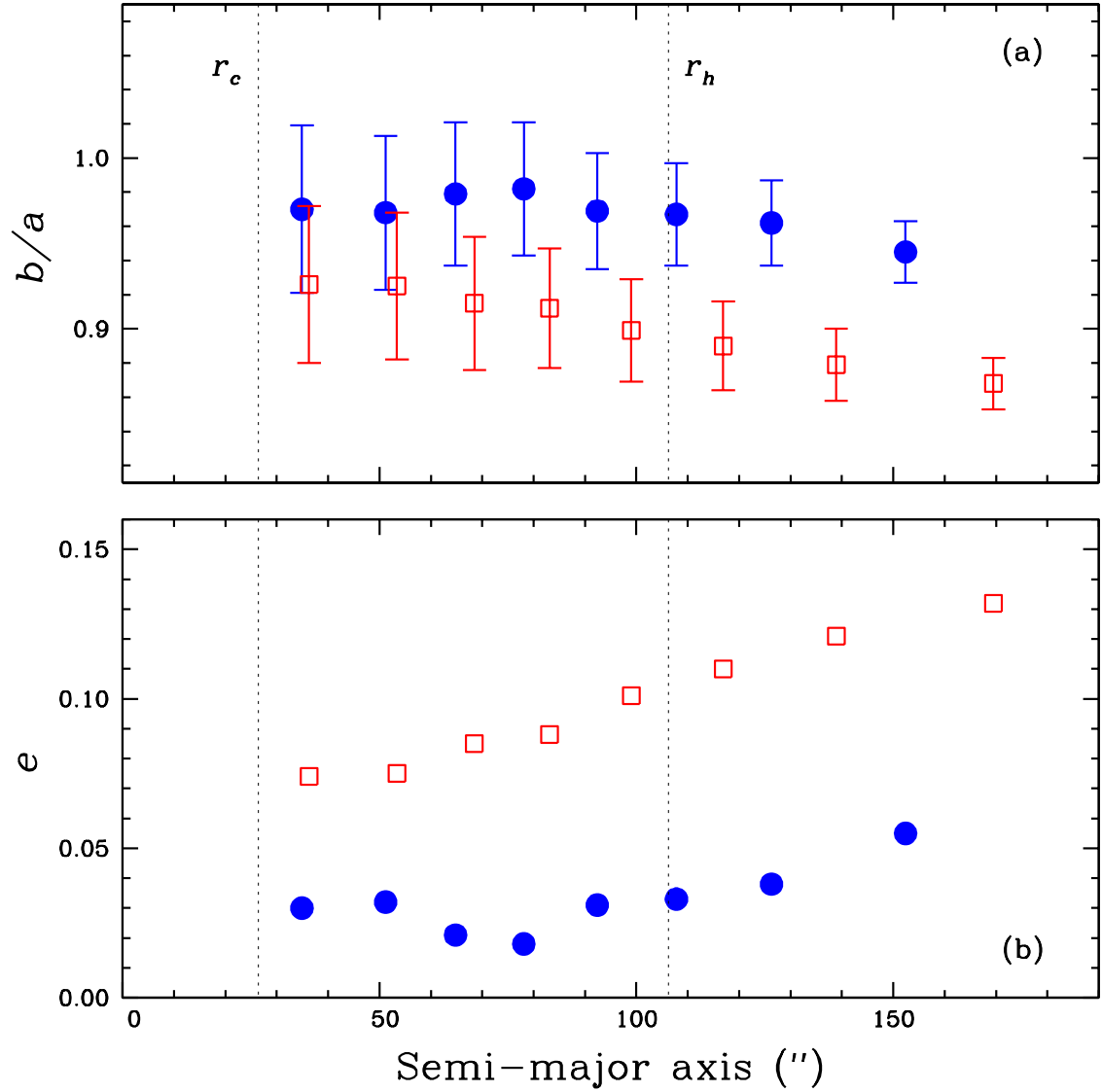}
\caption{The run of the axial ratio, $b/a$, the ellipticity, $e$ ($= 1 - b/a$)
of the \cnw\ (blue) and the \cns\ (red) populations 
against the major axis, $a$. 
Note that the ellipticity of the \cns\ population is significantly larger
than that of the \cnw, consistent with our results shown 
in Figure~\ref{fig:density}.
}\label{fig:ellip}
\end{figure}

The projected distribution of the \cns\ population
is more spatially elongated than that of the \cnw\ population.
The major axis of the projected density profile of the \cns\ population
is placed along the NW-SE direction.
In Table~\ref{tab:ellipse}, we show the axial ratio, $b/a$,
and the ellipticity, $e$ $(= 1 - b/a)$. 
The radial distributions of the axial ratio and the ellipticity 
can be found in Figure~\ref{fig:ellip}.
As shown, the ellipticity of the \cns\ population
is much larger than that of the \cnw\ population 
up to $\approx$ 2$r_h$ of the cluster.
This is not expected or difficult to anticipate 
from the cumulative radial distributions or the SBPs of each population
as shown in Figures~\ref{fig:rad} and \ref{fig:sbp}. 
Our results vividly demonstrate the importance of using the spatial
distribution at rather large radial distance
to examine the structural differences between the MSPs in GCs.
As will be shown below, it is believed that 
the elongated projected spatial distribution
of the \cns\ RGB population is inextricably linked with its fast 
projected rotation.

\subsubsection{Projected Rotations}
In our previous work \citep{lee15}, we showed that the peculiar GC M22
have different projected rotations between the two populations,
providing a strong constraint to the formation scenario of M22.
The rotation of M5 has not been reported yet, especially for
the individual stellar populations in M5.
Using the radial velocity measurements by \citet{carretta09},
we explore the projected rotation of RGB stars 
in each population.

We estimated the amplitude of the mean rotation of the cluster 
using the method described by \citet{lane09}.
Assuming an isothermal rotation, the mean rotation can be measured by dividing
the cluster in half at a given position angle and calculating differences
between the average velocities in the two halves.
During our calculation, we used the RGB stars within 3\arcmin\ 
from the cluster's center, which is about the largest radius of the ellipse 
that we showed in Figure~\ref{fig:ellip}.
We repeated this calculation by increasing the position angle of the boundary
of the two halves by 10$^\circ$.
Then the net rotation velocity is the half of the amplitude 
of the sinusoidal function in the differences between the average velocities
in the two halves.

\begin{figure}
\epsscale{1}
\figurenum{19}
\plotone{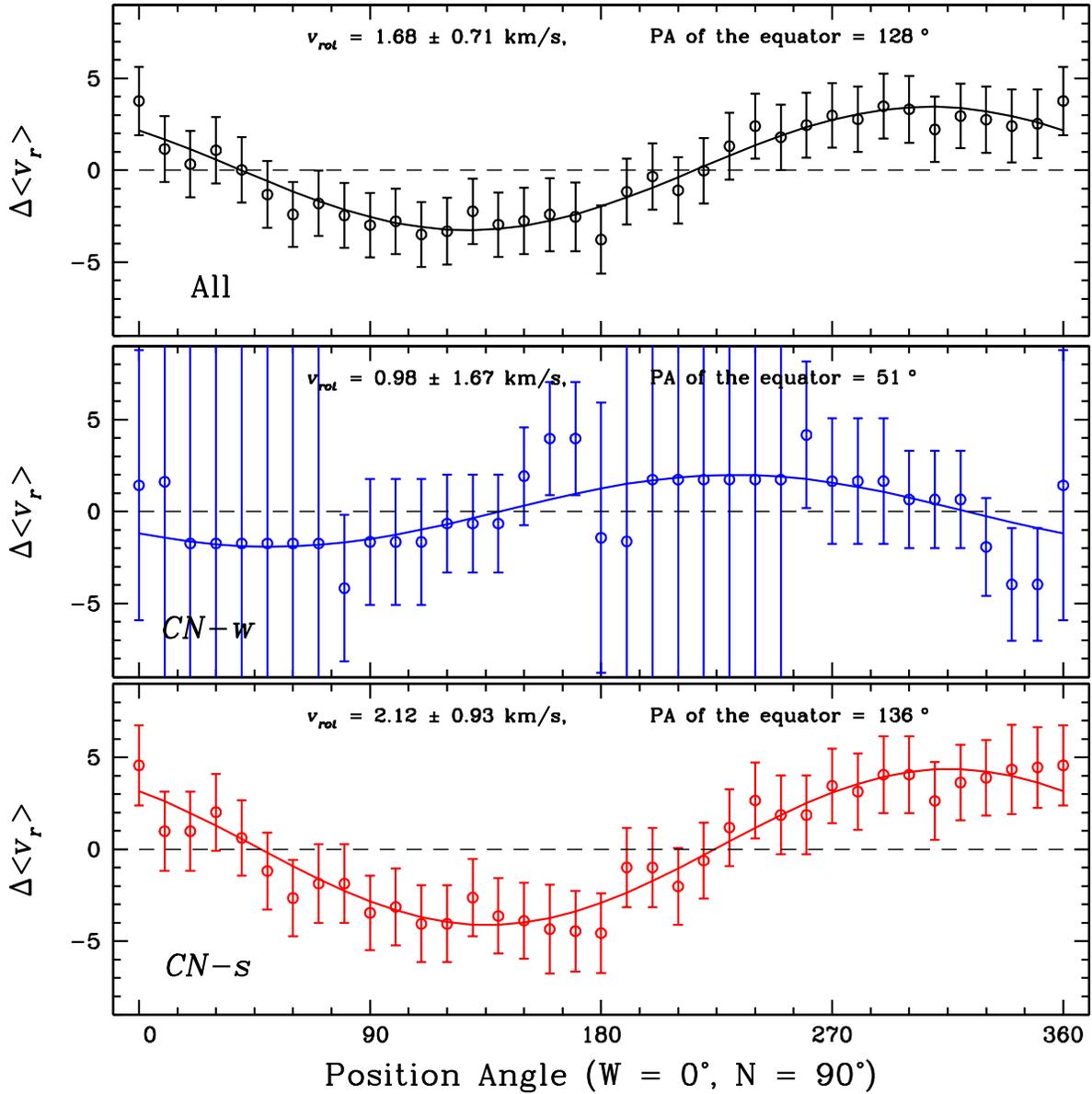}
\caption{The difference in the mean radial velocities between both 
hemispheres against its position angle, where the net rotation is the half
of the amplitude of the sinusoidal fit.
It is evident that the \cns\ population shows a substantial rotation, 
while the \cnw\ population does not.
The position angle refers to that of the equator and
the error bar is the error of the mean.
The position angle of the equator of the projected rotation for 
the \cns\ population, 136$^\circ$ (i.e., along the NE-SW direction), 
is consistent with the \cns\ RGB distribution shown in Figure~\ref{fig:density}.
}\label{fig:rot}
\end{figure}

In Figure~\ref{fig:rot}, 
we show the differences in the mean radial velocities as a function of
the position angle (West = 0$^\circ$ and North = 90$^\circ$)
along with the best-fitting sine function.
We obtained the mean rotation velocity of 1.7 $\pm$ 0.7 \kms\ 
for all RGB stars and the position angle of the equator of 128$^\circ$ 
(i.e., along the NE-SW direction).
In the middle and the bottom panels,
we show plots of differences in the mean radial velocities for the \caw\
and the \cas\ populations. 
We obtained the mean rotation of 1.0 $\pm$ 1.7 \kms\ with 
the position angle of 51$^\circ$ for the \cnw\ population and
2.2 $\pm$ 0.9 \kms\ with the position angle of 136$^\circ$ 
for the \cns\ population.
It is thought that the \cnw\ population does not appear to have
a net projected rotation, while \cns\ population shows a substantial net
projected rotation.
The position angle of the axis of the projected rotation for 
the \cns\ population is 136$^\circ$ (i.e., along the NE-SW direction), 
which is in accord with the projected spatial distribution
of the \cns\ population if its elongated shape is mainly 
due to its fast rotation.

In the future, more systematic radial velocity measurements of M5 RGB stars 
will be very desirable and will shed more light 
on the kinematic property of each population.

\subsection{Asymptotic giant stars}\label{s:agb}
Until recently, the AGB populations in the GC systems were rather neglected,
despite of their intrinsic brightness.
However, in the context of the MSP, the importance of the AGB star
became greater \citep[e.g.\ see][]{pilachowski96,sneden00}. 
For example, the AGB stars can provide a wonderful opportunity to understand 
the stellar structure and evolution of the low mass stars 
\citep[e.g., see][]{norris81,campbell13}.
For decades, it has been known that some GCs have different CN contents 
in the AGB phase than in the RGB phase.
In their pioneering study, \citet{norris81} found that
NGC~6752 does not appear to have \cns\ AGB stars,
while the \cns\ RGB stars are the major component,
which led them to propose that the \cns\ RGB stars failed to evolve into 
the AGB phase (the so-called \agbm\ stars).
Interestingly, M5 is in the opposite case. 
\citet{smith93} found a deficiency of the \cnw\ AGB stars in M5.

The trouble with the AGB stars is that they are much
rarer than their progenitors; the MS, RGB and HB stars.
For example, the typical number ratio of the AGB stars to the RGB stars
in Galactic GCs is very small, $n$(AGB)/$n$(RGB) $\approx$ 0.1 - 0.2,
depending on the definition of the AGB and the RGB phases
\citep[e.g., see][]{buzzoni83,gratton10}.
As a consequence, statistical fluctuations introduced from the small number 
statistics can not be avoided.
The situation in the spectroscopic study of AGB stars will be worse,
because the AGB stars in the central part of GCs would be very
difficult to observe, which is the case for M5 as we will show below.
In this regard, our \cnjwl\ index can open a new era on the MSP study 
of the AGB stars, because our approach is applicable to the central
part of the cluster, where the spectroscopic method has an ultimate
limitation that can not be avoided due to crowdedness.
If so, our results can provide the complete census on the true nature 
of the MSP of the AGB stars.

\subsubsection{The relative AGB frequency, $n$(AGB)/$n$(RGB), as an indicator
for the missing AGB population in GCs}
\begin{figure}
\epsscale{1}
\figurenum{20}
\plotone{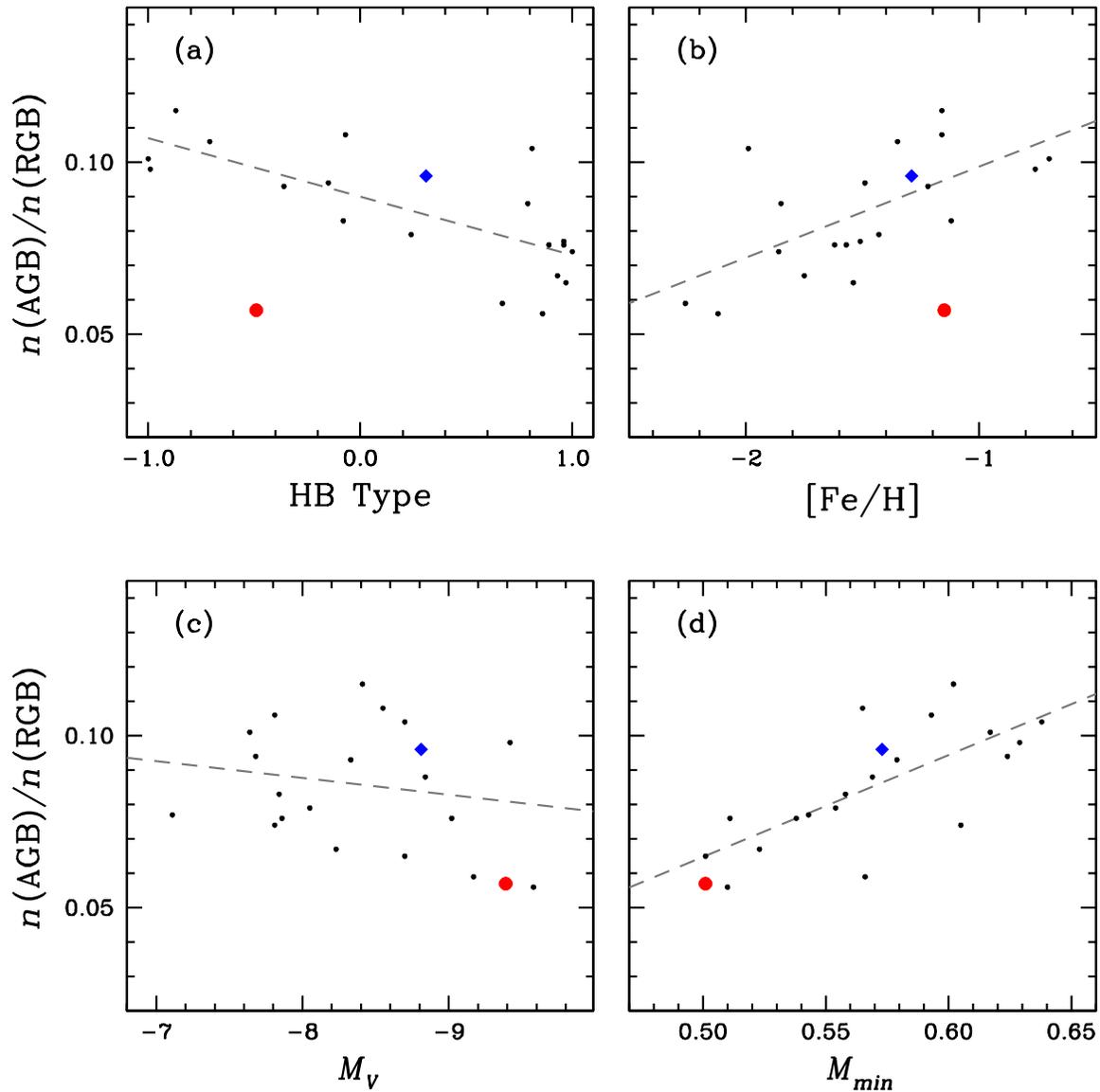}
\caption{The relative AGB frequencies for GCs studied
by \citet{gratton10}.
The blue diamonds and red circles denote M5 and NGC~2808, respectively.
Note that M5 is located near the upper limit of the relative AGB
frequency, indicating that the missing AGB population would be negligible
for M5.
The peculiar GC NGC~2808 appears to be an outlier and
it is not included in deriving linear correlations in each panel.
}\label{fig:gratton}
\end{figure}

In Figure~\ref{fig:gratton}, we show plots of the relative AGB frequencies
against the HB type (HBT), metallicity, absolute integrated magnitude
and the minimum mass along the HB  using the data by \citet{harris96}
and \citet{gratton10}.
In their recent study, \citet{gratton10} proposed that the relative AGB 
frequency  is correlated with the minimum mass along the HB and metallicity.
We calculated the linear correlations and show our results in 
Table~\ref{tab:gratton}, confirming what \citet{gratton10} proposed.

The relative AGB frequency as functions of the HBT or metallicity
may depend on various factors; the metallicity dependency on
the time spent on the RGB and the AGB evolutionary phases, 
the mass loss, and etc.
If our linear correlations between the relative AGB frequency
and other parameters are mainly related to the relative fraction
of the \agbm\ stars, our results can be used to estimate
the missing AGB populations in a given GC.
If so, Figure~\ref{fig:gratton} can provide a strong line of evidence that 
the missing AGB population owing to the presence of the \agbm, 
preferentially in the \cns\ population with enhanced helium abundances, 
would be negligible in M5,
since M5 is located near the upper limit of the relative AGB frequency.
In other word, the AGB stars in both populations in our study should
represent the complete sample in terms of the stellar evolution.

\subsubsection{Multiple AGB populations in M5}
It is fortunate that AGB stars in GCs are not hot enough to suppress 
the CN band formation in their atmospheres. 
Therefore, if there exists a spread 
or a variation in the CN abundance among AGB stars, 
we should be able to tell the abundance differences using our \cnjwl\ index.
However, at a fixed visual magnitude, the AGB stars are slightly 
warmer than the RGB stars by 100 - 200 K and, therefore, 
the CN band absorption strengths in the AGB stars
are expected to be weaker than those in the RGB stars
due to the temperature effect.\footnote{The AGB stars have
slightly larger surface gravities than the RGB stars do.
However, the CN band is rather insensitive to the change
in the surface gravity.}
Therefore, it is expected that if there exist MSPs in the AGB phases,
the separation in our \cnjwl\ index 
between \cnw\ and the \cns\ AGB populations would become
smaller than that in the RGB populations.

\begin{figure}
\epsscale{1}
\figurenum{21}
\plotone{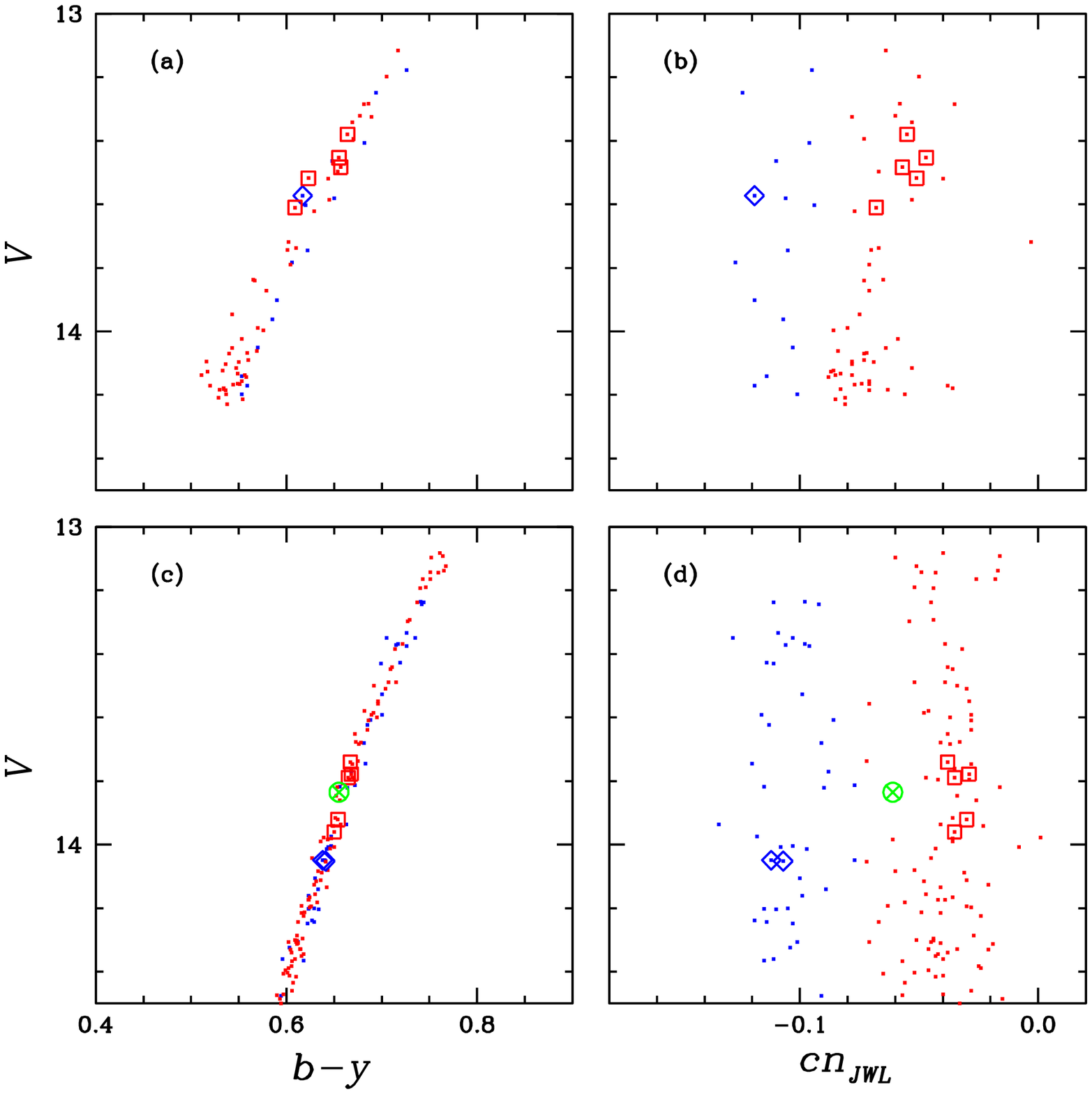}
\caption{(a) The \vby\ CMD for M5 AGB stars. We also show AGB stars
studied by \citet{smith93}.
The blue open diamond is the CN-weak AGB star (IV-26) and the red open squares
the CN-strong AGB stars by \citet{smith93}.
(b) The \vcn\ CMD for M5 AGB stars. 
Note the discrete double AGB sequences, as already shown in Figure~\ref{fig:deltas}.
The blue dots are the \cnw\ and the red dots are the \cns\ AGB stars
based on the EM estimator, consistent with the classification
made by \citet{smith93}.
(c)--(d) Same as (a)--(b) but for RGB stars. The green open circle
is for the CN-intermediate RGB star (II-50).
}\label{fig:agb}
\end{figure}

In Figure~\ref{fig:agb}(a-b), we show the \vby\ and the \vcn\ CMDs 
for AGB stars in M5, where the discrete double AGB sequences 
in the \cnjwl\ index can be clearly seen as in the RGB sequence.
Note that we do not include the AGB stars near the RGB tip, where
it becomes difficult to tell the AGB from the RGB sequences.
Also shown are the six AGB stars\footnote{Since \citet{smith93} did not 
provide the positions of the two AGB stars (S344 and S445)
in their Table~1, we were not able to match them in our results.} 
studied by \citet{smith93}, who performed
a low resolution spectroscopic study to investigate the CN distributions 
of the RGB and the AGB stars in M5.
Their \scn\ measurements for the RGB and the AGB stars are in excellent
agreement with our \cnjwl\ index, although the extent of the separation
between the two populations is smaller in the AGB phase, which is
thought to be the temperature effect as we mentioned before.
This was also noted by \citet{smith93}, who found a smaller spread
of CN band strengths in the AGB stars than in the RGB stars in M5. 
As shown in the figure, the \cnjwl\ values of the \cnw\ AGB and RGB stars
are in accord with each other, while the mean \cnjwl\ value 
of the \cns\ AGB stars are slightly smaller than that of the \cns\ RGB stars.

\begin{figure}
\epsscale{1}
\figurenum{22}
\plotone{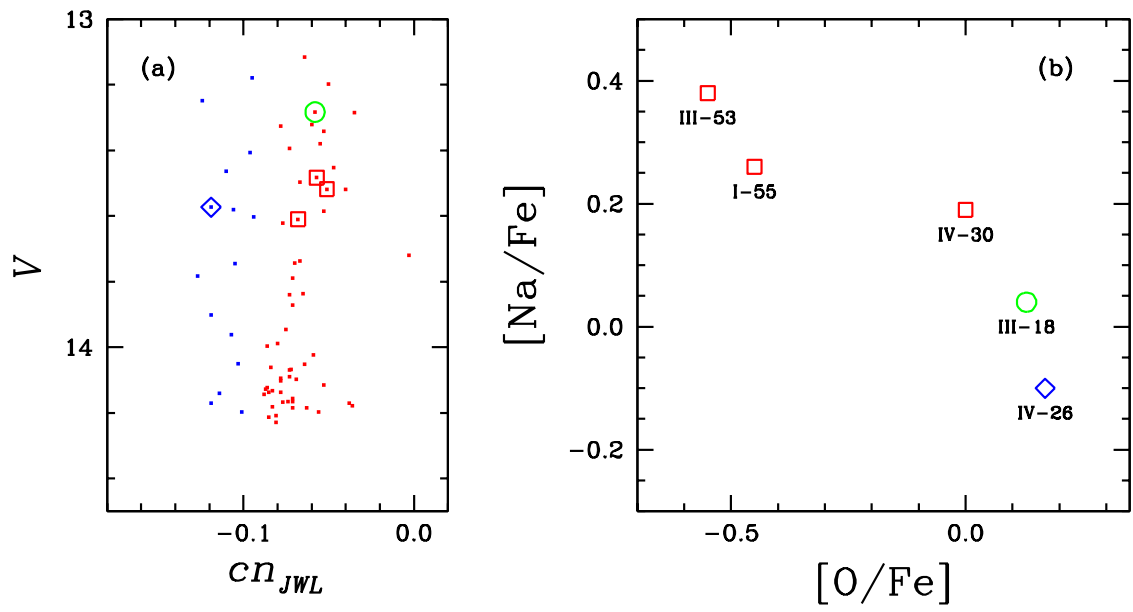}
\caption{(a) Same as Figure~\ref{fig:agb}(d), but for AGB stars
studied by \citet{ivans01}.
(b) The Na-O anticorrelation of M5 AGB stars.
The Na-O anticorrelation is closely related with the double AGB sequences
in M5, similar that can be seen in the RGB sequence.
}\label{fig:agbivans}
\end{figure}

The high resolution spectroscopic study of the AGB stars in M5 by
\citet{ivans01} also confirms that our \cnjwl\ index works nicely
for the MSPs of the AGB stars.
In Figure~\ref{fig:agbivans}, we show the \vcn\ CMD and 
the Na-O anticorrelation of the AGB stars studied by \citet{ivans01}.
Similar to what showed for the RGB stars in Figure~\ref{fig:nao},
the \cns\ AGB stars based on our \cnjwl\ index have lower oxygen
and higher sodium abundances than the \cnw\ AGB star do.
Therefore, the \cnw\ AGB stars shown in Figures~\ref{fig:agb} and 
\ref{fig:agbivans} are equivalent to the \cnw\ RGB stars.

\subsubsection{Radial distributions of AGB stars}
\citet{smith93} first noted that the majority of the AGB stars in their sample
(seven out of eight) belongs to the \cns\ population in M5
\citep[see also][]{smith13}.
As discussed before, the lack of the \cns\ AGB population in NGC~6752
can be naturally explained by the presence of the AGB-manqu\'e stars.
However, the deficiency of the \cnw\ AGB population in M5
can not be easily explained by the theory of the stellar evolution 
of the low-mass stars.
It is thought that the deficiency of the \cnw\ AGB stars is likely originated
from the stochastic truncation of the \cnw\ AGB stars in the outer region,
which ultimately lead the \cnw\ AGB population to be more centrally concentrated.

First, we calculate the number ratio between the two AGB populations. 
Assuming a bimodal distribution, we applied the EM algorithm for 
the two-component Gaussian mixture distribution model to calculate
the contributions from two groups of stars.
In an iterative manner, we derive the probability of individual AGB stars
for being the \cnw\ and the \cns\ AGB populations.
We obtained the number ratio of \nrgb\ = 21:79 ($\pm$ 7),
which is marginally consistent with that from RGB stars,
\nrgb\ = 29:71 ($\pm$ 2).
The relative fraction of the \cnw\ AGB stars by \citet{smith93}, \nrgb\ = 1:7,
is considered to be very small compared to ours.

\begin{figure}
\epsscale{1}
\figurenum{23}
\plotone{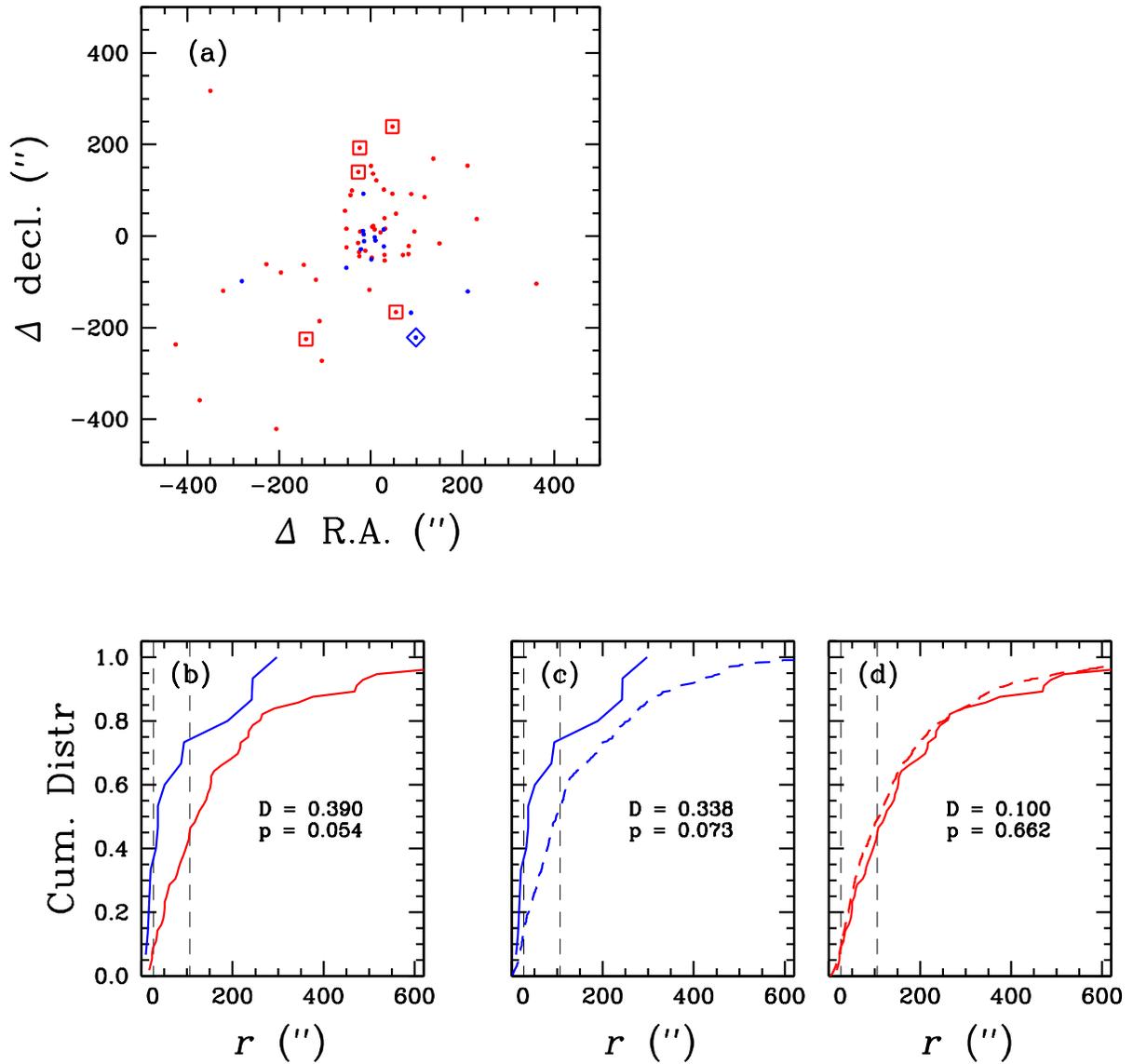}
\caption{(a) 
The spatial distributions of the \cnw\ (blue dots) and the \cns\ (red dots)
AGB stars in M5. The blue open diamond and the red open squares
are those studied by \citet{smith93}.
Note that the distribution of the \cns\ AGB stars is more elongated,
stretched from NW to SE, similar to that of the \cns\ RGB stars.
(b) The radial distribution of the M5 AGB stars. The blue solid line
denotes the \cnw\ AGB stars and the red solid line the \cns\ AGB stars.
The \cnw\ AGB stars are more centrally concentrated than the \cns\ AGB stars are.
The K-S test indicates that they are likely drawn from the different
parent distributions.
(c) A comparison of the radial distribution of the \cnw\ RGB (the dashed line)
and the \cnw\ AGB (the solid line) stars. 
The two groups of stars are likely drawn from the different parent populations.
(d) A comparison of the radial distribution of the \cns\ RGB (the dashed line)
and the \cns\ AGB (the solid line) stars. 
The two groups of stars are most likely drawn from the same parent population.
}\label{fig:agbdistr}
\end{figure}

In Figure~\ref{fig:agbdistr}(a), we show the spatial distributions 
of the \cnw\ and the \cns\ AGB stars, along with
the six AGB stars studied by \citet{smith93}.
In the figure, it is evident that the \cns\ AGB stars are preferentially 
located along the NW-SE direction,
consistent with the spatial distribution of the \cns\ RGB stars.
This also indicates that both the AGB and the RGB stars may
share the same structural and, furthermore, kinematical properties.
We also note that the \cnw\ AGB stars are more centrally 
concentrated than the \cns\ AGB stars are, which can be clearly seen
in Figure~\ref{fig:agbdistr}(b).
We performed a K-S test and we obtained the significance level
of being drawn from the identical population is 0.054, 
with a K-S discrepancy of 0.390, suggesting that they are likely drawn from 
the different parent distributions.

In Figure~\ref{fig:agbdistr}(c-d), we show comparisons of
the radial distributions of the AGB populations 
and those of the RGB populations.
For the \cnw\ population, the radial distribution of the AGB
stars does not agree with that of the RGB stars.
Our K-S test may suggest that the \cnw\ RGB and AGB stars
are likely drawn from different parent distributions,
with the the significance level of being drawn form the identical population
of 0.073.
On the other hand, the radial distribution of the \cns\ AGB stars
is in excellent agreement with that of the \cns\ RGB stars.
Our K-S test strongly suggest that the two populations are most 
likely drawn from the identical parent distribution, 
with the significance level of 0.662.
Again, our statistical tests indicate that the \cns\ AGB and RGB stars
share the same structural property, while the \cnw\ AGB and RGB stars do not.
As shown in Figures~\ref{fig:agb} and \ref{fig:agbivans}, the number
ratio of the \cnw\ AGB stars is small and the radial distribution
of the \cnw\ AGB stars may suffer from the unavoidable effect
from the small number statistics.

The cumulative distributions for the \cnw\ and \cns\ AGB stars look different,
as we already showed in Figure~\ref{fig:agbdistr}(b).
The use of the cumulative distribution has some advantages in the statistical 
evaluations of the data, however, some subtle differences can be missed 
with it.
In Table~\ref{tab:agbrgb}, we show the number ratios of the AGB to
the RGB stars for each population at different radial zones.
Note that these number ratios are not the same as those by \citet{gratton10},
as already shown in Figure~\ref{fig:gratton}, who adopted different definitions
of the AGB and the RGB phases in their calculations.
In the table, the number ratios of the AGB stars to the RGB stars for 
both populations are in excellent agreement in the central part of the cluster 
($r \leq r_h$), with the number ratios of 0.048 
with slightly different estimated errors. 
The number ratio of the \cns\ population in the outer part,
$r_h < r \leq 5r_h$ , is also in good agreement, with that of 0.056 $\pm$ 0.012.
However, that of the \cnw\ population in the outer part is significantly 
smaller, with the number ratio of 0.028 $\pm$ 0.014, and does not agree with others.
Therefore, what makes different cumulative distribution for the \cnw\ population is 
the outer part of the cluster, where the number of stars is very small.

\begin{figure}
\epsscale{1}
\figurenum{24}
\plotone{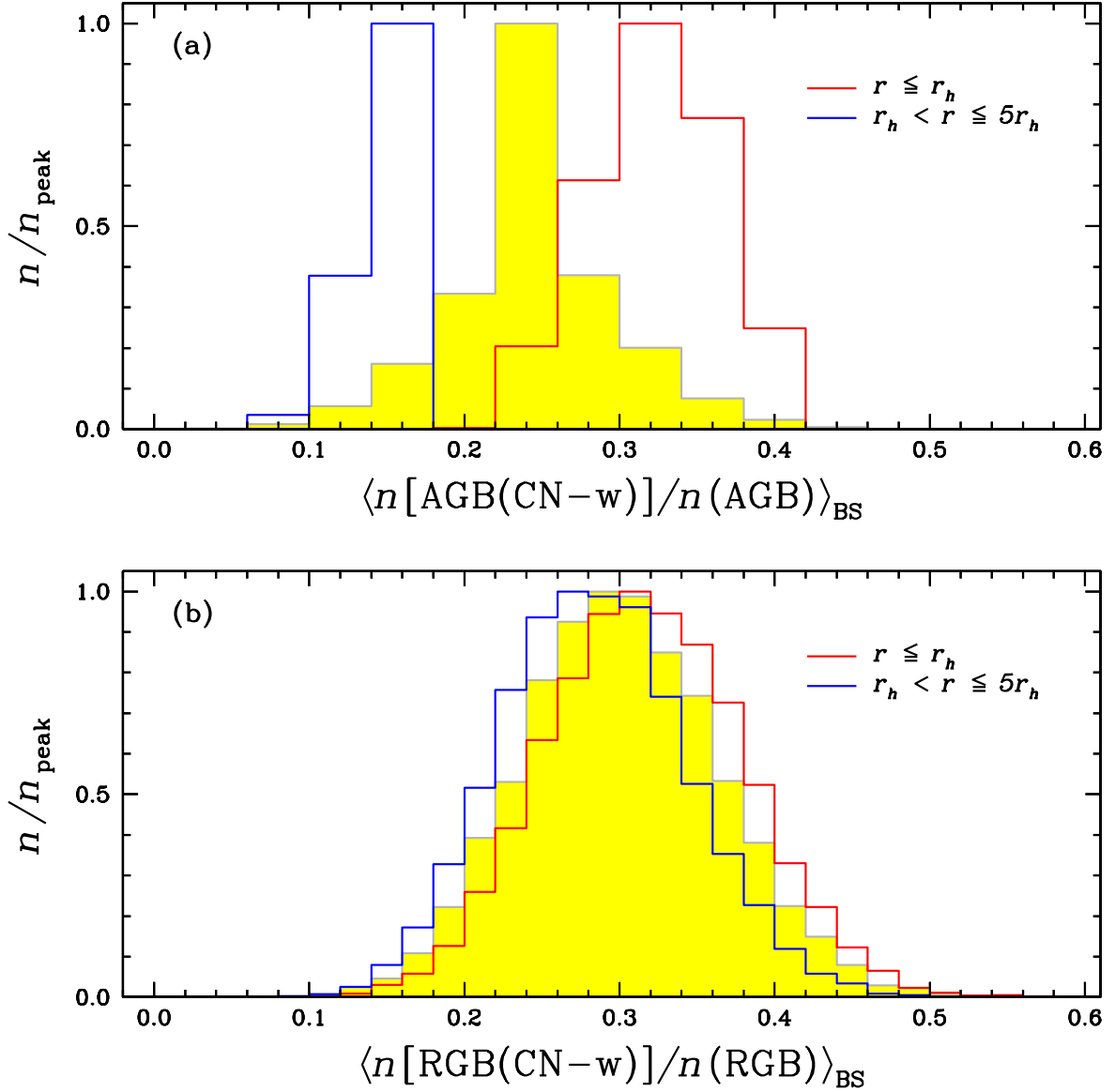}
\caption{
Empirical distributions of the mean value of relative fraction
of the \cnw\ populations
in the AGB (upper panel) and the RGB (lower panel) sequences from 
the bootstrap method.
The red solid lines are for the inner region ($r$ $\leq r_h$)
and the blue solid lines are for the outer region 
($r_h < r \leq 5r_h$) of the cluster.
The yellow shaded histograms are for all stars within 5$r_h$ from the center.
Note that the mean frequency of the \cnw\ AGB population in the outer part 
of the cluster is significantly smaller than that in the inner part,
most likely due to the stochastic truncation of the \cnw\ AGB stars
in the outer part of the cluster.
}\label{fig:boot}
\end{figure}

In order to see if the radial distribution of the \cnw\ AGB stars is 
intrinsically different from those of others, 
we also undertook a randomization test.
We calculated the empirical distributions of the mean value 
of relative fraction of the \cnw\ AGB and the \cnw\ RGB populations in
two radial zones, the inner region ($r \leq r_h$) 
and the outer region ($r_h < r \leq 5r_h$), and
we show our results in Table~\ref{tab:boot} and Figure~\ref{fig:boot}.
In the inner part of the cluster,
there appears to be no difference between the AGB and RGB stars
in the relative fraction of the \cnw\ population.
The mean values of the relative fraction of the \cnw\ population 
in the empirical distributions for the AGB and the RGB stars are
almost identical, 
0.302 $\pm$ 0.036 and 0.301 $\pm$ 0.063, respectively.
In the outer part of the cluster, however,
the the empirical distributions for the AGB and the RGB stars
are significantly different each other,
with the mean values of 0.123 $\pm$ 0.017 for the AGB
and 0.282 $\pm$ 0.061 for the RGB stars.
It should be reminded that our \cnw\ AGB fraction for the outer part
is exactly the same as what \citet{smith93} found for the cluster, \nrgb\ = 1:7.
The total distributions ($r \leq 5r_h$) for both phases
are marginally in agreement each other, with the mean values
of 0.221 $\pm$ 0.057 for the AGB and 0.291 $\pm$ 0.064 for the RGB stars.

The astrophysical origin of the deficiency of the \cnw\ AGB stars 
in the outer part of the cluster is not clear.
It is difficult to believe that the \cnw\ AGB stars in the outer region
of the cluster had been preferentially removed from the cluster
in the course of the relaxation process, for example.
It is naturally expected that the AGB stars simply do not have enough
time to be affected by the relaxation process due to their very short
life times on the AGB phase. 
Also they used to be more massive in the past than
the RGB stars at present time, which can lead the AGB stars to sink
to the central part of the cluster if the two-body relaxation took place.
The tidal effect from our Galaxy cannot explain the situation, 
since the tidal effect should have affected both the AGB and the RGB stars
at the same time, but the \cnw\ or the \cns\ RGB stars do not show any sign of
the deficiency in the outer region of the cluster.

Instead, we suspect that the fundamental and unavoidable stochastic 
truncation due to the small number of stars can explain the deficiency
of the \cnw\ AGB stars in the outer region of M5.
For example, adding three \cnw\ AGB stars in the outer region
of the cluster ($r_h < r \leq 5r_h$) can results in the mean value 
of 0.252 $\pm$ 0.083, dramatically mitigating the discrepancy.

Finally, our number statistics also indicates that the \agbm\ population in M5 
can be negligible, as already shown in Figure~\ref{fig:gratton}, 
seeing that the number ratios of the AGB to the RGB populations
for the \cns\ population is in excellent agreement 
with that of the \cnw\ population in the central part of the cluster.
Therefore, our results may set an observational constraint on the emergence
of the \agbm\ stars.

\section{SUMMARY AND DISCUSSIONS}
The understanding of the formation of the GCs with MSPs is one of 
the outstanding problems in the near field cosmology, however,
we are still in the period of the stamp collecting.
In this work, we provided very important observational lines of evidence to
shed more light on the true nature of the MSPs in GCs.

Throughout our decade long painstaking effort, 
we invented new filter systems, which allow us to measure the heavy 
and the CN abundances simultaneously.
Our new \cnjwl\ index can provide a very powerful means to probe
the MSPs in the RGB and the AGB stars in the Galactic GC systems.
Apparently, among other photometric indices being used in the ground-based
observations, our \cnjwl\ is the only reliable photometric index
that can accurately distinguishing MSPs in normal GCs.
Furthermore, our approach can complement the intrinsic weaknesses of
the powerful instruments that currently used in the field of the MSPs in GCs;
for example, the photometric study using the \hst\ (a small FOV) and 
the spectroscopic study using the \vlt\ (an incomplete sampling
in the central part of GCs).
The FOV of our typical instrument setup
is about 55 times larger than that from the WFC3/HST
and our method can easily be used in the central part of GCs.
The large FOV is critical in the statistical study of the MSPs in GCs
and in understanding the formation of such GCs.
For example, \citet{vesperini13} discussed that the local relative number
ratio between the MSPs can be different from the global number ratio
until the achievement of the complete mixing. 
Therefore the large area of the spatial coverage
is essential to ensure the correct population number ratios,
although M5 does not show any radial gradient 
in the population number ratio.

As a pilot work, we investigated the MSPs of the RGB and the AGB in M5, 
an archetype of GCs with a CN bimodality.
Our multi-color CMDs showed for the first time that our \cnjwl\ index 
exhibits the discrete double RGB and AGB sequences, while other indices 
known to trace the variations in the lighter elemental abundances, such as
$m1$ and $cy$, do only show some spreads and trends, 
but fail to trace the accurate light elemental abundances.
Using the EM algorithm, we obtained the number ratios of 
\nrgb\ = 29:71 ($\pm$ 2) and 21:79 ($\pm$ 7) for the RGB and the AGB, 
respectively, placing the \cns\ population the major component of the cluster.

By comparing with previous results from low resolution spectroscopy
by others \citep{briley92,smith93,smith13}, we showed that our \cnjwl\ index 
accurately traces the CN at \cnwave\ band absorption strengths so that
our \cnjwl\ index can be used to derive the photometric \scn\ or \ds\
and, furthermore, the photometric nitrogen abundances of RGB stars in M5.
Our statistical test showed that M5 most likely has 
a bimodal nitrogen distribution.
Using the linear correlation between our \cnjwl\ index versus [N/Fe] measurements
by \citet{briley92}, we derived the photometric nitrogen abundances
for individual RGB stars.
On the other hand, the photometric nitrogen abundances from other indices, 
$m1$ and $cy$, failed to reproduce the nitrogen abundance distribution 
by \citet{briley92}.

Comparisons with the previous high resolution spectroscopic studies
by \citet{ivans01} and \citet{carretta09} also confirmed that our \cnjwl\ index
is a very powerful means to explore the MSPs in the RGB and in the AGB.
The \cnw\ population from our photometry has higher oxygen
and lower sodium abundances both in the RGB and the AGB, while
the \cns\ population has lower oxygen and higher sodium abundances.
From the photometric point of view, there is no difference between 
the intermediate and the extreme populations defined by \citet{carretta09},
which was also pointed out by \citet{smith13} in their study of M5 
using the \ds.
Perhaps, this may be due to somewhat arbitrary definitions of
the intermediate and the extreme populations.

One of the astounding results is the discontinuities in 
the \cnjwl\ versus [O/Fe] and the \cnjwl\ versus [Na/Fe] relations 
(i.e., the discontinuous [N/Fe] versus [O/Fe]
and the [N/Fe] versus [Na/Fe] relations) between the \cnw\ and 
the \cns\ populations, while  no such discontinuity exists
in the Na-O anticorrelations in GCs \citep[e.g., see][]{carretta09}.
The large sample size in our study allows us to detect
these discontinuities more clearly than what \citet{smith13} 
first suspected in their limited number of sample.
Our careful re-examination of the results by \citet{yong08}
showed that NGC~6752 also exhibits such discontinuities 
in the [N/Fe] versus [O/Fe] and the [N/Fe] versus [Na/Fe] relations,
which will be presented in our forthcoming paper (Lee 2017a, in preparation).
The discontinuous chemical evolution in terms of the nitrogen abundances 
(while apparently continuous chemical evolution in terms 
of the oxygen and sodium abundances) in normal GCs
may pose a difficult but crucial constraint on the GC formation scenario.

In order to probe the helium abundance, which is less well-known in GC stars, 
we investigated the visual magnitude of the RGB bump,
finding that the \vbump\ magnitude of the \cns\ population
is slightly brighter than that of the \cnw\ population,
$\Delta$\vbump = 0.07 $\pm$ 0.04 mag.
If real, the difference in the helium abundance between the two populations
could be \dy\ = 0.028 $\pm$ 0.016, in the sense that the \cns\ population
is more helium enhanced. Then, the \cns\ RGB stars are likely the progenitor
of the blue HB populations in M5.

From the hiatus in our \cnjwl-[O/Fe] or the \cnjwl-[Na/Fe] relations,
it is thought that the \cns\ stars formed out of gas that already 
significantly experienced the proton capture processes at high temperature.
Regardless of the candidates, helium also should be supplied 
to the proto-stellar medium for the \cns\ population through such processes.
Up to date, no age differences between the two populations in M5
has been reported yet. 
However, judging from the currently available
HST photometry by \citet{anderson08}, the age difference between the two
populations should be very small, if any.

Our measurements of the centers of each population from the bright RGB stars
using various methods suggest that both populations appear to have  
nearly the same position of the center.
We examined the radial distributions of the two RGB populations, finding that
both populations are likely drawn from the same parent population,
which is also supported by the consistent SBPs of both populations.
However, the spatial distributions tell a different story.
The projected spatial distribution of the \cns\ RGB population
is more elongated along the NW-SE direction and, subsequently, 
has a significantly larger ellipticity than the \cnw\ RGB population.
The spatial distribution of the \cns\ RGB stars is apparently consistent 
with that of the \cns\ AGB stars, suggesting they share the same
structural and, furthermore, kinematical characteristics.

The more elongated spatial distribution of the \cns\ population
is in accord with the internal kinematics of the population.
We measured the projected rotations of each population and we obtained
that the \cns\ population has a substantial net projected rotation
while the \cnw\ population does not appear to show any net projected rotation.
The equator of the projected rotation (i.e., the axis perpendicular to
the rotation axis) and the major axis of the projected iso-density profile
of the \cns\ population are aligned well each other along the NW-SE direction,
suggesting that the rotation of the system is inextricably linked
to the ellipticity profile of the population \citep[see also][]{lee15}.

For the first time, we presented the complete view of
the MSPs of the AGB stars in M5.
Thanks to the capability of distinguishing MSPs
in the dense stellar environment,
our approach is easily applicable to the central part of the cluster,
where the spectroscopic method has an ultimate limitation due to crowdedness.
Furthermore, our method can open a new era for the MSPs of AGB populations
in GCs.

Using the number ratios of the AGB to the RGB stars by \citet{gratton10},
we showed that the probability of the emergence of the \agbm\ stars in M5
is negligibly small and, therefore, our measurements
for the relative AGB fraction will be a complete set of data.
We confirmed the earlier notion made by \citet{smith93}, who found
a deficiency of the \cnw\ AGB stars.
Our statistical tests indicated that the radial distribution of 
the  \cnw\ AGB stars are very different from other groups of stars
in the cluster.
We proposed that the stochastic truncation due to the small number
of the \cnw\ AGB stars in the outer part of the cluster is responsible
for the deficiency of the \cnw\ AGB stars in M5.

Unlike other peculiar GCs, like $\omega$ Cen and M22 
\citep{marino11ocen,marino11m22,lee15}, each stellar population 
in M5 based on our \cnjwl\ index does not exhibit its own Na-O anticorrelation.
Therefore, our results for the different spatial distributions and 
the projected rotations for the two populations in M5
may not support the merger scenario of the two normal GCs,
each of which has its own Na-O anticorrelation.
On the contrary, our results of the different physical properties
for each population can be understood from the view presented 
by \citet{bekki10}, who proposed that the SG of stars 
(i.e., the \cns\ population in our work in the context of the self-enrichment
scenario) formed from the gas expelled from the FG of stars 
(i.e., the \cnw\ population) in the central part of the cluster 
can have the flattened structure with a fast rotation.

The relaxation time scale at the half-light radius for M5
is less than 3 Gyr \citep{harris96} and it is significantly
smaller than the age of the cluster, 11.50 $\pm$ 0.25 Gyr \citep{vandenberg13}.
However, the recent $N$-body numerical simulations by \citet{vesperini13}, 
for example, showed that the time required to achieve the complete mixing 
can be about at least 20 half-mass relaxation time. 
During the course of the long-term dynamical evolution of GCs with MSPs,
any structural differences between the MSPs are expected 
to be gradually eliminated. From this view, the discrepancy 
between the radial and the spatial distributions of the MSPs
in M5 may pose a somewhat contradictory problem.
The similarity in the radial distributions of the both populations with
the flat number ratios against the radial distance and with the same positions
of the centers, can be a strong observational line of evidence 
of the complete mixing,
while the very different spatial distributions with different internal
kinematics may suggest that they are yet to be homogenized.

It is not clear if the homogenization in the radial distributions between
the MSPs works in different time scale than that in the spatial distributions
or the internal kinematics.
The rotation of the GC system is the least well-known subject.
Some numerical simulations suggest that the rotation of the GC system
gradually decreases with time due to the combined effect
of two-body relaxation and mass-loss \citep[e.g., see][]{wang16}.
The substantial net rotation of the \cns\ RGB population, while
no net rotation of the \cnw\ RGB population, could be 
a strong observational line of evidence of incomplete mixing. 

Perhaps, may the similarity in the radial distributions in both populations
tell that they were governed by the tidal effect induced mass-loss?
Our preliminary results for more than two dozen of normal GCs 
using our own ground-based photometry are against this hypothesis.
As will be presented in our forthcoming paper (Lee 2017b, in preparation),
we obtained the relative RGB number ratio between the MSPs of GC and
we found that there is no correlation between the relative number ratios
against the Galactocentric distances
and the spatial locations or the kinematic properties of individual GCs
in our Galaxy.

In the future, high precision photometry to examine the potential 
age differences between the two populations would be very desirable.
At the same time, more systematic and extensive radial velocity measurements 
would undoubtedly help to shed more light on understanding
the formation of M5.

\acknowledgements
J.-W.\ Lee acknowledges financial support from
the Basic Science Research Program (grant no. 2016-R1A2B4014741)
through the National Research Foundation of Korea (NRF)
funded by the Korea government (MSIP)
and from the Center for Galaxy Evolution Research (grant No. 2010-0027910).
J.-W.\ Lee thanks Drs. Sbordone, Piotto, Milone, Nardiello and Lardo
for providing synthetic spectra, \hst\ $UV$ and \sdss\  photometry for M5
and the anonymous referee for constructive comments.
Finally, special thanks must go to Bruce W. Carney, who inspired the author 
of the paper over the years.

\appendix

\section{COMPLETENESS TESTS}\label{ap:s:completeness}
In our earlier study of the peculiar globular cluster M22, we showed that 
our ground-based photometry is complete down to \vvhb\ $\approx$ 4.35 mag.
However, the apparent central crowdedness \citep[see][for the definition]{n6723} 
of  M5  is about 6 times larger than that of M22 \citep{lee15}
and the crowdedness of the central part of M5 could be a potential
problem for ground-based observations.

To examine the completeness of our photometry, we performed a series 
of artificial star experiments.
We generated 100 artificial star images using
a FORTRAN program to distribute 
200 stars in the inner region ($r \leq r_h$), 
1000 stars in the intermediate region ($r_h < r \leq 3r_h$)
and 1500 stars in the outer field ($3r_h < r \leq 5r_h$), 
by adopting the observed radial stellar number density and 
the luminosity profile in our study. 
The number of artificial stars added to the observed images is carefully
chosen so as not to dramatically change the crowdedness characteristics
between our data reduction procedures and our artificial star experiments.

We used the same data reduction procedures that we described 
in \S\ref{s:reduction} and we derived the completeness fractions 
as a function of $V$ magnitude.
Our results are shown in Figure~\ref{fig:completeness}.
Our experiments suggest that our photometry is complete down to $V$ = 18.0
and 19.0 mag for the intermediate and the outer regions of M5, respectively.
On the other hand, owing to the rather large apparent central crowdedness
of M5, our photometry for the inner part of the cluster becomes
incomplete at as bright as $V$ $\approx$ 16.0 mag, which is equivalent to 
\vvhb\ $\approx$ 1.0 mag.
The incomplete detection of stars in the very dense central part of GCs
is the intrinsic weakness of any ground-based photometry.
However, we emphasize that incomplete detection of stars in the central part of M5 
does not affect our results presented in this work, since 
the radial distribution of the number ratio between the \cnw\ and the \cns\ 
remains flat as we already showed in Figure~\ref{fig:rad}.

\section{NEW STRATEGY OF DATA REDUCTIONS OF THE GROUND-BASED PHOTOMETRY
USING THE PRIOR POSITIONAL INFORMATION FROM HST ACS PHOTOMETRY}\label{ap:s:acs}
\citet{anderson08} presented homogeneous \hst\ \acs\ photometry for 65
Galactic GCs. 
Assuming their photometry is more complete than any other ground-based 
photometry, their results may provide a very useful means to examine 
the validity of our completeness tests using artificial stars presented above.

In Figure~\ref{fig:withoutACS}, we show the CMDs and positions of stars in M5
from \hst\ \acs\ photometry by \citet{anderson08}.
In panels (a--b), we show stars matched with our ground-based observations,
while panels (c--d) are for stars not detected in our measurements.
We also derived the completeness fraction at each magnitude bins and
we show our results in panel (e).

\begin{figure}
\epsscale{1}
\figurenum{25}
\plotone{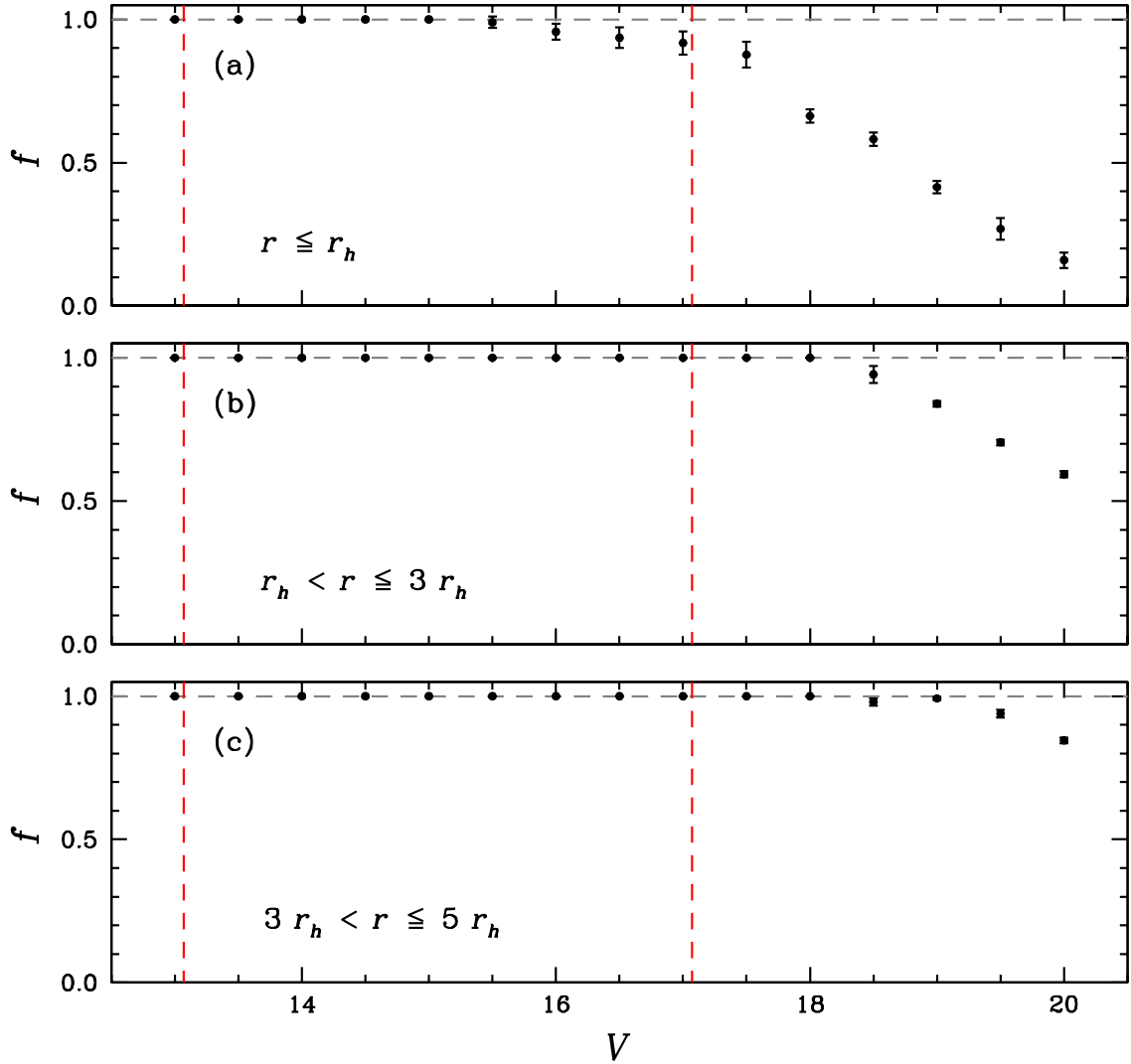}
\caption{
Completeness fractions at each magnitude bins measured from artificial
star experiments for the inner, intermediate and outer regions of M5.
The vertical dashed lines denote \vvhb\ = $\pm$2.0 mag,
with $V_{\rm HB}$ = 15.07 mag for M5.
Our experiments suggest that our photometry is complete down to $V$ = 18.0
and 19.0 mag for the intermediate and the outer regions of M5, respectively.
On the other hand, owing to the rather large apparent central crowdedness
of M5, our photometry for the inner part of the cluster becomes
incomplete at $V$ $\approx$ 16.0 mag, which is equivalent to 
\vvhb\ $\approx$ 1.0 mag.
}\label{fig:completeness}
\end{figure}

\begin{figure}
\epsscale{1}
\figurenum{26}
\plotone{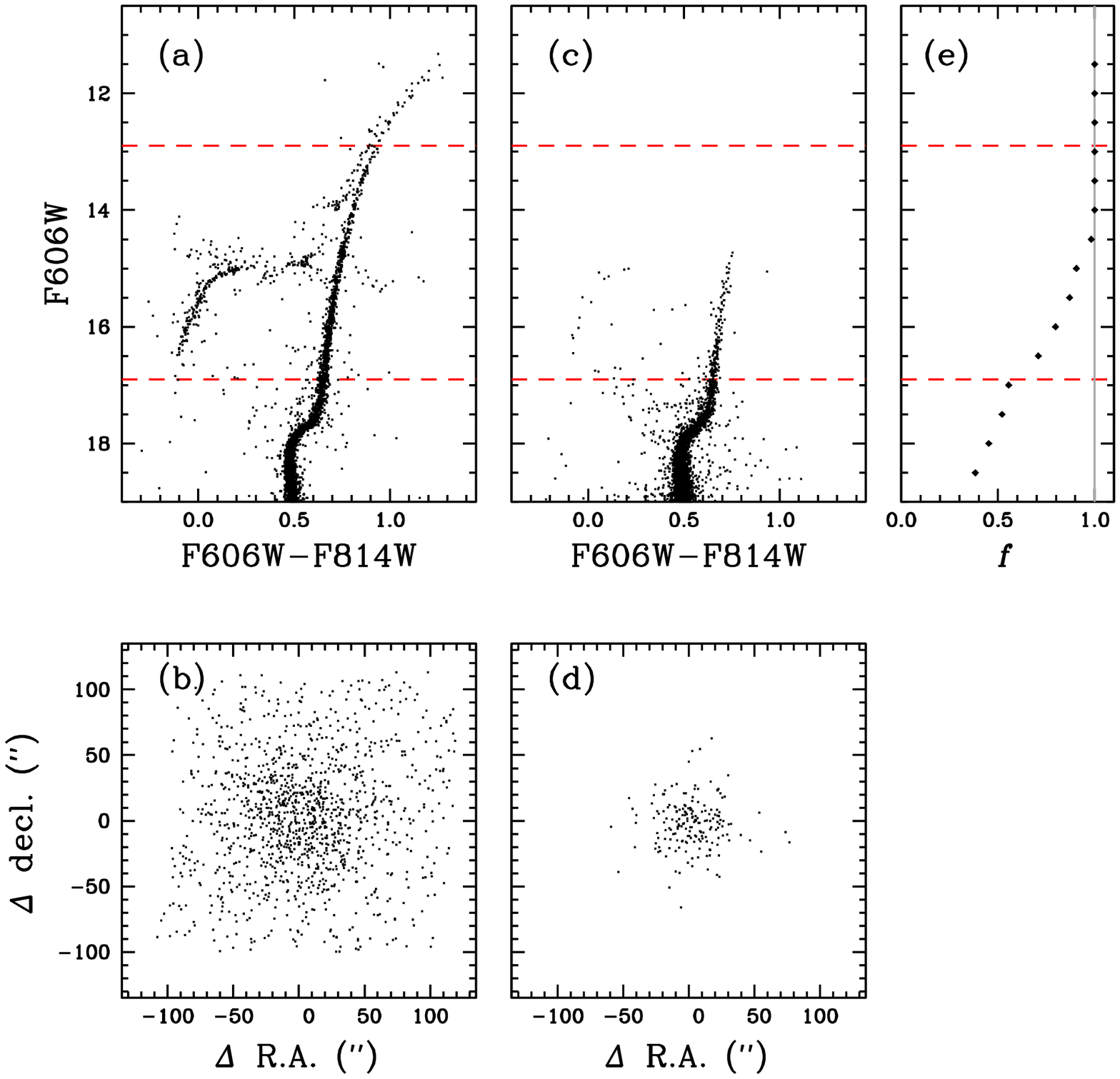}
\caption{
(a) The M5 \hst\ \acs\ CMD \citep{anderson08}
using stars detected in our ground-based photometry.
The horizontal red dashed lines denote \hsthb\ = $\pm$2.0 mag
with F606W$_{\rm HB}$ = 14.90 \citep{dotter10}.
(b) Positions of stars in (a) with $-2$ $\leq$ \hsthb\ $\leq$ +2 mag.
(c) The M5 \hst\ \acs\ CMD using stars not detected in our ground-based photometry.
(d) Positions of stars in (c) with $-2$ $\leq$ \hsthb\ $\leq$ +2 mag.
(e) Completeness fractions at each magnitude bins measured from 
the number of stars not detected in our ground-based photometry.
}\label{fig:withoutACS}
\end{figure}

Note that the effective FOV of the \hst\ \acs\ survey program is about
3$\times$3\arcmin, equivalent to the area with $r \lesssim r_h$ for M5.
If our artificial star experiments correctly reflect the completeness 
fraction in M5, the completeness fraction shown in Figure~\ref{fig:withoutACS}(e)
should be the same as Figure~\ref{fig:completeness}(a).
However, our results may indicate that this is not the case.
We carefully re-examined every step in our data reductions and the artificial 
star experiments, and we confirmed that our procedures are correct.
We suspect that the widely practiced artificial star experiments 
with the empirical point-spread functions
may overestimate the completeness fractions.

\begin{figure}
\epsscale{1}
\figurenum{27}
\plotone{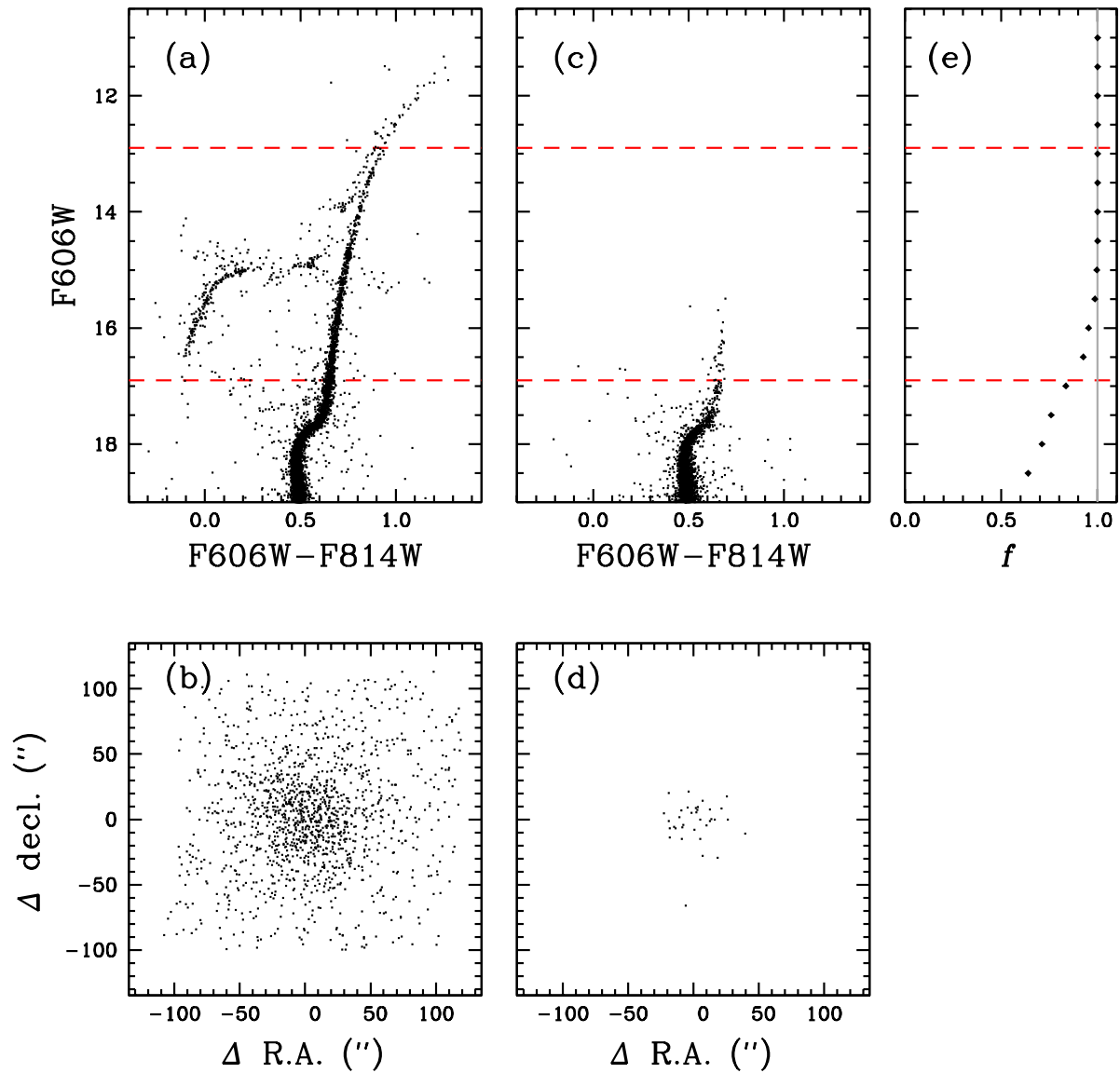}
\caption{
Same as Figure~\ref{fig:withoutACS}, but using our new ground-based
photometry making use of the positions of the stars of the \hst\ \acs\ photometry
by \citet{anderson08}.
Note that our approach can significantly improve the detection rate in the 
central part of the cluster.
Due to the unavoidable seeing effect in any ground-based observations,
our photometry is still incomplete at the magnitude of our interest, 
$-2 \leq$ \vvhb\ $\leq$ 2.0 mag. 
However, incomplete detection of stars in the central
part of the cluster does not affect our results presented in this work.
}\label{fig:withACS}
\end{figure}

In order to improve our photometry,
we updated the input star list by merging the star list returned from
our ALLFRAME run and the \hst\ \acs\ star list not detected in our
ground-based observations, i.e., stars in Figure~\ref{fig:withoutACS}(c).
We re-run ALLFRAME with this updated input star list and we show our
measurements in Figure~\ref{fig:withACS}.
As shown, our approach can significantly improve the detection rate 
in the central part of M5.
However, the detection rate is still not complete
in the magnitude level of our interest, $-$2 $\leq$ \vvhb\ $\leq$ 2 mag,
equivalent to 13.07 $\leq V \leq$ 17.07 mag.
It is thought that this incomplete detection of stars from the extensive
input star list mainly due the treatment of blended stars in ALLFRAME.
The critical separation parameter in ALLFRAME is set to be 0.375$\times$FWHM
and any stars within this distance are considered to be blended.
Therefore, the many stars added in our updated list are treated
to be blended with pre-detected stars from our previous ALLFRAME run.

\begin{figure}
\epsscale{1}
\figurenum{28}
\plotone{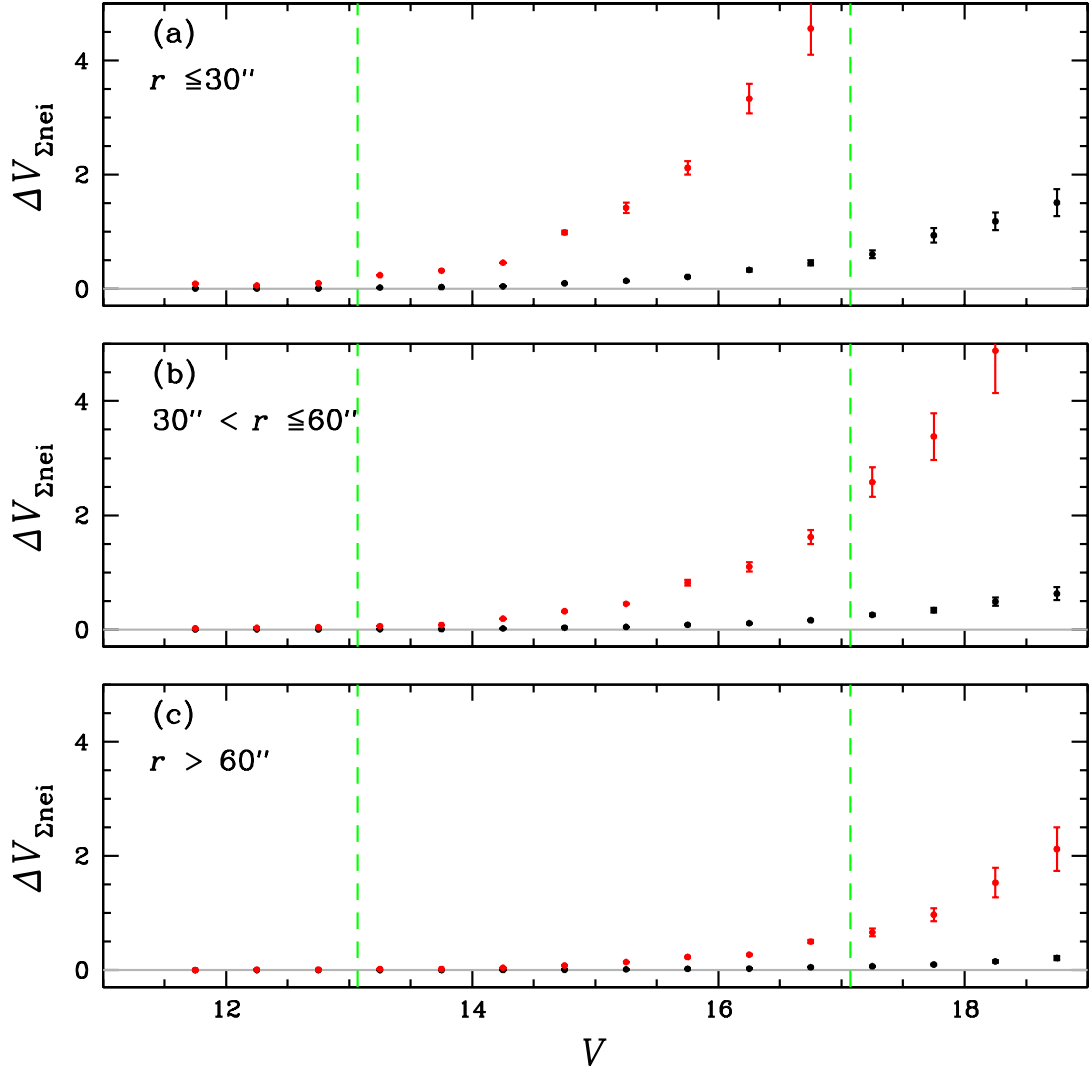}
\caption{
Estimation of the influence of the undetected nearby stars 
on the ground-based photometry.
The black dots and error bars denote the mean magnitudes with $\pm 1\sigma$ 
of the sum of the surface brightness
produced by nearby undetected stars in the ground-based observations.
The red dots and error bars are for those ten times of black ones.
The vertical green dashed lines denote \vvhb\ = $\pm$2.0 mag.
In the inner part of M5 ($r \leq 1\arcmin$), the residual flux from 
the undetected nearby stars supplemented from the \hst\ \acs\ photometry
could cause the excess in the $V$ magnitude of $\approx$ 0.08 mag
at \vhb\ in our ground-based photometry.
In practice, the undetected nearby stars do not appear to greatly affect
the magnitude measurements as we already showed in Table~\ref{tab:bump}.
Instead, they increase the background brightness level of the central part
and their net effect on the photometric measurements is almost nil
if a sky background subtraction is properly applied.
}\label{fig:magaffect}
\end{figure}

We estimated the influence of
the undetected stars on the ground-based photometry.
We calculated the contribution in the surface flux 
from the individual undetected nearby stars.
During our calculation we adopted a Moffat-type 
model point-spread function \citep{moffat69,pbs03},
\begin{equation}
 I(r) = \frac{I_0}{[1 + (\frac{r}{\alpha})^2]^2},
\end{equation}
where $\alpha$ = 0.7769$\times$FWHM and we adopted the FWHM of 1\farcs2.
As shown in Figure~\ref{fig:magaffect}, our simulations suggest that 
the residual flux from the undetected nearby stars
could cause a magnitude excess of \vexcess\ $\approx$ 0.01 mag
at \vvhb\ = $-$2 mag, $\approx$ 0.08 mag at \vhb, and
$\approx$ 0.35 at \vvhb\ = +2 mag.
However, some caution is advised when interpreting the results of our simulations.
The magnitude excess due to the undetected nearby stars, \vexcess,
shown in the figure is what one would 
get without a proper background brightness subtraction.
This could cause a serious and mostly irreparable problem in spectroscopy.
In photometry, however, the undetected nearby stars in a dense environment
are treated as diffuse background sources and, as a consequence, 
they increase the background brightness.
The sky subtraction algorithm implemented in ALLFRAME appears to work great,
judging by the fact that
our \vbump\ measurements do not show any significant radial fluctuations
as shown in Table~\ref{tab:bump}. 
It should be reminded that the \vbump\ magnitudes between
in the central and in the outer part of M5 are in excellent agreement
to within 0.01 mag.\footnote{Our previous study of the RR Lyrae variables
in NGC~6723 also confirms our result presented here. 
See Figure~17 of \citet{n6723} for the constancy of the RR Lyrae magnitudes 
in NGC~6723 against the radial distance from the center.}

\begin{figure}
\epsscale{1}
\figurenum{29}
\plotone{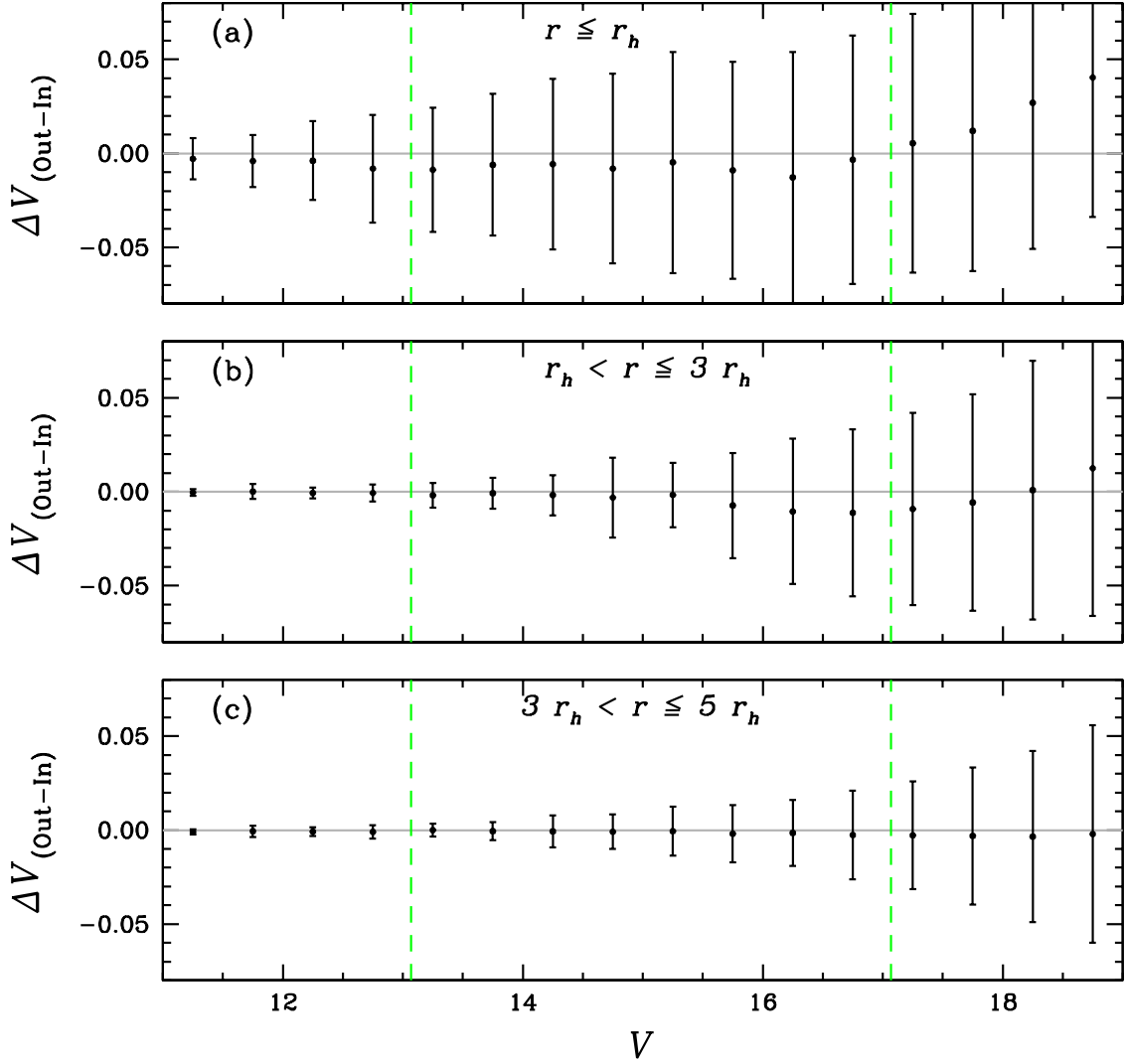}
\caption{
Comparisons of the input magnitudes to the output magnitudes from
our artificial star experiments.
The black dots and error bars denote the mean magnitudes with $\pm 1\sigma$ 
of the difference in magnitude.
The vertical green dashed lines denote \vvhb\ = $\pm$2.0 mag.
Our experiments show that the differences between 
the input and output magnitudes are spatially independent at the magnitude
of our interest, $-2 \leq$ \vvhb\ $\leq$ 2.0 mag, 
although the standard deviations of the mean in the inner part of M5 
($r \leq 1\arcmin$) are larger than those in the outer part.
}\label{fig:fakemag}
\end{figure}

Finally, a more robust test for the influence of the undetected stars
can be found in Figure~\ref{fig:fakemag},
where we show comparisons of the input magnitudes to the output magnitudes 
from our artificial star experiments.
Our experiments show that the differences between 
the input and output magnitudes are spatially independent at the magnitude
of our interest, $-2 \leq$ \vvhb\ $\leq$ 2.0 mag,
although the standard deviations of the mean in the inner part of M5 
($r \leq 1\arcmin$) are larger than those in the outer part.

We conclude that the incomplete detection of stars in the central part
of M5 does not affect the results presented here;
The number ratios between the two populations remain the same 
thanks to the absence of any radial variations of the number ratios.
At the same time, the residual flux from the undetected nearby stars
is well taken care of by ALLFRAME and does not hardly affect our magnitude
measurements, such as the \vbump\ magnitudes of the both populations.

\clearpage

\begin{landscape}
\begin{deluxetable}{ccrrrrrrrrrrrrrrrrr}
\tablecaption{Spectral resolving powers for selected UV filters.\label{tab:filter}}
\tablenum{1}
\tablewidth{0pc}
\tablehead{
\multicolumn{1}{c}{} &
\multicolumn{1}{c}{} &
\multicolumn{2}{c}{Johnson} &
\multicolumn{1}{c}{} &
\multicolumn{3}{c}{\hst} &
\multicolumn{1}{c}{} &
\multicolumn{2}{c}{\sdss} &
\multicolumn{1}{c}{} &
\multicolumn{1}{c}{Wash.} &
\multicolumn{1}{c}{} &
\multicolumn{5}{c}{Extended \str} \\
\cline{3-4}\cline{6-8}\cline{10-11}\cline{13-13}\cline{15-19}
\multicolumn{1}{c}{} &
\multicolumn{1}{c}{} &
\multicolumn{1}{c}{$U$} &
\multicolumn{1}{c}{$B$} &
\multicolumn{1}{c}{} &
\multicolumn{1}{c}{F275W} &
\multicolumn{1}{c}{F336W} &
\multicolumn{1}{c}{F438W} &
\multicolumn{1}{c}{} &
\multicolumn{1}{c}{$u$} &
\multicolumn{1}{c}{$g$} &
\multicolumn{1}{c}{} &
\multicolumn{1}{c}{$C$} &
\multicolumn{1}{c}{} &
\multicolumn{1}{c}{$u$} &
\multicolumn{1}{c}{$v$} &
\multicolumn{1}{c}{$Ca_{\rm new}$} &
\multicolumn{1}{c}{$JWL39$} &
\multicolumn{1}{c}{\cnjwl\tablenotemark{\dagger}} 
}
\startdata
$\lambda_{\rm c}$ & (nm) 
       & 367 & 436 && 275 & 337 &  432 && 354 & 477 && 391 && 349 &   411 &  395 & 390 &  388 \\
$\Delta\lambda$  & (nm) 
       & 66  & 94  &&  50 &  55 &   70 &&  57 & 137 && 110 && 30 &   19 &    9 &   18 &    9 \\
$\lambda_{\rm c}/\Delta\lambda$ & 
       & 5.5 & 4.6 && 5.5 & 6.1 &  6.1 && 6.2 & 3.5 && 3.6 && 11.6 & 21.6 & 43.8 & 21.7 &  43.1 \\
\enddata
\tablenotetext{\dagger}{\cnjwl = $JWL39 - Ca_{\rm new}$}
\end{deluxetable}
\end{landscape}

\clearpage

\begin{deluxetable}{cccccccccc}
\tablecaption{Integration times (s).\label{tab:obs}}
\tablenum{2}
\tablewidth{0pc}
\tablehead{
\multicolumn{5}{c}{CTIO\tablenotemark{*}} &
\multicolumn{1}{c}{} &
\multicolumn{4}{c}{New Filters} \\
\cline{1-5}\cline{7-10}
\multicolumn{1}{c}{$y$} &
\multicolumn{1}{c}{$b$} &
\multicolumn{1}{c}{$v$} &
\multicolumn{1}{c}{$u$} &
\multicolumn{1}{c}{$Ca$} &
\multicolumn{1}{c}{} &
\multicolumn{1}{c}{$y$\tablenotemark{\dagger}} &
\multicolumn{1}{c}{$b$\tablenotemark{\dagger}} &
\multicolumn{1}{c}{$Ca_{\rm new}$\tablenotemark{\dagger}} &
\multicolumn{1}{c}{$JWL39$\tablenotemark{\ddagger}} 
}
\startdata
2630 & 4010 & 2100 & 3900 & 9660 && 3785 & 8900 & 37400 & 13000 \\
\enddata
\tablenotetext{*}{4 $\times$ 4 inches filters provided by the CTIO, 
which have been used from 2007 to 2010 seasons.}
\tablenotetext{\dagger}{4 $\times$ 4 inches filters designed and owned by J.-W. Lee,
which have been used from 2011 to 2014 seasons.}
\tablenotetext{\ddagger}{A 4 $\times$ 4 inches filter designed and owned by J.-W. Lee,
which has been used from 2013 to 2014 seasons.}
\end{deluxetable}

\clearpage

\begin{deluxetable}{crrrrrr}
\tablecaption{
Coefficients and the goodness of the fit between 
$S(3839)$ [and \ds] by \citet{briley92} and \cnjwl, $cy$ and $m1$.
\label{tab:deltas_briley}}
\tablenum{3}
\tablewidth{0pc}
\tablehead{
\multicolumn{1}{c}{Index} &
\multicolumn{1}{c}{} &
\multicolumn{1}{c}{Slope\tablenotemark{\dagger}} &
\multicolumn{1}{c}{Intercept\tablenotemark{\dagger}}&
\multicolumn{1}{c}{$p$-value}&
\multicolumn{1}{c}{$\rho$\tablenotemark{\ddagger}}
}
\startdata
\cnjwl & \scn  & 4.201 $\pm$ 0.907 & 0.512 $\pm$ 0.063 & 0.000 & 0.757 \\
       & \ds   & 4.112 $\pm$ 0.369 & 0.463 $\pm$ 0.025 & 0.000 & 0.941 \\
&&&&& \\
$cy$   & \scn  & 0.051 $\pm$ 1.129 & 0.262 $\pm$ 0.242 & 0.965 & 0.011 \\
       & \ds   & 1.063 $\pm$ 0.848 & 0.432 $\pm$ 0.182 & 0.228 & 0.299 \\
&&&&& \\
\dcy\tablenotemark{1}
       & \scn  & 2.596 $\pm$ 0.766 & 0.572 $\pm$ 0.086 & 0.003 & 0.647 \\
       & \ds   & 1.471 $\pm$ 0.699 & 0.361 $\pm$ 0.079 & 0.052 & 0.466 \\
&&&&& \\
$m1$   & \scn  & 2.428 $\pm$ 0.613 & $-$0.222 $\pm$ 0.006 & 0.001 & 0.704 \\
       & \ds   & 0.924 $\pm$ 0.639 & 0.030 $\pm$ 0.129 & 0.172 & 0.337 \\
&&&&& \\
\dmo\tablenotemark{2}
       & \scn  & $-$0.491 $\pm$ 1.724 & 0.276 $\pm$ 0.096 & 0.780 & $-$0.071 \\
       & \ds   & 1.173 $\pm$ 1.328 & 0.150 $\pm$ 0.074 & 0.390 & 0.216 \\
\enddata
\tablenotetext{\dagger}{$S(3839)$ or \ds\ = Intercept + Slope $\times$ Index.}
\tablenotetext{\ddagger}{Pearson's correlation coefficient.}
\tablenotetext{1}{\dcy\ = $cy - (-4.030 - 0.480V - 0.015V^2$)}
\tablenotetext{2}{\dmo\ = $m1 - (3.050 - 0.329V + 0.009V^2$)}
\end{deluxetable}

\clearpage

\begin{deluxetable}{ccrrrrr}
\tablecaption{Coefficients and the goodness of the fit between 
selected color indices and the nitrogen abundance.
\label{tab:fit}}
\tablenum{4}
\tablewidth{0pc}
\tablehead{
\multicolumn{1}{c}{Index} &
\multicolumn{1}{c}{$V$ level} &
\multicolumn{4}{c}{[N/Fe]} \\
\cline{3-6}
\multicolumn{1}{c}{} &
\multicolumn{1}{c}{} &
\multicolumn{1}{c}{Slope\tablenotemark{\dagger}} &
\multicolumn{1}{c}{Intercept\tablenotemark{\dagger}}&
\multicolumn{1}{c}{$p$-value}&
\multicolumn{1}{c}{$\rho$\tablenotemark{\ddagger}}
}
\startdata
\cnjwl & bright & 11.887 $\pm$ 2.543 & 1.628 $\pm$ 0.183 & 0.009 & 0.919 \\
       & faint  & 14.544 $\pm$ 1.757 & 1.626 $\pm$ 0.124 & 0.000 & 0.940 \\ 
       & all    & 12.925 $\pm$ 1.576 & 1.562 $\pm$ 0.110 & 0.000 & 0.899 \\ 
 & & & & & \\
$cy$ & bright & 28.569 $\pm$ 9.798 & 7.607 $\pm$ 2.312 & 0.043 & 0.825 \\ 
     & faint  &  4.406 $\pm$ 3.308 & 1.569 $\pm$ 0.670 & 0.217 & 0.406 \\ 
     & all    &  3.433 $\pm$ 2.795 & 1.483 $\pm$ 0.599 & 0.237 & 0.294 \\ 
 & & & & & \\
$m1$ & bright & 13.134 $\pm$ 5.743 & $-$2.520 $\pm$ 1.489 & 0.084 & 0.753 \\ 
     & faint  &  6.860 $\pm$ 6.599 & $-$0.407 $\pm$ 1.072 & 0.326 & 0.327 \\
     & all    &  3.316 $\pm$ 2.075 &    0.113 $\pm$ 0.418 & 0.130 & 0.371 \\
\enddata
\tablenotetext{\dagger}{[N/A] = Intercept + Slope $\times$ color index.}
\tablenotetext{\ddagger}{Pearson's correlation coefficient.}
\end{deluxetable}

\clearpage

\begin{deluxetable}{rrrrrr}
\tablecaption{Coefficients and the goodness of the fit between 
$S(3839)$ [and \ds] by \citet{smith13} and \cnjwl.
\label{tab:deltas}}
\tablenum{5}
\tablewidth{0pc}
\tablehead{
\multicolumn{1}{c}{} &
\multicolumn{1}{c}{Slope\tablenotemark{\dagger}} &
\multicolumn{1}{c}{Intercept\tablenotemark{\dagger}}&
\multicolumn{1}{c}{$p$-value}&
\multicolumn{1}{c}{$\rho$\tablenotemark{\ddagger}}
}
\startdata
\scn   & 5.196 $\pm$ 0.417 & 0.738 $\pm$ 0.029 & 0.000 & 0.892 \\
\ds    & 5.609 $\pm$ 0.435 & 0.641 $\pm$ 0.030 & 0.000 & 0.898 \\
\enddata
\tablenotetext{\dagger}{$S(3839)$ or \ds\ = Intercept + Slope $\times$ \cnjwl.}
\tablenotetext{\ddagger}{Pearson's correlation coefficient.}
\end{deluxetable}

\clearpage

\begin{deluxetable}{crr}
\tablecaption{RGB bump $V$ magnitudes
\label{tab:bump}}
\tablenum{6}
\tablewidth{0pc}
\tablehead{
\multicolumn{1}{c}{} &
\multicolumn{1}{c}{\cnw} &
\multicolumn{1}{c}{\cns}
}
\startdata
$r \leq 1\arcmin$   & 15.032  $\pm$ 0.030 & 14.963 $\pm$ 0.030\\
$r > 1\arcmin$      & 15.043  $\pm$ 0.030 & 14.973 $\pm$ 0.030\\
all                 & 15.038  $\pm$ 0.030 & 14.970 $\pm$ 0.030\\
\enddata
\end{deluxetable}

\clearpage

\begin{deluxetable}{ccrrcrrcrr}
\tablecaption{Centers of M5 measured from stars within $r \leq 3r_h$.\label{tab:cnt}}
\tablenum{7}
\tablewidth{0pc}
\tablehead{
\multicolumn{1}{c}{} &
\multicolumn{1}{c}{} &
\multicolumn{2}{c}{Simple mean} &
\multicolumn{1}{c}{} &
\multicolumn{2}{c}{Half-sphere} &
\multicolumn{1}{c}{} &
\multicolumn{2}{c}{Pie-slice }\\
\cline{3-4}\cline{6-7}\cline{9-10}
\colhead{} & \colhead{} & 
\colhead{$\Delta\alpha$(\arcsec) } & \colhead{$\Delta\delta$(\arcsec)} &
\colhead{} &
\colhead{$\Delta\alpha$(\arcsec) } & \colhead{$\Delta\delta$(\arcsec)} &
\colhead{} &
\colhead{$\Delta\alpha$(\arcsec) } & \colhead{$\Delta\delta$(\arcsec)} }
\startdata
All    & &    0.6 &  2.7 &&   2.6 &   3.1 &&    0.6 &  2.5 \\
\cnw\  & &    8.2 &  5.1 &&   9.5 &   0.6 &&    6.5 &  0.5 \\
\cns\  & & $-$2.5 &  3.2 &&   0.6 &   1.5 && $-$4.2 &  1.3 \\
\enddata
\end{deluxetable}

\clearpage

\begin{deluxetable}{lrrrr}
\tablecaption{Ellipse fitting parameters for M5 RGB stars.
\label{tab:ellipse}}
\tablenum{8}
\tablewidth{0pc}
\tablehead{
\multicolumn{1}{c}{} &
\multicolumn{1}{c}{grid} &
\multicolumn{1}{c}{$\theta$} &
\multicolumn{1}{c}{$b/a$ } &
\multicolumn{1}{c}{$e$}}
\startdata
\cnw\   & 0.9 &   91.8 $\pm$  4.1 &  0.970 $\pm$  0.049 &  0.030 \\
        & 0.7 &   96.7 $\pm$  3.5 &  0.979 $\pm$  0.042 &  0.021 \\
        & 0.5 &   18.2 $\pm$  2.8 &  0.969 $\pm$  0.034 &  0.031 \\
        & 0.3 &   28.2 $\pm$  2.1 &  0.962 $\pm$  0.025 &  0.038 \\
        & & & &  \\
\cns\   & 0.9 &   22.2 $\pm$  4.0 &  0.926 $\pm$  0.046 &  0.074 \\
        & 0.7 &   23.2 $\pm$  3.4 &  0.915 $\pm$  0.039 &  0.085 \\
        & 0.5 &   25.0 $\pm$  2.7 &  0.899 $\pm$  0.030 &  0.101 \\
        & 0.3 &   28.7 $\pm$  1.9 &  0.879 $\pm$  0.021 &  0.121 \\
\enddata
\end{deluxetable}

\clearpage

\begin{landscape}
\begin{deluxetable}{ccrrrrrrrrrrr}
\tablecaption{Goodness of the fits between the relative AGB frequencies
and other parameters.
\label{tab:gratton}}
\tablenum{9}
\tablewidth{0pc}
\tablehead{
\multicolumn{1}{c}{} &
\multicolumn{1}{c}{$\langle \frac{n{\rm(AGB)}}{n{\rm(RGB)}}\rangle$} &
\multicolumn{2}{c}{HB Type} &
\multicolumn{1}{c}{} &
\multicolumn{2}{c}{[Fe/H]} &
\multicolumn{1}{c}{} &
\multicolumn{2}{c}{$M_V$} &
\multicolumn{1}{c}{} &
\multicolumn{2}{c}{$M_{\rm min}$} \\
\cline{3-4}\cline{6-7}\cline{9-10}\cline{12-13}
\multicolumn{1}{c}{} &
\multicolumn{1}{c}{} &
\multicolumn{1}{c}{$p$-value}&
\multicolumn{1}{c}{$r$\tablenotemark{\dagger}} &
\multicolumn{1}{c}{} &
\multicolumn{1}{c}{$p$-value}&
\multicolumn{1}{c}{$r$\tablenotemark{\dagger}} &
\multicolumn{1}{c}{} &
\multicolumn{1}{c}{$p$-value}&
\multicolumn{1}{c}{$r$\tablenotemark{\dagger}} &
\multicolumn{1}{c}{} &
\multicolumn{1}{c}{$p$-value}&
\multicolumn{1}{c}{$r$\tablenotemark{\dagger}}
}
\startdata
21 GCs\tablenotemark{1}  & 0.084 $\pm$ 0.017 
  & 0.005 & $-$0.589 && 0.015 & 0.522 && 0.213 & 0.284 && 0.000 & 0.750 \\
20 GCs\tablenotemark{2}  & 0.086 $\pm$ 0.017 
  & 0.000 & $-$0.731 && 0.003 & 0.635 && 0.421 & 0.191 && 0.000 & 0.714 \\
\enddata
\tablenotetext{\dagger}{Pearson's correlation coefficient.}
\tablenotetext{1}{21 GCs studied using the HST by \citet{gratton10}}
\tablenotetext{2}{Without NGC~2808}
\end{deluxetable}
\end{landscape}

\clearpage

\begin{deluxetable}{rcc}
\tablecaption{Number ratios of the AGB to the RGB populations.
\label{tab:agbrgb}}
\tablenum{10}
\tablewidth{0pc}
\tablehead{
\multicolumn{1}{c}{} &
\multicolumn{1}{c}{$\frac{n[{\rm AGB(\cnw)}]}{n{\rm[RGB(\cnw)]}}$}&
\multicolumn{1}{c}{$\frac{n[{\rm AGB(\cns)}]}{n{\rm[RGB(\cns)]}}$} 
}
\startdata
$r \leq r_h$        & 0.048 $\pm$ 0.015 & 0.048 $\pm$ 0.010 \\
$r_h < r \leq 5r_h$ & 0.028 $\pm$ 0.014 & 0.056 $\pm$ 0.012 \\
$r \leq 5r_h$       & 0.035 $\pm$ 0.009 & 0.053 $\pm$ 0.007 \\
\enddata
\end{deluxetable}

\clearpage

\begin{deluxetable}{rcc}
\tablecaption{Relative fraction of the \cnw\ population
from the bootstrap realization.
\label{tab:boot}}
\tablenum{11}
\tablewidth{0pc}
\tablehead{
\multicolumn{1}{c}{} &
\multicolumn{1}{c}{$\langle\frac{n[{\rm AGB(\cnw)}]}{n{\rm(AGB)}}\rangle_{\rm BS}$} &
\multicolumn{1}{c}{$\langle\frac{n[{\rm RGB(\cnw)}]}{n{\rm(RGB)}}\rangle_{\rm BS}$} 
}
\startdata
$r \leq r_h$        & 0.302 $\pm$ 0.036 & 0.307 $\pm$ 0.064 \\
$r_h < r \leq 5r_h$ & 0.123 $\pm$ 0.017 & 0.273 $\pm$ 0.060 \\
$r \leq 5r_h$       & 0.221 $\pm$ 0.057 & 0.291 $\pm$ 0.064 \\
\enddata
\end{deluxetable}

\clearpage

\clearpage

\clearpage

\clearpage

\clearpage

\clearpage

\clearpage

\clearpage

\clearpage

\clearpage

\clearpage

\clearpage

\clearpage

\clearpage

\clearpage

\clearpage

\clearpage

\clearpage

\clearpage

\clearpage

\clearpage

\clearpage

\clearpage

\clearpage

\clearpage

\clearpage

\clearpage

\clearpage

\clearpage

\end{document}